\documentclass[aps,preprint,showpacs,preprintnumbers,amsmath,amsbsy]{revtex4}

\usepackage{amsmath}
\usepackage{amsbsy}
\usepackage{pstricks,graphicx}
\newcommand{\bra}[1]{\ensuremath{\langle #1|}}
\newcommand{\ket}[1]{\ensuremath{|#1\rangle}}
\newcommand{\bracket}[2]{\ensuremath{\langle #1|#2\rangle}} 
\newcommand{\dirint}[3]{\ensuremath{\langle #1|#2|#3\rangle}}
\newcommand{\kett}[1]{\left|#1\right>}
\newcommand{\brac}[1]{\left<#1 \right|}
\newcommand{\brackett}[2]{\left< #1 | #2 \right>}

\newcommand{\bs}{\boldsymbol}

\newcommand{\ul}{\underline}
\newcommand{\lo}{\overline}
\newcommand{\half}{{\textstyle\frac{1}{2}}}
\newcommand{\sixth}{{\textstyle\frac{1}{6}}}
\begin{document}
\title{Review of biorthogonal coupled cluster representations for electronic excitation}
\author{J. Schirmer and F. Mertins}
\affiliation{Theoretische Chemie,\\ Physikalisch-Chemisches Institut,
Universit\"{a}t  Heidelberg,\\
D-69120 Heidelberg, Germany}

\date{\today}

\begin{abstract}

Single reference coupled-cluster (CC) methods for electronic excitation
are based on a biorthogonal representation (bCC) of the (shifted) Hamiltonian in terms 
of excited CC states, also referred to as correlated excited (CE) states, 
and an associated set of states biorthogonal to the CE states, the latter
being essentially configuration interaction (CI) configurations.
The bCC representation generates a non-hermitian secular matrix, the eigenvalues
representing excitation energies, while the corresponding spectral intensities
are to be derived from both the left and right eigenvectors. 
Using the perspective of the bCC representation, a systematic and comprehensive
analysis of the excited-state CC methods is given, extending and generalizing
previous such studies. Here, the essential topics are the truncation error 
characteristics and the separability properties, the latter being crucial for
designing size-consistent approximation schemes. 
Based on the general order
relations for the bCC secular matrix and the (left and right) eigenvector matrices,
formulas for the perturbation-theoretical (PT) order of the truncation errors (TEO) 
are derived for energies, transition moments, and property matrix elements of
arbitrary excitation classes and truncation levels.
In the analysis of the separability properties of the transition moments, the decisive
role of the so-called dual ground state is revealed.
Due to the use of CE states the bCC approach can be compared to so-called intermediate state 
representation (ISR) methods based exclusively on suitably orthonormalized CE states.   
As the present analysis shows, the bCC approach has  decisive advantages 
over the conventional CI treatment,
but also distinctly weaker TEO and separability properties in comparison 
with a full (and hermitian) ISR method. 

\end{abstract}

\maketitle
\newpage
\section{Introduction}

The extension of the coupled-cluster (CC) approach~\cite{coe58:421,ciz66:4256,
ciz69:35}, originally
devised for ground-states, to the treatment of electronic excitation 
has afforded the emergence of a variety of highly successful computational methods, 
excelling at the potential for both numerical efficiency and accuracy.
The excited-state CC methodology comprises  
three major developments referred to as coupled-cluster 
linear response (CCLR) theory~\cite{mon77:421,dal83:1217,tak86:1486,
koc90:3333,koc90:3345}, the 
equation-of-motion coupled-cluster (EOM-CC) 
approach~\cite{pal78:805,muk79:325,gho82:161,sek84:255,gee89:57,sta93:7029}, 
and the symmetry-adapted cluster configuration 
interaction (SAC-CI)~\cite{nak77:569,nak79:329,nak79:334}. 
While these developments vary in the
derivation of the CC equations, the resulting computational schemes
are largely equivalent. A notable difference, however, is the treatment of  
transition moments and excited-state properties, where, in contrast to
the CC-EOM and SAC-CI schemes, CCLR theory leads to size-consistent expressions.

The basic feature of the CC methods is the use of  
so-called correlated excited (CE) states as basis states in the expansion of
the exact excited states. These CE states are obtained
by applying physical excitation operators associated with single (S), double (D),
triple (T), $\dots$ electron excitations  
to the 
exact (correlated) CC ground-state, $\ket{\Psi^0_J} = \hat{C}_J \ket{\Psi^{cc}_0}$, rather than
to the Hartree-Fock (HF) ground state, $\ket{\Phi_J} = \hat{C}_J \ket{\Phi_0}$, 
establishing in the latter case 
the familiar configuration interaction (CI) basis states or CI configurations. 
Obviously, the CE states represent
intermediates of sorts, positioned in-between the simple CI configurations and 
the exact final states. Accordingly, methods based on the use of CE states  
have been referred to as intermediate state representations (ISR)~\cite{mer96:2140},
and the excited-state CC approach is closely related to the family of ISR methods.
However, the CE states are not orthonormal, and this problem
is dealt with by introducing in addition to the
CE states the associated set of biorthogonal states. Using the biorthogonal
states on the left side and the CE states on the right-hand side, one obtains a mixed
(biorthogonal) representation of the (shifted) Hamiltonian giving rise to a non-hermitian secular matrix.
The two sets of states differ distinctly in their intrinsic quality. In fact,  
the biorthogonal states can be identified essentially as CI configurations. This means 
that the bCC (biorthogonal CC) representation represents a hybrid approach, combining the CI and ISR 
concepts in equal measure.

What are the merits of an ISR approach as compared to the conceptually 
so much simpler CI treatment? 
The answer is that the ISR methods are not (or much less) affected by
two basic deficiencies of the CI approximation schemes, namely
the lack of size-consistency and the need for relatively large 
explicit configuration spaces. The size-consistency error 
inherent to limited CI treatments (as opposed to full(F) CI expansions) 
stems from the \emph{non-separable} structure of 
the CI secular equations. More precisely,
for a system composed of non-interacting fragments
there
is no \emph{a priori} decoupling of the CI secular matrix into corresponding 
fragment blocks. The inevitable and, moreover, uncontrollable size-consistency error 
associated with truncated CI expansions grows with the system size,
rendering the results for extended systems useless. This is why CI   
cannot rank as a genuine many-body method.
Secondly, large CI expansions (configuration spaces) are needed
to suppress the quite unfavorable truncation
error, that is, the error due to discarding higher excitation classes in 
the CI expansions. 
For example, a CI expansion extending through single and double excitations (CISD)  
induces a truncation error of second-order of perturbation theory (PT) in the  
excitation energies of singly excited states.
In the CC methods, by contrast, the corresponding (SD) truncation error  
is of third order. Yet more advantagous is the situation in full (hermitian) ISR   
methods, where the SD truncation error is already of fourth order.

The purpose of this paper is to review 
the separability properties and truncation error characteristics of the 
excited-state CC schemes,
aiming here at a more systematic and more comprehensive analysis than
available so far. 
Previous studies of this kind have been presented 
by J\o{}rgensen and collaborators within the context of CCLR 
theory, addressing the size-consistency of the CC excitation energies~\cite{koc90:3345}
and transition moments~\cite{koc94:4393}, and analyzing the 
truncation errors in the CC excitation energies~\cite{chr96:1451,hal01:671}.
Studies devoted to the size-consistency of the excited state CC equations have
also been presented by Mukhopadhyay \emph{et al.}~\cite{muk91:441} and
Stanton~\cite{sta94:8928}.
From a different perspective, the so-called order relations of the 
bCC secular matrix and the ensuing truncation errors in the CC excitation energies 
have been 
discussed by the present authors~\cite{mer96:2140} and by Trofimov \emph{et al.}~\cite{tro99:9982}.
However, the latter studies, devised essentially to enable 
comparison to the ADC (algebraic-diagrammatic construction) 
propagator methods~\cite{sch82:2395,tro95:2299,tro99:9982},
were somewhat limited and suffered, moreover, from certain misconceptions concerning the
CC transition moments.

The analysis given in this paper of the excited-states CC approach
will be formulated
entirely within the framework of the bCC concept, that is, a non-hermitian 
secular problem associated with a dual representation of the (shifted) Hamiltonian
in terms of two biorthogonal sets of basis states. While in the EOM-CC development,
the bCC representation is introduced 
more or less in an \emph{ad hoc} manner, depicting the bCC secular matrix
essentially as a CI-type representation of an effective (similarity transformed) Hamiltonian,  
the CCLR approach is based on 
response theory for time-dependent CC ground state expectation values.
Clearly, the CCLR derivation is highly original and instructive, and, in fact, 
transcends the simple bCC formulation in the case of the transition moments and
various reponse properties. However, the essence of the emerging computational scheme
can much easier be presented and understood using the bCC concept. Thus, it should be permissible 
and even advisable to abandon the original notations associated with linear response
theory and rather resort to a stringent wave function formulation adapted to 
the bCC concept. Not only will this make the excited state CC methods more amenable
to readers not familiar with the rather demanding time-dependent CC response theory,
it will also allow us to embed the CC approach quite naturally in the broader context
of ISR methods.    

%%%%%%%%%%%%%%%%%%%%%%%%%%%%%%%%%%%%%%%%%%%%%%%%%%%%
An outline of the paper is as follows. 
In Sec.~II, we briefly review the CI method   
with regard to the truncation error and separability properties.
Sec.~III presents the basics of the CC approach to electronic excitation.
This is followed in Sec.~IV by an analysis of the properties of the bCC representation
and the resulting excitation energies, transition moments, and excited state properties.
Sec.~V contrasts the bCC representation with a full (hermitian) ISR approach.
A summary and some conclusions are given in the final Sec.~VI.
Some important supporting material is presented in a tripartite Appendix.
In App.~A the proof of the bCC order relations is reviewed. In App.~B we derive the order relations
of both CI and bCC eigenvector matrices, which, in turn, allows us to generate 
general truncation error formulas. The CCLR forms of the right transition moments and
excited-state properties are reviewed in App.~C.
   
\clearpage
\section{A look at the CI method}

The justification of the intrinsically more complicated intermediate state
representations derives from basic shortcomings of the
standard configuration-interaction (CI) approach with regard to the 
truncation error of the CI expansions and the size-consistency of the results. 
To provide for a general background we begin with a brief recapitulation
of the CI problems.    
  
For the electronically excited states $\ket{\Psi_n}$ of an atom or molecule
the Schr\"odinger equation may be written in the form
\begin{equation}
\label{eq:seq}
(\hat{H} - E_0)\kett{\Psi_n} = \omega_n \kett{\Psi_n}  
\end{equation}
where $\hat{H}$ is the Hamiltonian of the system under consideration, and $E_0$ and 
$\omega_n =E_n -E_0$ denote the ground state energy and excitation energy, respectively.
In the CI treatment, being the standard quantum chemical method,  
the excited states are expanded according to
\begin{equation}
\label{eq:ciexp}
\kett{\Psi_n} = \sum_{J} X_{Jn} \kett{\Phi_J} 
\end{equation}
as a linear combination of CI states 
\begin{equation}
\kett{\Phi_J} = \hat{C}_J \kett{\Phi_0} 
\end{equation}
generated by applying ``physical'' excitation operators $\hat{C}_J$ to
the Hartree-Fock (HF) ground state, $\kett{\Phi_0}$. Using the notation of 
second quantization, the excitation operator manifold may be expressed 
as follows:
\begin{equation}
\label{eq:xops}
\{ \hat{C}_{J}\} \equiv \{ c_{a}^{\dagger}c_{k};
c_{a}^{\dagger}c_{b}^{\dagger}c_{k}c_{l}, a<b,k<l; \ldots \}
\end{equation}
Here $c^{\dagger}_p (c_p)$ denote creation (annihilition) operators 
associated with HF orbitals $\ket{\phi_p}$. 
Following a widely adopted convention, the 
subscripts $a,b,c,\dots$ and $i,j,k,\dots$ denote unoccupied (virtual) and occupied orbitals, 
respectively, while the indices $p,q,r,\dots$ will be used in the general case.
The capital indices $I,J,\dots$ are used as an abbreviation for strings of one-particle
indices, e.g., $I \equiv (abkl)$. 

The excitation operators in (\ref{eq:xops}) can be divided into classes of 
$p$-$h$ (single), $2p$-$2h$ (double), $\dots$ excitations. For brevity, these classes
will be numbered consecutively, that is, the class of $\mu p$-$\mu h$ 
excitations is referred to as class $\mu$. 
The class of a particular excitation $J$ will be denoted by $[J]$. 
For example, $[J] = 2$ means $J$ is a double excitation.      
The HF ground state $\kett{\Phi_0}$, being part of the CI expansions (\ref{eq:ciexp}),
constitutes a zeroth class ($\mu = 0$).

The excitation energies and expansion coefficients are obtained as the 
eigenvalues and eigenvector components, respectively, of the CI eigenvalue problem,
reading in a compact matrix notation
\begin{equation}
\bs{H} \bs{X} = \bs{X} \bs{\Omega},\;\; \bs{X}^{\dagger}\bs{X} = \bs{1}
\end{equation}
Here $\bs{H}$ is the (subtracted) CI secular matrix,
\begin{equation} 
H_{IJ} = \dirint{\Phi_I}{\hat{H} - E_0}{\Phi_J}
\end{equation}
$\bs{\Omega}$ denotes the diagonal matrix of excitation energies $\omega_n$,
and $\bs{X}$ is the matrix of (column) eigenvectors $\ul{X}_n$. Note that
the subtraction of the ground-state energy $E_0$ in the diagonal of the CI 
secular matrix
is a mere convention here, introduced for formal analogy to the bCC representation 
considered in Sec. III.

Approximate CI treatments are obtained by limited CI expansions as opposed to
full (FCI) expansions. In the following we will be concerned with 
systematic truncations of the CI expansions,
that is, expansions being complete through a given excitation class $\mu$.
These systematic truncation schemes can be examined
with respect to the perturbation-theoretical order of the induced error in the
CI results. For this purpose one has to inspect the ``order structure'' of the
CI secular matrix (Fig.~1), that is, a perturbation-theoretical classification 
of the subblocks $\bs{H}_{\mu\nu}$       
associated with a partitioning of $\bs{H}$ with respect to the excitation classes.
Fig.~1 shows the characteristic CI structure, where each excitation class is coupled
linearly in the Coulomb integrals (first order) to the next and next but one
excitation class. Owing to the (zeroth order) orbital energy contributions of the
diagonal matrix elements, the diagonal blocks are indicated by zeros.
  
The order structure of the CI secular matrix gives rise to characteristic
truncation errors in the excitation energies. As the most important case, let us consider
singly excited states, that is, states 
deriving perturbation-theoretically from $p$-$h$ configurations (CI basis states).
Due to the linear (first-order) coupling to triple ($3p$-$3h$) excitations (see Fig.~1),
there is a second-order energy contribution to the single excitation energies
arising from the admixture of triple excitations. This means that 
a second-order truncation error arises in the single-excitation energies if
the triple excitations are not taken into account. A stringent derivation
of the general CI truncation errors is given in App.~B.1. Specifically, 
the truncation error orders (TEO) for single excitation energies  
are given by the following formula:
\begin{equation}
\label{eq:teoci1}
O_{TE}(\mu) =  \begin{cases}
	\mu, &  \mu \text{ even } \\
 	\mu +1, & \mu  \text{ odd }  
      \end{cases}
\end{equation}
Here 
$\mu$ denotes the highest excitation class included in the CI expansion manifold.
In Table~1 the resulting TEOs for $\mu = 1,\dots,6$ are listed.

Besides the energies, the transition moments
\begin{equation}
\label{eq:tm1}
T_n = \brac{\Psi_n}\hat{D} \kett{\Psi_0} 
\end{equation}
are of interest, 
required to compute spectral intensities. 
Here $\hat{D}$ denotes a (one-particle) transition operator, e.g.,
a dipole operator component.
To evaluate the truncation error of the transition moments
one has to analyze the CI expression
\begin{equation}
\label{eq:tm1x}
T_n = \ul{X}^{\dagger}_n \bs{D} \ul{X}_0 
\end{equation}
with respect to the order relations of the CI eigenvectors
$\ul{X}_n$ and $\ul{X}_0$, respectively, and the CI representation of $\hat{D}$,
\begin{equation}
\label{eq:cirep1}
D_{IJ} = \dirint{\Phi_I}{\hat{D}}{\Phi_J}
\end{equation}
The order relations of the CI eigenvectors for singly excited states and for
the ground state, shown in Fig.~1, are part of the general order structure 
of the CI eigenvector matrix $\bs{X}$, derived in App.~B.1.  
Together with the trivial order structure of $\bs{D}$  
(zeroth-order diagonal and off-diagonal blocks, other matrix elements vanishing) 
one can deduce the following simple TEO expression for 
the transition moments of singly excited states: 
\begin{equation}
\label{eq:teoci2}
O_{TE}(\mu) =  \mu
\end{equation}
For odd values of $\mu$ the TEOs of the transition moments are smaller by 1 
as compared to the excitation energies.

The CI truncation errors are relatively large, which, in turn, implies that
large CI expansions are required to meet specific accuracy levels.  
For example, in order to treat singly excited
states consistently through second order of perturbation theory (PT), the 
CI configuration space must comprise the triple excitations ($\mu =3$). By contrast,
in the ISR methods a much smaller explicit configuration space, consisting of single 
and double excitations, affords the corresponding level of accuracy.

To analyze the size-consistency properties of a method, one usually resorts
to the separate fragment model, that is, a system $S$ consisting of two
strictly non-interacting fragments, $A$ and $B$. A method for 
treating electronic excitation 
is size-consistent (here, more specifically, size-intensive)
if for local excitations, say on fragment A, the computed excitation energies
and transition moments do not depend on whether the method is applied 
to the fragment or the composite system. It is well known that truncated
CI treatments do not fulfill this property. There is an uncontrollable 
size-consistency error in the treatment of the total system, corrupting
not only the results for the separate fragment model, but, 
more generally, any truncated CI treatment of larger molecules. This can 
be nicely demonstrated in the exactly solvable model of a chain of non-interacting 
two-electron two-orbital($2E$-$2O$) systems, such $1s^22s^0$ He atoms or minimal basis       
H$_2$ molecules (see, for example, Meunier and Levy~\cite{meu79:955}). 
Because of this deficiency the CI method
does not qualify for a genuine many-body method.

As a preparation for the analysis of the bCC
schemes let us briefly inspect the case of CI in some more detail. 
Obviously, the Hamiltonian for $S$ decomposes into
the sum of the fragment Hamiltonians, $\hat{H} = \hat{H}_A + \hat{H}_B$.
Moreover, the 
one-particle states (HF orbitals) of $S$  
can be classified as belonging either to fragment $A$ or $B$ (local on A or B, respectively).  
Accordingly, the CI states can be partitioned 
into three different sets, that is, 
local excitations $I_A$ on fragment $A$, local excitations $I_B$ on
 fragment $B$, and mixed (or non-local) excitations $I_{AB}$ involving both fragment
A and B. In the latter class we may disregard any excitations that do not conserve
the local electron number, e.g., $A^+B^-$ charge-transfer excitations. In the 
non-interacting fragment model, such charge transfer excitations are strictly decoupled from
the fragment-charge conserving excitations, which are of interest here.    

The CI configurations $\ket{\Phi_I}$ can be written as 
products of fragment states, e.g.,
\begin{eqnarray}
\nonumber
\ket{\Phi_0} &=& \ket{\Phi^{A}_0} \ket{\Phi^{B}_{0}}\\
\nonumber
\ket{\Phi_{I_A}} &=& \ket{\Phi^{A}_{I_A}} \ket{\Phi^{B}_{0}}\\
\label{eq:cibf}
\ket{\Phi_{I_AI_B}} &=&  \ket{\Phi^{A}_{I_A}} \ket{\Phi^{B}_{I_B}}
\end{eqnarray}
It should be noted that the neglect of full antisymmetrization of these product states
is irrelevant because in the non-interacting fragment model  
matrix elements are not affected by inter-fragment antisymmetrization.   
Fig.~2 shows
the partitioning of the CI secular matrix with respect 
to the three different types of configurations, that is, local excitations on fragment $A$,
local excitations on fragment $B$, and non-local excitations, respectively.
While the $A$ and $B$ states are strictly decoupled, $\bs{H}_{AB} = \bs{0}$, 
there is a coupling between the local and mixed excitations, e.g., 
\begin{equation}
H_{I_A,J_AJ_B} = \delta_{I_AJ_A} \dirint{\Phi^{B}_{0}}{\hat{H}_B}{\Phi^{B}_{J_B}}   
\end{equation}
as can easily be derived using the CI state factorization according to Eq.~(\ref{eq:cibf}).
An explicit example is the coupling matrix element 
\begin{equation}
H_{ak,bc'd'i'j'l} = \delta_{ab} \delta_{kl} V_{c'd'[i'j']}
\end{equation} 
for a single excitation $I_A = (ak)$ associated with fragment $A$ and a non-local
triple excitation $J_{AB} = (bl,c'd'i'j')$. Here the unprimed (primed) indices 
denote fragment $A$ (fragment $B$) one-particle states; $V_{pq[rs]}$ denotes
the anti-symmetrized Coulomb integral.

Given the structure as shown in Fig.~2, the CI secular matrix is said to be
\emph{non-separable} because there is no \emph{a priori} decoupling
of local excitations (say on fragment $A$) from non-local (or mixed) excitations. 
The CI treatment of the composite system $S$ aims in an inextricable way
at an optimal description of both fragments, that is, the excited state of fragment $A$ and 
the ground state of fragment $B$. In the exact (full) CI result, say for the 
energy  $E_n = E^A_n + E^B_0$ of the
locally excited system, the ground state energy $E^B_0$ of 
the unaffected fragment $B$ would cancel exactly upon subtraction of the 
exact ground-state energy $E_0 =  E^A_0 + E^B_0$ of $S$, so that, of course, the full CI
excitation energy, $E_n - E_0 = E^A_n -E^A_0$, is size-consistent. 
At the level of a limited CI expansion, however, neither the excited state energy
nor the ground-state energy are simply the sums of the fragment energies.

\clearpage

\section{Biorthogonal CC representation}

The bCC formulation (see, for example, Helgaker \emph{et al.}~\cite{Helgaker:2000}) 
is based on a mixed representation of the (subtracted)
Hamiltonian, $\hat{H}-E_{0}$, in terms of two sets of left and right expansion manifolds:
i) the CC states 
\begin{equation}
\label{eq:ccstates}
\kett{\Psi^0_{J}}=
\hat{C}_{J} \kett{\Psi^{cc}_0}=
\hat{C}_{J} e^{\hat{T}}\kett{\Phi_{0}} 
%\exp(\hat{T})\hat{C}_{J}\left|\Phi_{0}\right>
\end{equation}
on the right-hand side, and
ii)
the associated biorthogonal states
\begin{equation}
\label{eq:bostates}
\left<\lo{\Phi}_{I}\right|=
\left<\Phi_{0}\right|\hat{C}_{I}^{\dagger} e^{-\hat{T}} 
\end{equation}
on the left-hand side. 
Here $\hat{C}_{I}$ denote the physical excitation 
operators as specified in Eq.~(\ref{eq:xops}).
The familiar ground state CC parametrization (and normalization) 
\begin{equation}
\label{eq:ccgs}
 \left|\Psi_{0}^{cc}\right>=
 e^{\hat{T}}\left|\Phi_{0}\right >
\end{equation} 
is used, where 
$\left|\Phi_{0}\right>$ denotes the HF ground state, and
\begin{equation} 
\label{eq:ccop}
\hat{T}=\sum_{I} t_{I} \hat{C}_{I} 
\end{equation}  
is the cluster operator with the amplitudes $t_J$ determined by the ground-state CC equations.
The $\hat{T}$ operator, comprising physical excitation operators only, commutes with any 
(physical) $\hat{C}_{J}$ operator, so that the CC states of Eq.~(\ref{eq:ccstates}) can likewise be
written as
\begin{equation}
\label{eq:ccs}
\left|\Psi_{J}^{0}\right>=
e^{\hat{T}}\hat{C}_{J}\left|\Phi_{0}\right>
\end{equation}
The biorthonormality of the two sets of states, 
\begin{equation}
\label{eq:biorth}
\brackett{\lo{\Phi}_{I}}{\Psi_{J}^{0}} 
= \brac{\Phi_I} e^{-\hat{T}} e^{\hat{T}} \kett{\Phi_J} = \delta_{IJ}  
\end{equation}
is an obvious consequence of the orthonormalization of the CI configurations 
$ \kett{\Phi_J} = \hat{C}_J \kett{\Phi_0}$.

The bCC representation of $\hat{H}-E_{0}$ gives rise to a non-hermitian secular matrix
${\bf M}$ with the elements
\begin{eqnarray}
\label{eq:ccsm}
M_{IJ}&=&\brac{\lo{\Phi}_{I}}\hat{H}-E_{0}
\kett{\Psi_{J}^{0}}
\nonumber \\
&=&\brac{\Phi_{0}}\hat{C}_{I}^{\dagger} e^{-\hat{T}}
[\hat{H},\hat{C}_{J}] e^{\hat{T}}\kett{\Phi_{0}}.
\end{eqnarray}
In the latter form the (CC) ground-state energy $E_0$ no longer appears explicitly. 
Let us note that the bCC secular matrix can also be written as a CI representation  
\begin{equation}
\nonumber
M_{IJ}=\dirint{\Phi_I}{\lo{H}-E_{0}}{\Phi_J}
\end{equation}
of the similarity transformed Hamiltonian,
$\lo{H} = e^{-\hat{T}}\hat{H}e^{\hat{T}}$. However, this form is less transparent
then the bCC representation~(\ref{eq:ccsm}) and, thus, less useful for the 
analysis intended here. 

The (vertical) electronic excitation energies, $\omega_{n} = E_n -E_0$, can be identified
as the eigenvalues of the CC secular matrix ${\bf M}$. Because ${\bf M}$ is not hermitian 
one has to deal with 
right and left eigenvalue problems,
\begin{eqnarray}
\label{eq:ccsm1}
{\bf M X}&=& {\bf X \Omega} \\
\label{eq:ccsm2}
{\bf Y^{\dagger} M}&=& {\bf \Omega Y^{\dagger}}
\end{eqnarray}
where ${\bf \Omega}$ is the diagonal matrix of eigenvalues $\omega_{n}$, 
and  ${\bf X}$ and ${\bf Y}$ denote the matrices of the right and left 
eigenvectors, respectively. To obtain a definite normalization 
the two sets of secular equations have to combined
according to
\begin{equation}
{\bf Y^{\dagger} M\, X = \Omega, \qquad \ Y^{\dagger}  X = 1}
\end{equation}
so that the resulting right and left eigenvectors form two mutually biorthonormal sets. 
As a consequence, the corresponding right and left excited states, 
\begin{eqnarray}
\label{eq:rxs}
\kett{\Psi^{cc}_{n}}&=&\sum_{I} X_{In} \left|\Psi_{I}^{0}\right> \\ 
\label{eq:lxs}
\brac{\Psi^{(l)}_{m}}&=&\sum_{I}Y^{*}_{Im}
\left<\lo{\Phi}_{I}\right|.
\end{eqnarray}
are biorthonormal too, $\bracket{\Psi^{(l)}_{m}}{\Psi^{cc}_{n}} = \delta_{mn}$.

In general, the right excited states $\kett{\Psi^{cc}_{n}}$ are not yet 
 eigenstates of $\hat{H}$, because the underlying $\kett{\Psi_{I}^{0}}$ expansion manifold of 
 Eq.~(\ref{eq:ccstates}) is incomplete as long as 
the CC ground state $\kett{\Psi^{cc}_{0}}$ is not taken into account. 
Using the extended CC expansion manifold $\{R\}=\{\kett{\Psi^{cc}_0}, \kett{\Psi_I^0}\}$ on the  
right-hand side, and, likewise, the extended biorthogonal manifold 
$\{L\}=\{\brac{\Phi_0}, \brac{\lo{\Phi}_I}\}$
on the left side, one arrives at the full bCC representation of $\hat{H} - E_0$ 
associated with
the extended secular matrix
\begin{equation}
\label{eq:ccsmx}
{\bf M'}=\left( \begin{array}{cc}
0 & \underline{v}^{t} \\
\underline{0} & {\bf M}
\end{array} \right).
 \end{equation}
Here $\underline{v}^{t}$ is a transposed (row) vector with the elements 
\begin{equation}
v_{I}= \dirint{\Phi_0}{\hat{H}}{\Psi^0_I}
\end{equation}
that is, the coupling matrix elements between the HF ground state and the excited CC states.
Let us note that in the usage of the EOM-CC approach 
$\bf M'$ is denoted by $\lo{\bf H}$; in the CCLR context, on the other hand,
the (inner) bCC secular matrix $\bf M$ 
(Eq.~\ref{eq:ccsm}) is referred to as the CC Jacobian ${\bf A}$ and the coupling vector
$\underline{v}^{t}$ is denoted by $\bf{\eta}$.   

The two expansion manifolds used in the bCC representation are of quite different quality. 
The ``correlated excited states'' (CES) of the set $\{R\}$ are superior to the biorthogonal 
$\{L\}$ states, if more complex. 
Obviously, a CC state of class $[I]$ can be written according to
\begin{equation}
\label{eq:rset}
\kett{\Psi^0_I} = e^{\hat{T}} \kett{\Phi_I} =  
\kett{\Phi_I} +\sum_{K,\, [K]>[I]}  z^{(I)}_K \kett{\Phi_K}   
\end{equation}
as a linear combination of $\kett{\Phi_I}$ and CI configurations of \emph{higher} excitation classes,
$[K]>[I]$, extending through $N$-tuple excitations. 
By contrast, the CI expansion of a
biorthogonal $\{L\}$ set state reads 
\begin{equation}
\label{eq:lset}
\brac{\lo{\Phi}_I} = \brac{\Phi_I} e^{-\hat{T}} =  
\brac{\Phi_I} +\sum_{K,\, [K]<[I]}  z^{(I)}_K \brac{\Phi_K}   
\end{equation}
that is,  a linear combination of $\brac{\Phi_I}$ and \emph{lower} class CI excitations, $[K]<[I]$,  
including the zeroth class, $[K]=0$.
This follows from the observation that $z^{(I)}_K =\brac{\Phi_I} e^{-\hat{T}} \kett{\Phi_K}$ 
vanishes for $[K]>[I]$ (and $z^{(I)}_K=\delta_{IK}$ for $[K]=[I]$).

As is easily seen,  
the linear space spanned by 
the biorthogonal states through a given excitation class $\mu$ is identical with the corresponding
space of CI configurations:        
\begin{equation}
\label{eq:span}
span \{\brac{\Phi_I} e^{-\hat{T}},\,\, [I]=0,1,\dots,\mu\} =
span \{\brac{\Phi_I},\,\, [I]=0,1,\dots,\mu\} 
\end{equation}
This means that (truncated) expansions in terms of the biorthogonal ($\{L\}$ set) states 
are essentially of CI-type. Let us note that the SAC-CI equations are obtained 
by using the CI expansion manifold for the left eigenstates rather than 
the $\{L\}$  states~\cite{nak77:569,nak79:329,nak79:334}. 
 
As will be discussed below, the use of a CI-type expansion manifold on the left-hand side of the
bCC representation deteriorates the overall order relations and separability properties to a 
certain extent. 
One may wonder then why one could not simply use the CC states as the common expansion manifold 
for both sides of the secular matrix. However, in such an approach, referred to as 
variational or unitary CC version
(see Kutzelnigg~\cite{kut91:349} and Szalay \emph{et al.}~\cite{sza95:281}), 
there is a major problem associated with evaluating the secular (and overlap) matrix elements:
being of the form
$\brac{\Phi_I}e^{\hat{T}^{\dagger}} \hat{H} e^{\hat{T}} \kett{\Phi_J}$, 
there is no obvious truncation of higher excitation contributions (below $N$-tuple excitation level).

Let us now briefly inspect the eigenpair manifold of the extended bCC secular matrix ${\bf M'}$.
Obviously, there is one more eigenvalue, $\omega_0 = 0$, corresponding to the ground state, while
the excited state eigenvalues $\omega_n, n > 0$, of the subblock ${\bf M}$ are also eigenvalues of 
 ${\bf M'}$. The corresponding extended left and right eigenvectors can easily be determined. Let us
first consider the ground state solutions. Here the right eigenvector is trivial,
\begin{equation}
\label{eq:rgsv}
\underline{X}_0=\left( \begin{array}{c}
1\\
\underline{0}
\end{array} \right)
 \end{equation}       
which is consistent with the fact that  $\kett{\Psi^{cc}_{0}}$ is the exact ground-state. 
Less obvious is the left ground-state eigenvector
\begin{equation}
\label{eq:lgsv}
\underline{Y}_{0}^{' \dagger}=\left(1, \,\,\, \underline{Y}^{\dagger}_0\right)
\end{equation}
where the row vector $\underline{Y}^{\dagger}_{0}$ can explicitly be 
obtained from  ${\bf M}$ and $\underline{v}$ according to 
\begin{equation}
\label{eq:lgsvx}
\underline{Y}^{\dagger}_{0}=- \underline{v}^{t}\, {\bf M^{-1}} 
\end{equation}
The corresponding
representation of the ground state,
\begin{equation}
\label{eq:lgs}
\bra{\overline{\Psi}_{0}}=\left<\Phi_{0}\right|+
\sum_{I} Y_{I0}^{*}\left<\lo{\Phi}_{I}\right|
\end{equation}
in terms of the biorthogonal bCC states is referred to as the ``dual''
ground state~\cite{koc90:3333}. In the CCLR nomenclature, 
the dual ground state is denoted 
by $\bra{\Lambda}$. 
As will be discussed below, the dual ground-state 
is a non-separable CI-type representation 
of the ground state, leading to undesired features in the bCC transition moments.

The left excited state eigenvectors of ${\bf M'}$ are obtained as obvious extensions
according to
\begin{equation}
\label{eq:lxsv}
\underline{Y}'^{\dagger}_n = (0,\,\,\,\underline{Y}^{\dagger}_n)
\end{equation}
from the left eigenvectors of the  ${\bf M}$ subblock. This means that the excited states
$\bra{\Psi^{(l)}_n}$ of Eq.~(\ref{eq:lxs}) are proper eigenstates of $\hat{H}$.
In particular, they are orthogonal to the exact CC ground-state,
\begin{equation}
\label{eq:orth1}
\brackett{\Psi^{(l)}_{n}}{{\Psi^{cc}_0}} = 0
\end{equation}
which follows from $\brackett{\lo{\Phi}_{I}}{{\Psi^{cc}_0}} = 0$.

In general, that is, if not forbidden by symmetry,
the right extended eigenvectors acquire non-vanishing zeroth components $x_n = X'_{n0}$,
and the extended eigenvectors take on the form 
\begin{equation}
\label{eq:rxsv}
\underline{X}_{n}^{'}=\left(
\begin{array}{c}
x_n\\ \underline{X}_{n}
\end{array}
\right)
\end{equation} 
where $\underline{X}_{n}$ is the ($n$-th) eigenvector of the ${\bf M}$ subblock, and the 
zeroth component $x_n$ is given by 
\begin{equation}
\label{eq:zn1}
x_n = \omega_n^{-1} \underline{v}^{t}\, \underline{X}_n
\end{equation} 
Since ${\bf M} \underline{X}_n =  \omega_n \underline{X}_n$ 
the following relations hold: 
\begin{equation}
\label{eq:zn2}
x_n = \underline{v}^{t}\, {\bf M}^{-1} \underline{X}_n = - \underline{Y}^{\dagger}_0 \underline{X}_n
\end{equation} 
Here Eq.~(\ref{eq:lgsvx}) has been used to arrive at the last expression.

As a consequence, the right expansion of an excited eigenstate takes on the form
\begin{equation}
\label{eq:rrxs}
\kett{\Psi^{(r)}_n} = x_n \kett{\Psi^{cc}_0} + \kett{\Psi^{cc}_n} 
\end{equation}
where $\kett{\Psi^{cc}_n}$ is given by Eq.~(\ref{eq:rxs}). 
Let us note that $\bracket{\Phi_0}{\Psi^{cc}_n} = 0$ since $\bracket{\Phi_0}{\Psi^{0}_I} = 0$,
so that the relation
\begin{equation}
\label{eq:rrxs1}
x_n = \bracket{\Phi_0}{\Psi^{(r)}_n}
\end{equation}
can be established. The excited eigenstates $\ket{\Psi^{(r)}_n}$ are manifestly
orthogonal to the dual ground state:
\begin{equation}
\label{eq:orthx}
\brackett{\overline{\Psi}_0}{\Psi^{(r)}_n} = x_n + \underline{Y}^{\dagger}_0 \underline{X}_n = 0
\end{equation}
where the relations $\bracket{\lo{\Phi}_I}{\Psi^{(r)}_n} = X_{In}$ and Eq.~(\ref{eq:zn2}) 
have been used.

For spectral intensities the squared moduli $|T_n|^2$ of the transition moments
(\ref{eq:tm1})
are required, involving normalized ground and excited states.
In the bCC representation a properly normalized expression for $|T_n|^2$
is obtained according to~\cite{sta93:7029,koc94:4393}  
\begin{equation}
\label{eq:tramo}
|T_{n}|^{2}=\brac{\overline{\Psi}_{0}}\hat{D}\kett{\Psi^{(r)}_{n}}
           \brac{\Psi^{(l)}_{n}}\hat{D}\kett{\Psi^{cc}_{0}}      
\end{equation}
using both the left and right transition moments,
\begin{eqnarray}
\label{eq:tramol}
T_{n}^{(l)}&=&\brac{\Psi^{(l)}_{n}}\hat{D}\kett{\Psi^{cc}_{0}}\\
\label{eq:tramor}
T_{n}^{(r)}&=&\brac{\overline{\Psi}_{0}}\hat{D}\kett{\Psi^{(r)}_{n}}   
\end{eqnarray}
Individually, the left and right transition moments have no significance because 
the respective ground and excited states are not normalized. However,
the biorthonormality relations
$\brackett{\overline{\Psi}_{0}}{\Psi^{cc}_{0}}=\bracket{\Psi^{(l)}_n}{\Psi^{(r)}_n} =1$
ensure the combined normalization in the product (\ref{eq:tramo}). 
In contrast to the left transition moments,   
the ordinary bCC form (\ref{eq:tramor}) for the right transition moments is not separable, 
so that the results obtained at truncated bCC levels are not size-intensive~\cite{koc94:4393}. 
Within the CCLR framework this shortcoming is avoided, as here   
a separable, if more elaborate expression is employed for the right 
transition moments~\cite{koc90:3333,koc94:4393}. 
The different treatment of the spectral intensities is a distinguishing feature of the otherwise
equivalent CCLR and EOM-CC methods. In Sec. IV.D and App.~C the CCLR expression
for the right transition moments will briefly be reviewed.

As will be discussed below, the problem in the right transition moments does not result from the 
right eigenstates but rather from the use of the dual ground state $\brac{\overline{\Psi}_{0}}$.
As can be concluded from Eqs.~(\ref{eq:lset},\ref{eq:lgsvx},\ref{eq:lgs}), the dual ground state
is a CI-type expansion of the form
\begin{equation}
\label{eq:dgsx}
\brac{\overline{\Psi}_{0}} = \brac{\Phi_0} + \sum _{K > 0} \tilde{z}_K \brac{\Phi_K}  
\end{equation}
where the expansion coefficients depend (via Eqs.~\ref{eq:lgsvx},\ref{eq:lgs})
on the $t$-amplitudes of the CC ground state. 
For the CC ground-state energy the following relation holds:
\begin{equation}
\label{eq:dgsen}
 E^{cc}_0 = \brac{\overline{\Psi}_{0}} \hat{H} \kett{\Psi^{cc}_{0}} 
\end{equation}
It should be noted that this expression applies not only to the exact CC ground state
(where  $E^{cc}_0 =  E_0$),
but also to CC approximations based on truncations of the expansion manifolds, such as in CCSD.
This can be seen by writing the rhs of Eq.~(\ref{eq:dgsen}) more explicitly as
\begin{equation}
\label{eq:dgsex}
\brac{\overline{\Psi}_{0}} \hat{H} \kett{\Psi^{cc}_{0}}
= \brac{\Phi_0} \hat{H} \kett{\Psi^{cc}_{0}} + \sum_I Y^*_{I0} 
\brac{\lo{\Phi}_I} \hat{H} \kett{\Psi^{cc}_{0}}
\end{equation}
The first term on the rhs is the CC energy equation, while the second (summation) term vanishes 
in compliance with the CC amplitude equations,
$\brac{\lo{\Phi}_I} \hat{H} \kett{\Psi^{cc}_{0}} = 
\brac{\Phi_I} e^{-\hat{T}} \hat{H} e^{\hat{T}} \kett{\Phi_0} = 0$.

What is the relation of the dual ground state to the CI ground state in the case of 
truncated expansions? Obviously, the 
energy expectation value of the dual ground state
will always be greater than (or equal to) the corresponding CI energy, 
\begin{equation}
\label{eq:vp}   
\frac{\brac{\overline{\Psi}_{0}} \hat{H} \kett{\overline{\Psi}_{0}}}
{\brackett{\overline{\Psi}_{0}}{\overline{\Psi}_{0}}} \geq E^{CI}_0
\end{equation}
because the dual state expansion coefficients are non-variational. This can be
nicely demonstrated in the exactly solvable model of 2 (or more) non-interacting
2E-2O systems (He atoms, or H$_2$ molecules).

It should be noted that the right and left eigenvalue problems~(\ref{eq:ccsm1},\ref{eq:ccsm2})
for the secular matrix $\bf{M}$ (Eq.~\ref{eq:ccsm}) follow from a variational 
principle, $\delta\dirint{\lo{\Phi}}{\hat{H}}{\Psi} = 0$,
under the constraint  
$\brackett{\lo{\Phi}}{\Psi} = 1$. Independent variations 
$\bra{\delta \lo{\Phi}}$ and $\ket{\delta \Psi}$ on the left and right side of the 
energy and overlap matrix elements 
lead directly to the right and left eigenvalue equations, respectively.
\clearpage

\section{Analysis of bCC excitation energies and intensities}
\subsection{Order relations and separability of the bCC secular matrix}

Fig. 3 shows the order structure of the bCC secular matrix $\bf{M}$. For the partitioning
according to excitation classes, $\mu =1,2,\dots$, the lowest (non-vanishing)
PT orders are given here in the respective $\bf{M}_{\mu \nu}$ subblocks. In the upper right (UR) triangular
part we recover the characteristic CI structure of Fig. 1. This outcome can readily be
understood by inspecting the general expression for the secular matrix elements, 
\begin{equation}
\label{eq:or1}
\brac{\lo{\Phi}_{I}}\hat{H}\kett{\Psi_{J}^{0}}
=\brac{\Phi_{I}}\hat{H}\kett{\Phi_{J}} + \sum_{[K]<[I]}\,\sum_{[L]>[J]}z^{(I)}_K z^{(J)}_L 
                \brac{\Phi_{K}}\hat{H}\kett{\Phi_{L}}
\end{equation}
obtained by using the expansions (\ref{eq:rset},\ref{eq:lset}) for the CC and biorthogonal states,
respectively. For the UR matrix elements with $[I] < [J]$, the sums on the rhs of Eq.~(\ref{eq:or1})
do not contribute, because the excitation classes of the double summation indices $K$ and $L$ 
differ at least by a triple excitation, $[L]-[K]\geq 3$, so that the Hamiltonian matrix
elements $\brac{\Phi_K}\hat{H}\kett{\Phi_L}$ vanish. This means that the bCC and CI secular matrix
elements are identical, $M_{IJ} = H_{IJ}$, for $[I] < [J]$. For the diagonal blocks $\bf{M}_{\mu \mu}$,
of course, the lowest non-vanishing PT order is zero because the perturbation expansions of the 
diagonal matrix elements $M_{II}$ begin with zeroth-order (HF) excitation energies.

By contrast, the lower left (LL) triangular part, $[I]\geq [J]$  
gives rise to the remarkable ``canonical'' order relations, reading
\begin{equation}       
\label{eq:or2}
M_{IJ}\sim O([I]-[J]),\,\,\,\ [I]\geq [J]
\end{equation}
This means that the lowest non-vanishing contribution in the 
PT expansion of the matrix element $M_{IJ},[I]\geq [J]$ is of the order $[I]-[J]$.
Likewise we will use the notation $O[M_{IJ}] = [I]-[J]$.
These order relations were first specified 
by Christiansen \emph{et al.}~\cite{chr95:7429}, quite explicitly, 
for the lowest 5 excitation classes (singles through pentuples) and later
by Hald \emph{et al.}~\cite{hal01:671} for general levels of excitation.
A first general proof of the bCC order relations was given in  
Ref~\cite{mer96:2140}. A brief recapitulation of this proof is given in App.~A.

%%%here remark on original proof ?%%%
It should be noted that
these canonical order relations are highly non-trivial indeed. Let us consider, for example,
the $\bs{M}_{31}$ matrix elements, being of the order $2$. This means that the apparent first-order
contribution arising from the leading (CI) term on the rhs of Eq.~(\ref{eq:or1}) is exactly
cancelled by other first-order contributions from the summation part. 

The bCC order structure gives rise to specific truncation errors of the excitation energies,
which can be analyzed
by inspecting formal perturbation-theoretical (PT) interaction paths on the subblocks of 
the bCC secular matrix. 
As an example, let us consider the following path:     
\begin{equation}
\nonumber
M_{11}(0) \rightarrow M_{13}(1) \rightarrow  M_{33}(0) \rightarrow M_{31}(2) \rightarrow M_{11}(0)
\end{equation}
This path allows one to specify formally
a PT contribution in the excitation energy of a single excitation
(class 1) arising from the admixture of triple excitations (class 3). The
PT order of this path is 3, and there is no lower-order path involving class 3.  
This means
that the PT contributions to single excitation energies arising from triple 
excitations are of the order 3. Stated differently,     
the truncation error for single excitation energies due to the neglect of triple excitations in
the explicit bCC expansion manifold is of third order, as compared to a second-order truncation
error in the CI treatment. 

A more general and stringent derivation of the  
bCC truncation errors is to be based on the order relations
of the bCC eigenvector matrices, as is discussed 
in App.~B. In short, 
the order relations in the bCC secular matrix induce corresponding 
order relations for the left and right eigenvector matrices $\bs{Y}$ and $\bs{X}$, 
respectively. 
As shown in Fig.~10, the bCC eigenvector matrices 
combine canonical and CI-type behaviour: 
the LL part of $\bs{X}$ and the UR part of $\bs{Y}$  exhibit canonical order relations,
whereas CI-type order relations (Fig.~9) apply to the UR part of $\bs{X}$ 
and the LL part of $\bs{Y}$.
The eigenvector order relations, in turn, allow one to analyze the
truncation errors in the excitation energies, transition moments, and 
excited state properties. For the singly excited states, the 
truncation error orders (TEO) in the bCC excitation energies are 
given by the following formula (deriving from Eq.~\ref{eq:teccen}) 
\begin{equation}
\label{eq:teo1}
O_{TE}(\mu) =  \begin{cases}
	\frac{3}{2} \mu, &  \mu \text{ even}\\
 	\frac{3}{2} \mu + \frac{1}{2}, &  \mu \text{ odd}
      \end{cases}
\end{equation} 
Here $\mu$ denotes the highest excitation class in the expansion manifold.  
In Table 1, the TEOs in the excitation energies of singly excited states
are listed for the first 6 truncation levels.

To discuss the \emph{separability properties} of the bCC 
schemes~\cite{koc90:3345,koc94:4393, Helgaker:2000} 
(see also Refs.~\cite{muk91:441,sta94:8928}) 
we revisit the separate fragment model, $S \equiv A + B$, considered in Sec.~2. 
Like the Hamiltonian, $\hat{H} = \hat{H}_A + \hat{H}_B$, also the CC operator can be written
as the sum of the fragment operators, $\hat{T} = \hat{T}_A + \hat{T}_B$. As a consequence,   
both the CC states $\kett{\Psi^0_I}$ and the biorthogonal states $\brac{\overline{\Phi}_I}$
used in the right and left expansion manifolds, respectively, can be written as 
products of fragment $A$ and $B$ states:
\begin{eqnarray}
\nonumber
\kett{\Psi^{0}_{I_A}} &=& \kett{\Psi^{A}_{I_A}} \kett{\Psi^{B}_{0}},\;\,
\kett{\Psi^0_{I_AI_B}} =  \kett{\Psi^{A}_{I_A}} \kett{\Psi^{B}_{I_B}}\\
\label{eq:ccfac}
\brac{\overline{\Phi}_{I_A}} &=&  \bra{\overline{\Phi}^{A}_{I_A}} \bra{\Phi^{B}_0},\;\;\;\; 
\brac{\overline{\Phi}_{I_AI_B}} = \bra{\overline{\Phi}^{A}_{I_A}}\bra{\overline{\Phi}^{B}_{I_B}}
\end{eqnarray}
Here the notation of the fragment states has been somewhat simplified by 
omitting the superscripts $0$ and $cc$: for example, 
$\kett{\Psi^{A}_{I_A}} \equiv \kett{\Psi^{A,0}_{I_A}}$ and
$\kett{\Psi^{B}_{0}} \equiv \ket{\Psi^{B,cc}_{0}}$.  
Fig.~4 shows
the partitioning of the bCC secular matrix with respect 
to the three different types of configurations, that is, local excitations on fragment $A$,
local excitations on fragment $B$, and non-local (or mixed) excitations.
Let us note once again that 
in the latter set we can disregard any charge-transfer type configurations.

The separability properties of the bCC secular matrix, as shown in Fig.~4, can be readily derived
by using the factorization (Eqs.~\ref{eq:ccfac}) of the bCC basis functions. 
As an explicit example let us derive 
that the block $\bs{M}_{AB,A}$ vanishes: 
\begin{eqnarray}
\nonumber
M_{I_AI_B,J_A} &=& \dirint{\overline{\Phi}_{I_AI_B}}{\hat{H}_A + \hat{H}_B}{\Psi^{0}_{J_A}}\\
             &=& \dirint{\overline{\Phi}^A_{I_A}}{\hat{H}_A}{\Psi^{A}_{J_A}} 
                  \bracket{\overline{\Phi}^{B}_{I_B}}{\Psi^{B}_0}   
                + \bracket{\overline{\Phi}^{A}_{I_A}}{\Psi^{A}_{J_A}} 
             \dirint{\overline{\Phi}^{B}_{I_B}}{\hat{H}_B}{\Psi^{B}_0} = 0
\end{eqnarray}
Here the first term on the rhs vanishes because of the orthogonality relation 
$\bracket{\overline{\Phi}^{B}_{I_B}}{\Psi^{B}_0} = 0$.
The second term vanishes,
\begin{equation}
\label{eq:cceq}
\dirint{\overline{\Phi}^{B}_{I_B}}{\hat{H}_B}{\Psi^{B}_0} = E^B_0 \, 
                      \bracket{\overline{\Phi}^{B}_{I_B}}  
{\Psi^{B}_0} = 0
\end{equation}
because 
$\ket{\Psi^{B}_0}$ is an eigenfunction of $\hat{H}_B$, and 
$\bra{\overline{\Phi}^{B}_{I_B}}$ and $\ket{\Psi^{B}_0}$ are orthogonal. In fact, 
this result is not predicated on  
the exact fragment $B$ ground state, but applies also to any (systematic) CC approximation,
since $\dirint{\overline{\Phi}^{B}_{I_B}}{\hat{H}_B}{\Psi^{B}_0} = 0$ is satisfied
as CC amplitudes equation for fragment $B$.

In a similar way one may establish that the bCC secular matrix for fragment $A$
is identical to the ($A,A$)-block of the composite system secular matrix, 
\begin{equation}
\label{eq:smsp1}
\bs{M}_{AA}= \bs{M}^A
\end{equation}
In deriving this result, one utilizes the CC equation for the ground-state of fragment $B$, 
$\dirint{\Phi^{B}_0}{\hat{H}_B - E^B_0}{\Psi^{B}_0} = 0$.  
The matrix elements in the non-vanishing ($A,AB$) coupling block read
\begin{equation}
\label{eq:smsp2}
M_{I_A,J_AJ_B} = \delta_{I_AJ_A} v^{B}_{J_B}
\end{equation} 
where
$v^{B}_{J_B} = \dirint{\Phi^{B}_{0}}{\hat{H}_B}{\Psi^{B}_{J_B}}$.
Finally, the non-local diagonal block matrix elements are given by
\begin{equation}
\label{eq:smsp3}
M_{I_AI_B,J_AJ_B} = \delta_{I_BJ_B} M^A_{I_AJ_A} + \delta_{I_AJ_A} M^B_{I_BJ_B}
\end{equation}
It should be emphasized once again that these results do not presuppose 
exact CC ground states but apply as well to the (systematic) CC approximations.

The separability structure of $\bs{M}$ 
reflects once more the different quality of 
the left and right bCC expansion manifolds. While the LL triangular is separable, the
UR triangular displays the non-separable CI-type structure of Fig.~2. What are the 
consequences for excitation energies and eigenvectors? 
Notwithstanding the apparently non-separable block structure of $\bs{M}$,
the bCC excitation energies are obtained in a separable way~\cite{koc90:3345}.
It is 
readily seen that the characteristic polynomial for the fragment secular matrix
$\bs{M}^A$ is a factor in the full  
characteristic polynomial associated with $\bs{M}$. Accordingly, the eigenvalues 
(excitation energies) of fragment $A$ are a subset of the eigenvalues of the full
secular matrix $\bs{M}$. This means that the energies of local excitations 
are separable quantities: the bCC results 
do not depend on whether the 
method is applied to the fragment or the composite system. 

For the eigenvectors one will expect different  
separability properties in  
the left and right manifolds. 
In fact, the right eigenvectors are separable, as 
shown in Fig.~4. For a local excitation $n$, say on fragment $A$, the only non-vanishing
components are fragment-$A$ components $\ul{X}_{An}$, and since
$\bs{M}^A \ul{X}_{An} = \omega_n \,\ul{X}_{An}$ 
 it is readily established that the fragment-$A$ part of $\ul{X}_{n}$ is equal to 
the corresponding fragment-$A$ eigenvector, $\ul{X}_{An} = \ul{X}^A_{n}$. 
More explicitly,
the separability properties of right eigenvectors may be written as 
\begin{equation}
\label{eq:revsp}
X_{I_A n} = X^A_{I_A n},\; X_{I_B n} = X_{I_{AB} n} = 0,  
\end{equation}
assuming here a fragment-$A$ excitation, $n=n_A$.
 
For the left eigenvectors $\ul{Y}_n$ the 
fragment-$A$ eigenvector is recovered by the $A$ part of the full eigenvector, 
$\underline{Y}_{An} = \underline{Y}^A_{n}$. The local fragment-$B$ components vanish, 
$Y_{I_B n} =0$, but 
there are non-vanishing non-local components,  
$Y_{I_{AB},n} \neq 0$. These non-local components 
are related to the local ones according to 
\begin{equation}
\label{eq:levmc}
\underline{Y}^{\dagger}_{AB,n} =  \underline{Y}^{\dagger}_{An} \bs{M}_{A,AB}
                           (\omega_n - \bs{M}_{AB,AB})^{-1}  
\end{equation}
This means that for a local excitation, say on $A$ ($n=n_A$), the left eigenstate
will take on the form
\begin{equation}
\label{eq:genexle}
\left< \Psi^{(l)}_{n}\right| = \sum_{I_A} Y^{*}_{I_An}
 \bra{\overline{\Phi}^{A}_{I_A}}\bra{\Phi^{B}_0} + \sum_{I_AI_B} Y^{*}_{I_AI_B,n} 
  \bra{\overline{\Phi}^{A}_{I_A}}\bra{\overline{\Phi}^{B}_{I_B}}
\end{equation}
where $ Y_{I_An} = Y^A_{I_An}$. 
In the exact (full) bCC result this transforms into the product 
of the excited fragment-$A$ state and dual ground-state of 
fragment $B$ (see Sec.~IV.E).

\subsection{Transition moments: truncation errors}
Now we are in the position to examine the truncation errors in 
the transition moments. Let us first consider the \emph {left transition
moments} (Eq.~\ref{eq:tramol}),
\begin{equation}
\label{eq:ltm1}
T^{(l)}_n = \dirint{\Psi^{(l)}_n}{\hat{D}}{\Psi^{cc}_0}
\end{equation}
which may be written more explicitly as scalar products 
\begin{equation}
\label{eq:ltm2}
T^{(l)}_n =  \ul{Y}^{\dagger}_{n} \ul{F}^{(l)}
\end{equation}
of the left eigenvector $\ul{Y}_n$ and a vector $\ul{F}^{(l)}$
of basis set transition moments, 
\begin{equation}
\label{eq:ltm3}
F^{(l)}_I = \dirint{\overline{\Phi}_I}{\hat{D}}{\Psi^{cc}_0}
\end{equation}
associated with the biorthogonal (left) basis states.
The basis set transition moments, being part of the general 
bCC representation of the one-particle operator $\hat{D}$, 
exhibit canonical order relations 
\begin{equation}
\label{eq:ortml}
 O[\ul{F}^{(l)}_{\mu}] = \mu - 1
\end{equation}
as shown in Fig.~5b. A proof of these order relations is given in App.~A. 
The scalar product (\ref{eq:ltm2}) combines the left eigenvector
components with the left basis set transition moments. This means 
that the (lowest) PT order associated with a specific class $\mu$ of
eigenvector components is given by 
$O[\ul{Y}^{\dagger}_{\mu}\ul{F}^{(l)}_{\mu}] = O[\ul{Y}_{\mu}] 
+ O[\ul{F}^{(l)}_{\mu}]$. Accordingly, the truncation after class $\mu$ leads to an error
of the order  $O[\ul{Y}_{\mu + 1}] 
+ O[\ul{F}^{(l)}_{\mu + 1}]$.  
In the case of singly excited states, the CI-type order relations of the 
left eigenvectors (Fig.~3b) lead to the TEO formula (\ref{eq:teo1}).
However, so far we have disregarded the effect of truncations in the CC 
ground state. For example, there are first-order
contributions in $\ul{F}^{(l)}_{1}$ (that is the $p$-$h$-components 
of $\ul{F}^{(l)}$) associated with the $2p$-$2h$ cluster operator $\hat{T}_2$
in the CC expansion of $\ket{\Psi^{cc}_0}$. In conjunction with  $\ul{Y}_{1}$,
being of zeroth-order, this gives rise to 
a first-order truncation error in the left transition moments if the ground-state
CC expansion does not comprise the $2p$-$2h$ cluster operator. 
This can readily be generalized, noting that the $\hat{T}_{\mu}$ cluster 
operators are of the PT order $\mu -1$. This means that in the 
class $\mu$ components of the F vector, $\ul{F}^{(l)}_{\mu}$, there are
$\mu$-th order contributions arising from the T operators of class $\mu +1$.
Depending on the respective lowest order in the eigenvector components
of class $\mu$, $\ul{Y}_{\mu}$, this leads to an additional
truncation error, the order of which is by 1 smaller than that arising
from the eigenvector truncation. The overall truncation errors
for singly excited states are given by the following formula:
\begin{equation}
\label{eq:teo2x}
O_{TE}(\mu) =  \begin{cases}
	\frac{3}{2} \mu, &  \mu \text{ even}\\
        \frac{3}{2} \mu	- \frac{1}{2}, &  \mu \text{ odd}
      \end{cases}
\end{equation} 
In a similar way, one may analyze the bCC truncation errors for 
doubly and higher excited states; general TEO formulas are given in App.~B.2.

%%%%%%%%%%%%%%%%%%%%%%%%%%%%%%%%%%%%%%%%%%%%%%%%%

Likewise, the   
\emph {right transition moments} (Eq.~\ref{eq:tramor})
\begin{eqnarray}
\nonumber
T^{(r)}_n & = & \dirint{\overline{\Psi}_0}{\hat{D}}{\Psi^{(r)}_n}\\
\label{eq:rtm1}
         & = & x_n \dirint{\overline{\Psi}_0}{\hat{D}}{\Psi^{cc}_0} + \sum_{I} F^{(r)}_I X_{In}
\end{eqnarray}
can be written as  scalar products of the extended right eigenvectors $\ul{X}'_n$ 
(see Eq.~\ref{eq:rxsv}) 
and a vector $\ul{F}^{(r)}$ of right basis set transition moments,
\begin{equation}
\label{eq:rtm2}
 F^{(r)}_I = \dirint{\overline{\Psi}_0}{\hat{D}}{\Psi^{0}_I}
\end{equation}
associated here with the CC states of right expansion manifold. 
In Eq.~(\ref{eq:rtm1}), the ground-state contribution ($I=0$) is written separately.

Let us consider the second term on the rhs of Eq.~(\ref{eq:rtm1}), which 
will be seen to determine the overall truncation errors.
Using the expansion~(\ref{eq:lgs}) of the dual ground-state,
this term can be written as
\begin{equation}
\label{eq:rtmor1}
\sum_{I} F^{(r)}_I X_{In} = \sum_{I} \dirint{\Phi_0}{\hat{D}}{\Psi^{0}_I} X_{In} +
\ul{Y}^{\dagger}_0 \bs{D} \ul{X}_n
\end{equation}
where  $\bs{D}$ is the 
bCC representation of $\hat{D}$,
\begin{equation}
\label{eq:1pbcc}
D_{IJ} = \dirint{\overline{\Phi}_I}{\hat{D}}{\Psi^{0}_J}
\end{equation}
The order relations of $\bs{D}$ are shown in Fig.~5 (see App.~A for a proof). 
The summation in the first term on the rhs of Eq.~(\ref{eq:rtmor1}) is 
restricted to $p$-$h$ components, $[I]=1$, because 
$\dirint{\Phi_0}{\hat{D}}{\Psi^{0}_I} = 0$ for $[I] > 1$, and, thus, does
not cause a truncation error whatsoever. 
The second term can be written in the more explicit form
\begin{equation}
\nonumber
\ul{Y}^{\dagger}_0 \bs{D} \ul{X}_n = \sum_{\kappa,\lambda}\ul{Y}^*_{\kappa 0}
\bs{D}_{\kappa, \lambda} \ul{X}_{\lambda n} 
\end{equation}
reflecting the underlying partitioning with respect to excitation classes. 
To determine the truncation errors (at the level $\mu$) one has to 
analyze the PT orders of the contributions with $\kappa = \mu +1, \lambda \leq \mu +1$ 
and $\lambda = \mu +1, \kappa \leq \mu +1$. This is described in App.~B,
where general TEO formulas are derived. For singly excited states, $[n]=1$,
the TEOs of $\ul{Y}^{\dagger}_0 \bs{D} \ul{X}_n$  
are given by Eq.~(\ref{eq:teo2x}).

It remains to inspect the first term on the rhs of  
Eq.~(\ref{eq:rtm1}), being the product of the ground-state admixture coefficient,
$x_n = -\underline{Y}^{\dagger}_0 \underline{X}_{n}$, and the ground-state expectation value     
$\dirint{\overline{\Psi}_0}{\hat{D}}{\Psi^{cc}_0}$. 
The truncation errors in the $x_n$ coefficient have been specified in 
Eq.~(\ref{eq:teoxn}) of App.~B.2, which become 
\begin{equation}
\label{eq:teo2xx}
O_{TE}(\mu) =  \begin{cases}
	\frac{3}{2} \mu + 1, &  \mu \text{ even}\\
        \frac{3}{2} \mu	+ \frac{1}{2}, &  \mu \text{ odd}
      \end{cases}
\end{equation} 
in the case of singly excited states, $[n]=1$.
To determine the truncation errors in the 
factor $\dirint{\overline{\Psi}_0}{\hat{D}}{\Psi^{cc}_0}$,
we expand 
the dual ground state according to Eq.~(\ref{eq:lgs}), which yields 
\begin{equation}
\label{eq:rtmor3}
\dirint{\overline{\Psi}_0}{\hat{D}}{\Psi^{cc}_0} = \dirint{\Phi_0}{\hat{D}}{\Psi^{cc}_0}
                         + \underline{Y}^{\dagger}_0 \underline{F}^{(l)}
\end{equation} 
Here the first term is of zeroth order, and does not induce a truncation error. 
The second term can be analyzed in a similar way as the left TMs above. Here
the TEOs are given by 
\begin{equation}
\label{eq:teo2xxx}
O_{TE}(\mu) =  \begin{cases}
	\frac{3}{2} \mu, &  \mu \text{ even}\\
        \frac{3}{2} \mu	+ \frac{1}{2}, &  \mu \text{ odd}
      \end{cases}
\end{equation} 
Since for singly excited states ($[n] = 1$) $x_n$ is (at least) of 2nd
order, the overall truncation errors in the first term on the rhs of 
Eq.~(\ref{eq:rtm1}) are given by that of $x_n$ (Eq.~\ref{eq:teo2xx}), 
exceeding those of the second term (Eq.~\ref{eq:teo2x}) by one. 
We note that for singly excited states the TEOs in the left and right
transition moments are the same (see Table 1).

\subsection{Transition moments: separability}
The separability of the left transition moments (Eqs.~\ref{eq:ltm1}-\ref{eq:ltm3})
is easily established. According to
\begin{equation}
\label{eq:tmsc1}
F^{(l)}_{I_A} =  \bra{\lo{\Phi}^A_{I_A}} \bra{\lo{\Phi}^B_0} \hat{D}_A  + \hat{D}_B \kett{\Psi^{A}_0}
\kett{\Psi^{B}_0}
=  \dirint{\overline{\Phi}^A_{I_A}}{\hat{D}_A}{\Psi^{A}_0} = F^{(l)A}_{I_A}
\end{equation}
it is seen that for local excitations $I_A$ 
the basis set transition moments for fragment A are identical to the corresponding
moments of the composite system.
Moreover, the basis set transition moments for mixed excitations vanish,    
\begin{equation}
\label{eq:ltmjab}
F^{(l)}_{I_AI_B} = 0
\end{equation}
which means that the non-separable part of the left eigenvector does not come into play at all.
Accordingly, for a local (fragment-$A$) excitation, $n=n_A$, we may write 
\begin{equation}
T^{(l)}_n = \sum_{I_A} Y^{*}_{I_An}\, F^{(l)A}_{I_A} = T^{(l)A}_n 
\end{equation}
where $T^{(l)A}_n$ is the left transition moment for fragment $A$. In deriving this result,
we have used that 
the local components for the composite system eigenvectors
are identical to the components of the corresponding fragment 
eigenvectors, $Y_{I_An} =  Y^{A}_{I_An}$.

%%%%%%%%%%%%%%%%%%%%%%%%%%%%%%%%%%%%%%%%%%%%%%%%%%%%%%%%%%%%%%%%%%

Whereas the left transition moments are
separable notwithstanding the non-separable left eigenvectors, 
the right transition moments (Eqs.~\ref{eq:rtm1},\ref{eq:rtm2}), involving
separable eigenvectors, prove not to be separable.
The problem here arises from   
the use of the dual ground state $\brac{\overline{\Psi}_0}$, more precisely, from the 
fact that a factorization of the dual ground state,   
\begin{equation}
\label{eq:dgs1}
\brac{\overline{\Psi}_0} = \bra{\overline{\Psi}^A_0} \bra{\overline{\Psi}^B_0} 
\end{equation}
is attained only in the exact (full bCC) treatment. 
To better understand the problem let us first assume factorization of $\brac{\overline{\Psi}_0}$ 
and inspect $T^{(r)}_n$ for a local excitation, $n=n_A$:
\begin{equation}
\label{eq:rtm3}
T^{(r)}_n = x_n \dirint{\overline{\Psi}_0}{\hat{D}}{\Psi^{cc}_0} + 
            \sum_{I_A} F^{(r)}_{I_A}\, X^A_{I_A n}   
\end{equation}
Here the separability properties (\ref{eq:revsp}) of the right eigenvector $\underline{X}_n$ 
have been used; $F^{(r)}_I$ denote (right) basis set transition moments (Eq.~\ref{eq:rtm2}). 
The ground state component $x_n$ of the right eigenvector in the first term on the rhs
is always separable, since
\begin{equation}
\label{eq:xnsep} 
x_n = -\underline{Y}^{\dagger}_0\, \underline{X}_n =  
  - \underline{Y}^{A \dagger}_0\, \underline{X}^A_n = x^A_n
\end{equation}
By contrast, the ground-state expectation value   
\begin{equation}
\label{eq:gsev}
\dirint{\overline{\Psi}_0}{\hat{D}}{\Psi^{cc}_0} 
                 = \dirint{\overline{\Psi}^A_0}{\hat{D}_A}{\Psi^{A}_0}
                 + \dirint{\overline{\Psi}^B_0}{\hat{D}_B}{\Psi^{B}_0}
\end{equation}
is a non-local quantity, involving both fragment $A$ and $B$. It should be noted
that the separation of the bCC ground-state expectation value into the sum of
fragment expectation values not only holds
for the exact (factorizing) dual ground state, but also for truncated expansions
according to Eq.~(\ref{eq:dgsexp}) below. (In this sense the 
bCC ground-state expectation values themselves are separable quantities.)  
Being a product of a local and a non-local factor, however,
the first term on the rhs of Eq.~(\ref{eq:rtm3}) is
not separable. This means that the non-separable contribution 
$x_n \dirint{\overline{\Psi}^B_0}{\hat{D}_B}{\Psi^{B}_0}$ in the first term 
of Eq.~(\ref{eq:rtm3}) 
must be cancelled by a corresponding contribution in the second term, to
be identified in the following.
For a local configuration, $I = I_A$, the right basis set transition moments
(Eq.~\ref{eq:rtm2}) become
\begin{equation}
\label{eq:rbstm}
 F^{(r)}_{I_A} = \dirint{\overline{\Psi}^A_0}{\hat{D}_A}{\Psi^{A}_{I_A}}
              + \bracket{\overline{\Psi}^A_0}{\Psi^{A}_{I_A}}
                \dirint{\overline{\Psi}^B_0}{\hat{D}_B}{\Psi^{B}_0}
\end{equation}
Obviously, $F^{(r)}_{I_A}$ is not separable. While the first term on the 
rhs is the fragment-$A$ transition moment, $F^{(r)A}_{I_A} = 
\dirint{\overline{\Psi}^A_0}{\hat{D}_A}{\Psi^{A}_{I_A}}$, the second term is a 
non-separable contribution involving fragment $B$. 
In the full bCC treatment the two non-separable contributions in $T^{(r)}_n$
cancel each other. This is readily seen by inserting Eq.~(\ref{eq:rbstm}) in 
Eq.~(\ref{eq:rtm3}) and using that     
$\bracket{\overline{\Psi}^A_0}{\Psi^{A}_{I_A}} = Y^*_{I_A0}$
and
\begin{equation}
\sum_{I_A}  Y^*_{I_A0} X^A_{I_A n} = - x^A_n 
\end{equation}
For truncated bCC expansions, on the other hand, the 
non-separable contributions will not compensate each other,
giving rise to size-consistency errors in the computational results.

The non-separability of the right transition moments 
can be further elaborated by inspecting the general form 
of the dual ground-state,
\begin{equation}
\label{eq:dgsexp}
\brac{\overline{\Psi}_0} = \bra{\Phi^A_0}\bra{\Phi^B_0}
                          + \sum_{J_A} Y^*_{J_A0} \bra{\overline{\Phi}^{A}_{J_A}} \bra{\Phi^{B}_0} 
                          + \sum_{J_B} Y^*_{J_B0} \bra{\Phi^{A}_0} \bra{\overline{\Phi}^{B}_{J_B}} 
                          + \sum_{J_AJ_B} Y^*_{J_AJ_B,0} 
                    \bra{\overline{\Phi}^{A}_{J_A}}\bra{\overline{\Phi}^{B}_{J_B}} 
\end{equation}
applying both to truncated and full expansions.
Here the local expansion coefficients are separable, that is, $Y_{J_A0} = Y^A_{J_A0}$
and  $Y_{J_B0} = Y^B_{J_B0}$.
This follows from
Eq.~(\ref{eq:lgsvx}) and the separability properties of $\bs{M}$ and $\ul{v}$
(see below). 
Using the expansion (\ref{eq:dgsexp}), 
the right basis set transition moments take on the form
\begin{equation}
F^{(r)}_{I_A} =  F^{(r)A}_{I_A} +  Y^*_{I_A0}\dirint{\Phi^B_0}{\hat{D}_B}{\Psi^{B}_0}
                        + \sum_{J_B}  Y^*_{I_AJ_B,0}\dirint{\overline{\Phi}^B_{J_B}}{\hat{D}_B}{\Psi^{B}_0}
\end{equation}
With the help of Eqs.~(\ref{eq:xnsep}) and (\ref{eq:gsev}) we finally
arrive at the expression
\begin{equation}
\label{eq:rtm5}
T^{(r)}_n = T^{(r)A}_n +  x^A_n \left(\dirint{\overline{\Psi}^B_0}{\hat{D}_B}{\Psi^{B}_0} 
                             - \dirint{\Phi^B_0}{\hat{D}_B}{\Psi^{B}_0} \right)
            + \sum_{I_AJ_B} X_{I_An}  Y^*_{I_AJ_B,0}\dirint{\overline{\Phi}^B_{J_B}}{\hat{D}_B}{\Psi^{B}_0} 
\end{equation}
for the right transition moments. 
Here $ T^{(r)A}_n$ denotes the right transition moment for fragment $A$.
The non-separable contributions are identified as 
the second and third term on the rhs of Eq.~(\ref{eq:rtm5}). These terms cancel each other
if the non-local eigenvector components factorize,that is,
\begin{equation}
\label{eq:gsevfac}
 Y_{I_AJ_B,0} =  Y_{I_A0} Y_{J_B0}
\end{equation}
This can be seen by recalling that $\sum Y^*_{I_A 0} X_{I_A n} = - x^A_n$ and 
$\bra{\overline{\Psi}^B_0} = \brac{\Phi^B_0} + \sum Y^*_{J_B 0} \bra{\overline{\Phi}^B_{J_B}}$.
It should be clear, however, that the factorization (\ref{eq:gsevfac}) of the non-local 
eigenvector components is equivalent to the factorization (\ref{eq:dgs1}) of the 
dual ground state (\ref{eq:dgsexp}), applying only to the full bCC expansion.

It may be of interest to see how the 
factorization of the exact dual ground-state eigenvector components
derives from the explicit 
expression (Eq.~\ref{eq:lgsvx}),  
\begin{equation}
\nonumber
\underline{Y}^{\dagger}_0 = - \underline{v}^{t}\bf{M}^{-1}
\end{equation}
As is readily established, 
the local contributions to $\underline{v}$ are separable,
\begin{equation}
v_{I_A} = \dirint{\Phi^A_0}{\hat{H}_A}{\Psi^A_{I_A}} = v^{A}_{I_A}
\end{equation}
and the mixed components vanish,
\begin{equation}
v_{I_AI_B} = 0
\end{equation}
The inverse of the bCC secular matrix is given by
\begin{equation}
\label{eq:minv}
\bf{M}^{-1}=\left(
\begin{array}{ccc}
\bs{M}^{-1}_{AA} & - & \bs{P}_{A,AB}\\
 -  & \bs{M}^{-1}_{BB} & \bs{Q}_{A,AB} \\
 -  &    - & \bs{M}^{-1}_{AB,AB}
\end{array}
\right)
\end{equation} 
where
\begin{eqnarray}
\bs{P}_{A,AB} &=& - \bs{M}^{-1}_{AA} \bs{M}_{A,AB} \bs{M}^{-1}_{AB,AB}\\
\bs{Q}_{A,AB} &=& - \bs{M}^{-1}_{BB} \bs{M}_{B,AB} \bs{M}^{-1}_{AB,AB} 
\end{eqnarray}
so that, according to Eq.~(\ref{eq:lgsvx}), the non-local eigenvector components 
can be written as 
\begin{equation}
\label{eq:symb1}
\underline{Y}_{AB,0} = (\underline{v}^{t}_A \bs{M}^{-1}_{AA} \bs{M}_{A,AB} 
                    + \underline{v}^{t}_B \bs{M}^{-1}_{BB} \bs{M}_{B,AB}) \bs{M}^{-1}_{AB,AB}
\end{equation}
To proceed the secular matrix blocks $\bs{M}_{AB,AB}$, $\bs{M}_{A,AB}$, and $\bs{M}_{B,AB}$
have to be further evaluated. Using a  
somewhat symbolic but largely self-explanatory notation these blocks may be written as
\begin{eqnarray}
\label{eq:symb2}
\bs{M}_{AB,AB} = \bs{1}_B \times  \bs{M}_{AA} + \bs{1}_A \times  \bs{M}_{BB} \\
\label{eq:symb3}
\bs{M}_{A,AB} = \bs{1}_A \times \underline{v}^{t}_B, \,\,\, 
\bs{M}_{B,AB} = \bs{1}_B \times \underline{v}^{t}_A, 
\end{eqnarray}
Proceeding at this symbolic level
the desired result is readily obtained as follows:    
\begin{eqnarray}
\nonumber
\underline{Y}_{AB,0} &=& \underline{v}^{t}_A  \times \underline{v}^{t}_B 
                    (\bs{M}^{-1}_{AA} + \bs{M}^{-1}_{BB}) \bs{M}^{-1}_{AB,AB}\\
\nonumber
                  &=& \underline{v}^{t}_A  \times \underline{v}^{t}_B 
                      \bs{M}^{-1}_{AA}\bs{M}^{-1}_{BB} 
                 (\bs{1}_B \times  \bs{M}_{AA} +  \bs{1}_A \times  \bs{M}_{BB})\bs{M}^{-1}_{AB,AB}\\
\nonumber
                  &=& \underline{v}^{t}_A  \bs{M}^{-1}_{AA} \times  
                      \underline{v}^{t}_B  \bs{M}^{-1}_{BB}\\
\label{eq:symb4}
                  &=& \underline{Y}_{A0} \times \underline{Y}_{B0} 
\end{eqnarray}
In a more stringent manner, the preceding computation can be performed 
on the matrix-element level, that is, by explicitly expanding all matrix multiplications.
Here, the symbolic treatment according to Eqs.~(\ref{eq:symb1}-\ref{eq:symb4})  
may serve as a guidance.

Let us recall once again that a factorization of the dual ground state
cannot be expected if the configuration space is truncated. For example, assume
a configuration space extending through double excitations and let 
$I_A$ and $J_B$ denote double excitations on fragment $A$ and $B$, respectively.  
Then the factorization according to Eq.~(\ref{eq:gsevfac}) would require that 
the configuration space of the system as a whole comprises quadruple excitations of the
type $I_AJ_B$, which, however, are not available in the truncated configuration manifold.

\subsection{CCLR form of right transition moments}

The derivation of the excited state CC equations in the framework of
the linear response theory leads to the following separable, if
 more involved expression for the
right transition moment~\cite{koc90:3333}: 
\begin{equation}
\label{eq:rtmlrf}
T^{(r)}_n = \dirint{\overline{\Psi}_0}{[\hat{D}, \hat{C}_n]}{\Psi^{cc}_0}
- \sum_{I,J} \dirint{\overline{\Psi}_0}{[[\hat{H}, \hat{C}_I],\hat{C}_n]}{\Psi^{cc}_0}
            (\bs{M} + \omega_n)^{-1}_{IJ} \dirint{\overline{\Phi}_J}{\hat{D}}{\Psi^{cc}_0}
\end{equation}
Here
\begin{equation}
\hat{C}_n = \sum X_{Kn}\hat{C}_K 
\end{equation}
denotes an excitation operator associated with the  
the $n$-th (right) excited state:   
$\ket{\Psi^{(r)}_n} = x_n \ket{\Psi^{cc}_0} + \hat{C}_n \ket{\Psi^{cc}_0}$.
Since the CCLR derivation starts out from a separable expression for a
time-dependent ground-state expectation value, one may expect that
the separability properties will be maintained in the further development. 
Nevertheless, it is reassuring to see
directly that
the CCLR form of the right transition moments is separable~\cite{koc94:4393}.  
Moreover, one will expect that the ordinary bCC~(\ref{eq:rtm1}) 
and the CCLR~(\ref{eq:rtmlrf}) expressions, while
being of quite different form, must somehow become equivalent
in the exact (full) bCC treatment. The absolutely non-trivial proof
of this equivalence has been accomplished  
by Koch \emph{et al.}~\cite{koc94:4393}.
In the following we will briefly review the separability of the CCLR 
right transition moments.  
The equivalence of the two transition moment expressions is addressed in App.~C. 

To show the separability of the right CCLR transition moments we will suppose
the general expansion~(\ref{eq:dgsexp}) of the dual ground state,
which holds both for approximate and 
exact (full) bCC treatments. Let $n$ be a local excitation on 
fragment $A$ ($n=n_A$). According to the
separability of the right eigenvector, $\hat{C}_n$ consists only of local excitation operators,
$\hat{C}_n = \sum X_{I_A} \hat{C}_{I_A}$. Now it is easy to see that  
the first (commutator) term on the rhs of Eq.~(\ref{eq:rtmlrf}) is 
separable. Since $[\hat{D}, \hat{C}_{I_A}] = [\hat{D}_A, \hat{C}_{I_A}]$,
the commutator becomes a local (fragment-$A$) operator, say $\hat{O}_A$. As a consequence, 
in the matrix
element $\dirint{\overline{\Psi}_0}{\hat{O}_A}{\Psi^{cc}_0}$ the fragment $B$ and non-local 
(AB) contributions in the expansion of $\bra{\overline{\Psi}_0}$ are projected out, that is,
 $\dirint{\overline{\Psi}_0}{\hat{O}_A}{\Psi^{cc}_0} = \dirint{\overline{\Psi}^A_0}{\hat{O}_A}{\Psi^A_0}$.
Now let us consider the second term on the rhs of Eq.~(\ref{eq:rtmlrf}), involving a
double summation running over generic configuration indices $I,J$. The 
double commutator, involving the fragment-$A$ excitation operator $\hat{C}_n$
and excitation operators $\hat{C}_I$, leads to a restriction upon the indices $I$:  
there are no (non-vanishing) contributions for $I = I_B$ 
(fragment $B$ excitations), since 
$[\hat{H},\hat{C}_{I_B}] = [\hat{H}_B,\hat{C}_{I_B}]$ and 
$[[\hat{H}_B,\hat{C}_{I_B}], \hat{C}_{n_A}] = 0$. But what about contributions 
associated with non-local configurations $I = I_{AB}$, not excluded
by the double commutator term? To proceed let us inspect the matrix elements of 
$(\bs{M} + \omega_n)^{-1}$. According to the separability structure of this matrix
(see Eq.~\ref{eq:minv}) the only non-vanishing matrix elements of the type $(I_{AB}, J)$ 
are those where the $J$ index is non-local too, $J = J_{AB}$. However, the  non-local
left basis set transition moments, $F^{(l)}_J = \dirint{\overline{\Phi}_J}{\hat{D}}{\Psi^{cc}_0}$,
appearing as factors on the rhs of Eq.~(\ref{eq:rtmlrf}), vanish for 
non-local (mixed) configurations $J_{AB}$ (see Eq.~\ref{eq:ltmjab}), which 
means that non-local configurations can be excluded in both the $J$  
and the $I$ summations. To conclude: for a local excitation $n = n_A$, the double summation
on the rhs of  Eq.~(\ref{eq:rtmlrf}) runs only over local fragment $A$ configurations
$I_A, J_A$. With this restriction it is readily established that the three 
ingredients on the rhs of Eq.~(\ref{eq:rtmlrf}), i.e., the double commutator matrix elements,
the matrix inverse, and the left basis set transition moments, are separable: 
they give the same results irrespective of being computed for the entire system or for
fragment $A$ only.

\subsection{Excited state properties and transition moments}

So far we have discussed  
ground-to-excited state transition moments required to compute spectral 
intensities. Now we will turn to excited-state expectation values (properties)
for physical quantities of interest, e.g., excited-state dipole moments, and, 
more generally, transition moments associated with transitions between two excited states. 
 
In bCC form the general expression,
\begin{equation}
T_{nm} = \dirint{\Psi_n}{\hat{D}}{\Psi_m}
\end{equation}
for excited-state transition moments  
becomes
\begin{eqnarray}
\nonumber
T_{nm} & = & \dirint{\Psi^{(l)}_n}{\hat{D}}{\Psi^{(r)}_m}\\
\label{eq:esp1}
       & = & x_m \dirint{\Psi^{(l)}_n}{\hat{D}}{\Psi^{cc}_0} + \ul{Y}^{\dagger}_n \bs{D} \ul{X}_m
\end{eqnarray}
Here $\bs{D}$ denotes the bCC representation (\ref{eq:1pbcc}) of 
a given operator $\hat{D}$. The order structure of $\bs{D}$ is shown in Fig.~5,
supposing here that $\bs{D}$ is a one-particle operator; for a proof of
these order relations see App.~A.

The truncation errors of the $T_{nm}$ matrix elements  
are governed by the 2nd term on the rhs of Eq.~(\ref{eq:esp1}).
The secondary role of the first term can be seen
in a similar way as in the right transition moments discussed in Sec.~IV.B.
We here skip the corresponding analysis of the first term, noting only
that the order relations of two constituents, that is,  
the ground-state admixture coefficient $x_m$ and the left (ground-to-excited state) 
transition moment $T^{(l)}_n$, have already been established in Sec.~4.B.
The truncation errors associated with the second term, being of the form 
of a vector$\times$matrix$\times$vector product,  
can be derived from   
the order relations of $\bs{D}$ and the respective left and right eigenvectors,
as described more detailed in App.~B.2.
In the case of singly excited states ($[n] = [m]= 1$), the general
formula (\ref{eq:teccpm}) in App.~B.2 simplifies to    
\begin{equation}
\label{eq:teccpms}
O_{TE}(\mu) =  \begin{cases}
         \frac{3}{2}\mu - 1, &  \mu \text{ even}\\
	 \frac{3}{2}\mu - \frac{1}{2}, & \mu \text{ odd} 	
      \end{cases}
\end{equation}
As the comparison with Eq.~(\ref{eq:teo1}) shows, 
the excited-state transition moments for 
singly excited states (and a one-particle transition operator)
have larger truncation errors (i.e. lower TEOs)
than the excitation energies and ground-to-excited state transition moments.

To discuss the separability we consider once more 
local excitations on fragment $A$, that is, $n=n_A, m=m_A$.
In the exact case, where the left and right excited states can be written as
fragment state products,
\begin{eqnarray}
\nonumber
\brac{\Psi^{(l)}_n} = \brac{\Psi^{(l)A}_n} \bra{\overline{\Psi}^B_0}\\
\nonumber 
\kett{\Psi^{(r)}_m} = \kett{\Psi^{(r)A}_m} \kett{\Psi^B_0}
\end{eqnarray}
the excited-state transition moments take on the manifestly 
separable form,
\begin{equation}
\label{eq:esp2}
T_{nm} = T^A_{nm} + \delta_{nm}\dirint{\lo{\Psi}^B_0}{\hat{D}_B}{\Psi^B_0} 
\end{equation}
where 
\begin{equation}
T^A_{nm} = \dirint{\Psi^{(l)A}_n}{\hat{D}_A}{\Psi^{(r)A}_m}
\end{equation}
is the transition moment for fragment $A$. Note that for diagonal (property)
matrix elements ($n=m$) there is a
contribution $\dirint{\lo{\Psi}^B_0}{\hat{D}_B}{\Psi^B_0}$, corresponding to 
the ground-state expectation value of $\hat{D}$ for fragment $B$.

Now let us analyze Eq.~(\ref{eq:esp1}) in the case of a truncated bCC representation.
The first term on the rhs of Eq.~(\ref{eq:esp1}) is 
separable,
\begin{equation}
\label{eq:estm0} 
x_m T^{(l)}_n = x^A_m \,T^{(l)A}_n
\end{equation}
as has already been shown 
in the preceding subsection. In the 
second term, 
\begin{equation}
\label{eq:estm1}
T'_{nm} = \ul{Y}_n^{\dagger} \bs{D} \ul{X}_m 
\end{equation}
the separability structure of $\bs{D}$ comes into play. As in the case
of the bCC secular matrix (see Section IV.A), the separability structure 
of $\bs{D}$ can easily be derived. The result is shown in Fig.~6. The relevant 
subblocks $\bs{D}_{AA}$ and $\bs{D}_{AB,A}$ are given by
\begin{equation}
\label{eq:espsc1}
\bs{D}_{AA} = \bs{D}^A_{AA} + \bs{1}_A \dirint{\Phi^B_0}{\hat{D}_B}{\Psi^B_0}
\end{equation}
\begin{equation}
\label{eq:espsc2}
\bs{D}_{AB,A} =  \bs{1}_A \times \ul{F}^{(l)}_B 
\end{equation}
Here $\ul{F}^{(l)}_B = \ul{F}^{(l)B}_B$ is the vector of left basis set transition
moments for fragment $B$, as in Eq.~(\ref{eq:tmsc1}).
Using these expressions, Eq.~(\ref{eq:estm1}) becomes 
\begin{equation}
\label{eq:espsc3}
T'_{nm} = \ul{Y}^{\dagger}_{An} \bs{D}_{AA} \ul{X}_{Am} 
                 + \ul{Y}^{\dagger}_{AB,n} \bs{D}_{AB,A} \ul{X}_{Am}
\end{equation}
where 
\begin{equation}
\ul{Y}^{\dagger}_{AB,n} = \ul{Y}^{\dagger}_{An} \bs{M}_{A,AB}(\omega_n -  \bs{M}_{AB,AB})^{-1}
\end{equation}
is the non-local part of the left eigenvector $\ul{Y}^{\dagger}_{n}$, 
as specified by Eq.~(\ref{eq:levmc}). 
Let us now first consider the non-diagonal case, $n \neq m$,
where the left and right eigenvectors are orthogonal.
Using the result
\begin{equation}
\ul{Y}^{\dagger}_{An} \bs{D}_{AA} \ul{X}_{Am} = \ul{Y}^{A\dagger}_{n} \bs{D}^A \ul{X}^A_{m}   
\end{equation}
for the first term on the rhs of Eq.~(\ref{eq:espsc3}), 
as well as Eq.~(\ref{eq:estm0}), leads to 
the following expression:
\begin{equation}
\label{eq:esp7}
T_{nm} = T^A_{nm} +  \ul{Y}^{\dagger}_{AB,n} \bs{D}_{AB,A} \ul{X}_{Am}
\end{equation}
This means that $T_{nm}$ is not separable due to the second term on the 
rhs arising from the non-local left eigenvector contributions.
In the exact (full bCC) treatment these non-local eigenvector components factorize according to 
\begin{equation}
\label{eq:eslev}
Y_{I_AI_B,n} = Y_{I_An} Y_{I_B0}
\end{equation}
that is, they form a products of 
excited state and ground-state eigenvector components for fragment $A$ and 
$B$, respectively. Then the non-separable term 
vanishes due to the orthogonality of the
fragment eigenvectors, $\ul{Y}^{A \dagger}_{n} \ul{X}^A_{m} =0$ and the form of 
$\bs{D}_{AB,A}$ (Eq.~\ref{eq:espsc2}).

For the diagonal case, $n=m$, the result is
\begin{equation} 
T_{nn} = T^A_{nn} + \sum_{I_AI_B} Y^*_{I_AI_B} X_{I_A} 
           \dirint{\overline{\Phi}^B_{I_B}}{\hat{D}_B} {\Psi^B_0}
\end{equation}
where again Eq.~(\ref{eq:espsc2}) has been used.
Only upon factorization of the non-local eigenvector components (Eq.~\ref{eq:eslev}) 
the correct result of Eq.~(\ref{eq:esp2}) is obtained.
%%%%%%%%%%%%%%%%
%%%%%%%%explain further
%%%%%%%%%%%%%%%%

Again we may perform a brief 
symbolic calculation to demonstrate the factorization (\ref{eq:eslev}) of the 
exact excited-state eigenvector components 
The starting point is Eq.~(\ref{eq:levmc}), where we may replace $\bs{M}_{A,AB}$
according to
\begin{equation}
\label{eq:symbx1} 
\bs{M}_{A,AB} = \bs{1}_A \times \ul{v}_{B}^t = - \bs{1}_A \times \ul{Y}^{\dagger}_{B0}\bs{M}_{BB}   
\end{equation}
to give 
\begin{equation}
\label{eq:levmc1} 
\ul{Y}^{\dagger}_{AB,n} =  - \ul{Y}^{\dagger}_{An}
                      \times \ul{Y}^{\dagger}_{B0} \bs{M}_{BB} (\omega_n -  \bs{M}_{AB,AB})^{-1}
\end{equation}
Here the general relation (\ref{eq:lgsvx}), specialized to fragment $B$,
\begin{equation} 
\ul{Y}^{\dagger}_{B0} = -  \ul{v}_{B}^t \bs{M}^{-1}_{BB}
 \end{equation}
has been used in Eq.~(\ref{eq:symbx1}).
Since $\ul{Y}_{An}$ is an eigenvector of $\bs{M}_{AA}$ it follows that 
\begin{eqnarray} 
\ul{Y}^{\dagger}_{An}(\omega_n -  \bs{M}_{AB,AB}) 
      & = & \ul{Y}^{\dagger}_{An}(\bs{M}_{AA} -  \bs{M}_{AB,AB})\\ 
      & = &  - \ul{Y}^{\dagger}_{An} \times \bs{M}_{BB}
\end{eqnarray}
and, as a consequence,
\begin{equation}
\ul{Y}^{\dagger}_{An}(\omega_n -  \bs{M}_{AB,AB})^{-1} 
= - \ul{Y}^{\dagger}_{An} \times \bs{M}_{BB}^{-1}
\end{equation}
Using the latter result in Eq.~(\ref{eq:levmc1}) gives  
\begin{equation}
\label{eq:levmcx}
\ul{Y}^{\dagger}_{AB,n}  =  \ul{Y}^{\dagger}_{An} \times \ul{Y}^{\dagger}_{B0}                   
\end{equation}
The factorization of the non-local eigenvector components is of course equivalent
to the factorization $\bra{\Psi^{(l)}_{n}} = \bra{\Psi^{(l)A}_{n}}\bra{\lo{\Psi}^B_0}$
of the expansion (\ref{eq:genexle}).

%%%%%%%%%%%%%%%%%%%%%%%%%%%%%%%%%%%%%%%%%%%%%
%%%%%%%%%%%%%%%%%%%%%%%%%%%%%%%%%%%%%%%%%%%%
As we have seen, the ordinary bCC expression~(\ref{eq:esp1})  
for the excited state-transition moments and properties is non-separable,
which here is due to the non-separable components in the left 
excited-state eigenvectors. 
Again, the CCLR approach results in an alternative
separable expression~\cite{koc90:3333}, reading
\begin{equation}
\label{eq:estmcclr}
T_{nm} = \dirint{\overline{\Psi}^{(l)}_n}{[\hat{D},\hat{C}_m]}{\Psi^{cc}_0}
- \sum_{I,J} \dirint{\overline{\Psi}^{(l)}_n}{[[\hat{H}, \hat{C}_I],\hat{C}_m]}{\Psi^{cc}_0}
            (\bs{M} + \omega_{mn})^{-1}_{IJ} \dirint{\overline{\Phi}_J}{\hat{D}}{\Psi^{cc}_0}
        + \delta_{nm} \dirint{\overline{\Psi}_0}{\hat{D}}{\Psi^{cc}_0}
\end{equation}
where $\omega_{mn} = \omega_{m} - \omega_{n}$. The separability of this form can be shown
in the same manner as in the case of the ground-to-excited
state transition moments (Sec.~IV.D). 
The equivalence of the CCLR form (\ref{eq:estmcclr}) and the ordinary bCC
expression 
of the excited-state transition moments is briefly addressed in App.~C.   

\clearpage

\section{Hermitian intermediate state representation}

The bCC representation is a mixed or hybrid representation 
made up from the CC states and the associated biorthogonal states. Whereas the 
(correlated excited) CC states are genuine intermediate states, being based on the  
exact ground state, the biorthogonal states are essentially of CI-type,
that is, excited HF configurations. As was analyzed
in the previous section,
the use of the biorthogonal CI-type states as the left expansion manifold 
downgrades to a certain extent the truncation errors and separability properties 
of the bCC computational schemes. For the purpose of 
comparison, we will briefly 
inspect the properties of a hermitian intermediate state representation (ISR),
specifically the ADC-ISR approach~\cite{sch82:2395,tro95:2299,mer96:2140,tro99:9982,
sch04:11449}, in the following.

As the bCC representation, the ADC-ISR approach starts from
the correlated excited states (Eq.~\ref{eq:ccstates}) 
\begin{equation}
\label{eq:excorstat}
\ket{\Psi^0_{J}}=
\hat{C}_{J} \ket{\Psi_0}
\end{equation}
where $\ket{\Psi_0}$ now refers to the normalized ground state
rather than to the  CC parametrization.
The correlated excited states (CES) can then be transformed into 
orthonormal intermediate states (IS),
\begin{equation}
\nonumber
\ket{\Psi^0_{J}} \longrightarrow \ket{\tilde{\Psi}_{J}} 
\end{equation}
via a (formal) Gram-Schmidt orthogonalization procedure, in which
successively higher CES classes $\mu$ are orthogonalized 
with respect to the already constructed lower IS classes $\nu = 1,2, \dots,\mu-1$.
Within a given excitation class, symmetric orthonormalization is adopted.
All the states are explicitly orthogonalized to the exact ground state,
forming a zeroth excitation class ($\mu = 0$).
As a result of this (so far purely formal) procedure, one 
obtains an orthonormal set of intermediate states $\ket{\tilde{\Psi}_{J}}$,
\begin{equation}
\bracket{\tilde{\Psi}_{I}}{\tilde{\Psi}_{J}} = \delta_{IJ}
\end{equation}
being, moreover, orthogonal to the exact ground-state, $\bracket{\tilde{\Psi}_{J}}{\Psi_0} = 0$.
 
Representing the (shifted) Hamiltonian 
$\hat{H}-E_{0}$ in terms of these intermediate states  
gives rise to a hermitian secular matrix ${\bf M}$,
\begin{equation}
\label{eq:adcsm}
M_{IJ}= \dirint{\tilde{\Psi}_{I}}{\hat{H}-E_{0}}{\tilde{\Psi}_{J}}
\end{equation}
and the associated hermitian eigenvalue problem,   
\begin{equation}
\label{eq:adcevp}
{\bf M X} = {\bf X \Omega},\qquad {\bf X^{\dagger} X = 1} \\
\end{equation}
Here ${\bf \Omega}$ is the diagonal matrix of excitation energies,
$\omega_{n} = E_n - E_0$, 
and ${\bf X}$ denotes the matrix of (column) eigenvectors.
The $n$-th excited state can be expanded as
\begin{equation}
\label{eq:adcexp}
 \ket{\Psi_n}= \sum_{J} X_{Jn} \ket{\tilde{\Psi}_{J}}
\end{equation}
in terms of the intermediate states and the eigenvector
components $X_{Jn}$. The transition moments
take on the form
\begin{equation}
\label{eq:isrtm}
T_n = \dirint{\Psi_n}{\hat{D}}{\Psi_0} = \sum_J F_J X^*_{Jn} 
\end{equation}
where
\begin{equation}
F_J = \dirint{\tilde{\Psi}_{J}}{\hat{D}}{\Psi_0}
\end{equation}
are denoted as IS transition moments.

To obtain practical computational schemes, the Gram-Schmidt procedure
is used together with Rayleigh-Schr\"odinger (RS) perturbation theory for
$\ket{\Psi_0}$ and $E_0$, generating explicit
perturbation expansions for the secular matrix
\begin{equation}
\label{eq:smpe}
\bs{M} = \bs{M}^{(0)} + \bs{M}^{(1)} + \bs{M}^{(2)} + \dots
\end{equation}
and the IS transition moments
\begin{equation}
\ul{F} = \ul{F}^{(0)} + \ul{F}^{(1)} + \ul{F}^{(2)} + \dots
\end{equation}
By truncating the IS manifolds and the perturbation expansions for
the secular matrix elements and IS transition moments in a systematic and consistent 
manner, one arrives at a hierarchy of ADC($n$) approximations, where $n$ indicates
that both the energies and transition moments of the lowest excitation class
(singly excited states) are treated consistently through order $n$. 
An alternative and, beyond 2nd order, preferable 
derivation of the ADC-ISR perturbation expansions is the original ADC
formulation~\cite{sch82:2395,tro99:9982} 
based on diagrammatic perturbation theory for the 
polarization propagator~\cite{Fetter:1971}.

As a distinctive feature of the ADC-ISR, the canonical order structure~\cite{mer96:2140}
applies to the entire secular matrix (see Fig.~7), 
\begin{equation}
M_{IJ} \sim O(|[I] - [J]|)
\end{equation}
and, as a consequence, also 
to the eigenvector matrix $\bs{X}$:
\begin{equation}
\label{eq:corevm}
X_{IJ} \sim O(|[I] - [J]|)
\end{equation}
In analogy to the last paragraph of App.~B.1, one readily obtains 
the truncation error formula
\begin{equation}
O^{[n]}_{TE}(\mu) =  2 O[\ul{X}_{\mu + 1,n}] = 2 (\mu - [n] +1), \;\; \mu \geq [n]
\end{equation}
for the excitation energies. Here $[n]$ denotes the class of the excited state $n$
(as established by the PT parentage), and $\mu$ specifies the truncation level
of the ISR expansion manifold.
In a similar way, the 
canonical order relations 
\begin{equation}
F_{J} \sim O([J] - 1)
\end{equation}
for the IS transition moments lead (via Eq.~\ref{eq:isrtm}) 
to the following expression for the truncation error 
in the transition moments: 
\begin{equation}
O^{[n]}_{TE}(\mu) =   O[\ul{F}_{\mu + 1}] +  O[\ul{X}_{\mu + 1,n}] = 2 \mu - [n] + 1,  
\;\; \mu \geq [n]
\end{equation}
For the lowest excitation class of the singly excited states ($[n] = 1$),
the truncation error is $2\mu$, both for the excitation energies and the
transition moments. In Table 1, the errors for the six lowest truncation levels
are compared to the corresponding CI and bCC values.

The treatment of 
excited state properties and transition moments, 
\begin{equation}
T_{nm} = \dirint{\Psi_n}{\hat{D}}{\Psi_m} = \ul{X}^{\dagger}_n \bs{D} \ul{X}_m 
\end{equation}
is based on the ISR of a general one-particle operator,
\begin{equation}
D_{IJ}= \dirint{\tilde{\Psi}_{I}}{\hat{D}}{\tilde{\Psi}_{J}}
\end{equation}
Like the secular matrix (Eq.~\ref{eq:smpe}), the
IS property matrix $\bs{D}$ is subject to a perturbation expansion,
\begin{equation}
\bs{D} = \bs{D}^{(0)} + \bs{D}^{(1)} + \bs{D}^{(2)} + \dots
\end{equation}
Here the ``shifted canonical'' order relations
\begin{equation}
D_{IJ} \sim O(|[I] - [J]| - 1), \;\; |[I] - [J]|\geq 1
\end{equation}
apply, reflecting that a one-particle operator 
can couple HF (zeroth order) excitations of successive excitation classes.
The product $\ul{Z}_n = \bs{D}\ul{X}_n$ modifies the canonical order relations
of an ADC-ISR eigenvector accordingly, that is, 
\begin{equation} 
Z_{In} \sim O([I] - [n] - 1), \;\; [I] - [n] \geq 1
\end{equation}
This leads readily to the 
expression
\begin{equation}
O^{[n]}_{TE}(\mu)  =  O[\ul{X}_{\mu + 1,n}] +  O[\ul{Z}_{\mu + 1,n}] = 2 (\mu - [n]) + 1,  
\;\; \mu \geq [n]
\end{equation}
for the truncation error of excited state property matrix elements $T_{nn}$.

The ADC-ISR secular matrix is fully 
separable~\cite{sch96:329, mer96:2140}: it has a 
diagonal partitioning structure as shown in Fig.~8, and the 
diagonal blocks are identical with the corresponding fragment secular matrices,
that is, $\bs{M}_{AA} = \bs{M}^{(A)}$. 
For a local excitation, say on fragment $A$,
the fragment and entire system treatments give the same excitation energy, and the
fragment eigenvector is part of the entire system eigenvector, in which all non-local  
components vanish ($X_{I_B,n} = X_{I_AB,n} = 0$). The IS transition moments are
separable as well, $F_{I_A} = F^{(A)}_{I_A}$. Together with the separable eigenvectors
this ensures size-consistent results for the ADC-ISR transition moments. The case of 
the excited state properties and transition moments has been discussed 
in Ref.~\cite{sch04:11449}.

\section{Concluding remarks}

The basic concept
underlying the CCLR and EOM-CC methods for electronic excitation in atoms and molecules
consists in a specific biorthogonal representation 
of the (shifted) Hamiltonian in terms of two distinct sets of states:
on the one hand, the set of excited CC states based the CC ground state,
and, on the other hand, the set of their biorthogonal counterparts. 
This results in a non-hermitian (bCC) secular matrix. The excitation energies are obtained
as the eigenvalues of the bCC secular matrix, while both the left and right eigenvectors
enter the calculation of the spectral intensities.
The two sets of states are of quite different quality. While the latter (biorthogonal) basis
states are essentially of CI-type, that is, excited Hartree-Fock (HF) configurations, the 
excited CC states, by contrast, are formed by applying physical excitation operators to the 
exact $N$-electron ground state. The resulting correlated excited (CE) states
are expected to be superior to the simple CI states as they already account
for a major part of electron correlation. The intuitive idea here is that electron correlation 
in excited states should not be completely different from that in the ground state.
In fact, the use of CE intermediate states warrants distinctive advantages over the
simple CI treatment. Foremost, this concerns the truncation error associated with limited expansion 
manifolds. To attain comparable accuracy, the manifold of CE states can be truncated
at distinctly smaller excitation levels than the CI expansions. Moreover, the excited CC states
are intrinsically separable with regard to (hypothetical) non-interacting fragments,
which, in contrast to CI, allows to devise size-consistent approximation schemes
based on truncated expansion manifolds.

However, due to the equitable use of the biorthogonal set of states being essentially
of CI-type, the bCC representation must be viewed as a CI-ISR hybrid rather than a full ISR
approach. This becomes strikingly manifest in the split order structure of the
bCC secular matrix, being canonical and of CI-type in the LL and UR parts, respectively. 
As a result, the truncation error and separability properties are 
clearly superior to those of CI, but also
weaker than those of a full ISR method such as the ADC-ISR presented in Sec.~V. 
This is reflected in the
general truncation error order (TEO) formulas derived here 
for excitation energies, transition moments, and 
property matrix elements. In the case of single excitations at truncation level 2
(that is, neglecting triply and higher excited configurations), the TEOs in the 
excitation energies and transition moments are 2, 3, and 4 for CI, bCC, 
and ADC, respectively. At higher truncation levels the gap between CI and bCC, as well as
that between bCC and ADC widens. At the (already somewhat academic) truncation level 6
the respective TEOs are 6, 9, and 12. 
Of course, a given approximation may not exhaust the margin afforded by the 
respective TEO.   
For example, the error in the transition moments of the CC2 scheme 
is of PT order 2
(due to the first-order approximation used for the $T_2$ amplitudes), while the
truncation error is of PT order 3. The ADC(2) approximation allows for a consistent 
treatment of (single excitation) energies and transition moments through 2nd order,
the TEO being 4.
Let us note that the PT order of the overall error
is only an indicator for the quality of an approximation scheme.     
A large error order does not by itself guarantee accurate results,  
but rather must be seen as a necessary condition for accuracy: a certain accuracy level 
can only be attained in compliance with a corresponding PT order of the characteristic error.

The hybrid character of the bCC representation is also reflected in the 
separability properties. The excitation energies, given as the roots of the 
characteristic polynomial, are separable, which, in principle, ensures size-consistent
results at approximative levels beneath the full bCC treatment. It should be
noted, however, that the separability of the eigenvalues applies strictly speaking only
to the (fictitious) separated fragment model. Allowing for a small interaction
between the fragments $A$ and $B$, which is a more realistic simulation 
of an extended system,
the separability becomes blurred because the coupling block $\bs{M}_{AB,A}$ (see Fig.~4)
no longer vanishes, and local and non-local excitations may mix. 
What is problematic here is that the 
small matrix elements of $\bs{M}_{AB,A}$, associated with the weak physical coupling 
of the fragments, form products, e.g. in the characteristical polynomial, 
with the large (non-local coupling) matrix elements of  
$\bs{M}_{A,AB}$. The mixing can become substantial when local and non-local
excitations are nearly degenerate. 
This problem has been addressed by Helgaker \emph{et al.} (see p. 684), but a 
thorough dedicated study seems to be still outstanding.

The bCC eigenvectors
do not perform uniformly, as the left and right  
eigenvectors are non-seperable and 
seperable, respectively. In the associated 
left and (ordinary) right transition moments,
both of which are needed for the computation of spectral intensities, the
separability properties are reversed: separable left TMs, as opposed to non-separable right TMs. 
This reversal is due to the use of the dual ground-state and the CC ground-state
in the right and left TMs, respectively. As a result, the spectral intensities 
based on the ordinary bCC TMs are not size-consistent. 
This problem does not arise within the CCLR framework, as here a separable, though
more involved  
expression for the right TM is used. In the full bCC limit the CCLR expression and 
the ordinary one, used in the EOM-CC methods, are equivalent. At approximate
levels, however, the consistency of the left (ordinary) and right (CCLR) TMs
may become an issue. At the simple CCS (singles) level, for example, both the
left and ordinary right TMs are consistent through zeroth order (due to the use
of the HF ground state). The right TM in the CCLR formulation, however, is consistent
through first order. This means that the CCLR spectral intensity expression  
combines a zeroth-order left TM and a first-order right TM.
One might be inclined to see this as an improvement,
but it should be noted that a result comprising incomplete first-order terms may 
be inferior to that of a consistent zeroth-order approximation. The consistency problem
emerges also in the
finding that for transitions with low spectral strengths the signs of the
left and right CCLR transition moments may differ, leading to unphysical (i.e. negative) 
intensities. This is to say that the CCLR results are not necessarily more accurate than
those of EOM-CC as long as size-consistency does not play a role. A 
conclusive comparative
test of EOM-CC and CCLR intensity results for 
smaller and medium-sized molecules would be highly desirable. Such a study
should also comprise excited state properties for which both  
ordinary non-separable bCC or separable CCLR expressions are available.

Of course, the most
obvious drawback of the bCC methods   
is the non-hermiticity of the respective secular matrices $\bs{M}$. 
As already discussed, both the right and left eigenvectors are needed
when spectral intensities and property matrix elements are to be
determined, which requires an additional effort compared to the
hermitian eigenvalue problem of a full ISR method.  
In the case of degenerate eigenvalues, care has to be taken to 
ensure the proper biorthogonalization of the associated 
sets of left and right eigenvectors. More disturbing is the 
possibility of complex eigenvalues.   
Even though the 
underlying Hamiltonian is hermitian, so that the excitation energies
obtained as eigenvalues of $\bs{M}$ must ultimately (in the full bCC limit) 
be real quantities, 
one may encounter problems in actual computations. As first noticed 
and analyzed by
H\"{a}ttig~\cite{hae05:37},
complex eigenvalues can occur in the vicinity of conical intersections
of two excited state energy surfaces (or hyper-surfaces). 
K\"{o}hn and Tajti\cite{koe07:044105} have developed some ideas of how to 
deal with that situation, but so far there remain open questions.

%%%%%%%%%%%%%%%%%%%%%%%%%%%%%%%%%%%%%%%%%%%%%%%%%%%%%%%%%%%%%%%%%%%%%%%%%

The present analysis of the bCC methods for (neutral) electron excitations in an 
$N$-electron system can readily be extended to the case of generalized
excitations, such as ($N$-$1$)- or ($N$+$1$)-electron 
excitations used in the treatment of ionization or electron attachment
processes, respectively. Corresponding CC 
methods have been referred to as 
IP-EOM-CC and EA-EOM-CC~\cite{gho81:173,sta94:8938,noo95:3629,hir00:459}.
In the case of 
(single) ionization the neutral operators (\ref{eq:xops})  
 have to be replaced
by the manifold
\begin{equation}
\label{eq:ionops}
\{ \hat{C}_{J}\} \equiv \{c_{k};
c_{a}^{\dagger}c_{k}c_{l},k<l;\ldots \}
\end{equation}
of physical $1h$-, $2h$-$1p$-, $3h$-$2p$-,$\dots$ operators. As in the case of neutral 
excitations, the successive ($N$-$1$)-electron excitation classes of 
$\mu h$-($\mu$-1)$p$ excitations
are labeled by $\mu = 1,2,\dots$; a corresponding classification, $[n]=1,2,\dots$,
applies to the cationic energy eigenstates $\ket{\Psi^{N-1}_n}$, indicating the respective
PT parentage. In the ($N$-$1$)-electron case (as in the other 
generalized excitations) the 
complication arising from an admixture of the $N$-electron ground state in the neutral 
excitations (Eqs.~\ref{eq:rxsv}-\ref{eq:rrxs})  
does not apply, which somewhat simplifies the ionic bCC equations. 
With a few obvious adjustments, 
the discussion and the findings for the neutral excitations can readily be 
transferred to the case of ($N$-$1$)-electron (and the other generalized) bCC schemes.
It should be noted that here
generalized transition moments 
$T_n = \dirint{\Psi_0}{\hat{D}}{\Psi^{N-1}_n}$ come into play,
defined with respect to a suitable electron removal (or attachment) operator of the type 
$\hat{D} = \sum_p d_p c^{\dagger}_p$. This means that in the
bCC representation 
$D_{IJ} = \dirint{\overline{\Phi}_I}{\hat{D}}{\Psi^{0}_J}$ to be used
in the analogs of Eqs.     
(\ref{eq:rtmor1},\ref{eq:1pbcc}), the states $\bra{\overline{\Phi}_I}$ on the 
left side of the matrix elements are the    
$N$-electron biorthogonal states (Eq.~\ref{eq:bostates} based on the neutral
operators \ref{eq:xops}).
In the discussion of ionic state properties and transition moments according to Sec.~IV.E,
again a particle-number conserving operator $\hat{D}$ is to
be considered.
The corresponding bCC representation, 
$D_{IJ} = \dirint{\overline{\Phi}_I}{\hat{D}}{\Psi^{0}_J}$, is a pure ($N$-$1$)-electron
representation, where
both $\bra{\overline{\Phi}_I}$ and
$\ket{\Psi^{0}_J}$ are based on the operators (\ref{eq:ionops}).    

%%%%%%%%%%%%%%%%%%%%%%%%%%%%%%%%%%%%%%%%%%%%%%%%%%%%%%%%%%%%%%%%%%%%%%%%%%%%%%%%%%%

The analysis given here of the EOM-CC and CCLR methods from the perspective of the bCC
representation has shown decisive advantages over the conventional CI treatment,
but also distinctly weaker TEO and separability properties than those of a 
full ISR approach such as the ADC-ISR. It should be noted, however, that
the latter approach is manifestly based on perturbation theory for the 
secular matrix elements and effective transition coefficients, behaving essentially 
like the Rayleigh-Schr\"{o}dinger (RS) PT expansions of the ground state energies and
CI expansion coefficients. This means that both the ADC methods and ground-state PT 
have the same condition of applicability, namely a sufficiently large energy gap
between the occupied and virtual HF orbital energies. When the energy gap becomes too small,
for example at bond breaking nuclear conformations, 
PT based methods are bound to fail.  
The bCC quantities (secular matrix elements and basis set transition moments), on the 
other hand, are based on the T-amplitudes of the CC ground state, which can be
determined in a completely non-perturbative way. Yet this edge over methods involving PT
must be relativized, as the CC approach 
breaks down as well in situations where the ground-state is no longer adequately 
described by a dominant single reference configuration  
(see Bartlett and Musial~\cite{bar07:291}, Sec.VI.C, and references therein). 
The reason is that the usual single-reference CC ansatz is ill-suited to deal 
with the so-called static correlation. As a remedy for this deficiency, 
much effort has been devoted
to developing multi-reference (MR) CC 
schemes~\cite{lin79:3827,hos79:3827,jez81:1668,haq84:5058,lin87:93,pal88:4357,
muk89:561}
(for a more complete list of references
and an introduction into the vast field
of MRCC methods the reader is referred to Sec. IX in the recent review article by 
Bartlett and Musial~\cite{bar07:291}).
However, the MRCC approach to ground and excited states
is far more complex than the single reference bCC representation considered here,
and it has to be seen whether really effective computational schemes will emerge.

\clearpage

\section*{Acknowledgements}

One of us (JS) is indebted to Anthony Dutoi for illuminative
discussions on various aspects of the ground-state coupled-cluster
method.

\appendix
\renewcommand{\theequation}{A.\arabic{equation}}
\setcounter{equation}{0}
\section*{Appendix A: Order relations of bCC representations}

A general proof of the canonical order relations in the lower left (LL) triangle
of the bCC secular matrix
can be found in Ref.~\cite{mer96:2140}. A brief review of the derivation of 
these order relations is given in the following.

Let us first consider the simpler case of a one-particle operator $\hat{D}$,
reading in second-quantized notation 
\begin{equation}
\hat{D} = \sum d_{pq} c^{\dagger}_p c_q
\end{equation}
where $d_{pq} = \dirint{\phi_p}{\hat{d}}{\phi_q}$ denote the 
one-particle matrix elements associated with $\hat{D}$. 
The bCC representation of $\hat{D}$, 
\begin{eqnarray}
\nonumber
D_{IJ} & = & \dirint{\overline{\Phi}_I}{\hat{D}}{\Psi^{0}_J}\\
\label{eq:1pbccmat}
       & = & \dirint{\Phi_I}{e^{-\hat{T}} \hat{D}\, e^{\hat{T}}}{\Phi_J}
\end{eqnarray}
was encountered in the treatment  
of transition moments and excited state properties, as discussed in Secs.~IV.B and E 
(see Eq.~\ref{eq:1pbcc}).  

The bCC representation matrix $\bs{D}$ has an order structure associated with
the partitioning according to excitation classes, as shown in Fig.~5.
In the upper right (UR) triangle one finds the familiar CI structure 
for a one-particle operator. This result follows along the lines of the first paragraph
in Sec.~IV.A. In the lower left (LL) triangle the 
canonical order relations
\begin{equation}
\label{eq:or4d}
  D_{IJ} = O([I]- [J] - 1),\;\; [I] > [J]
\end{equation}
apply, which is to be shown in the following.

The operator in the bCC matrix element (\ref{eq:1pbccmat})
has a finite Baker-Hausdorff (BH) expansion,  
\begin{equation}
\label{eq:BHexp}
e^{-\hat{T}} \hat{D}\, e^{\hat{T}} = \hat{D} + [\hat{D}, \hat{T}] + \half \, 
[[\hat{D}, \hat{T}],\hat{T}] 
\end{equation}
terminating here already after the double commutator term because 
\begin{equation}
\hat{T} = \sum t_I \hat{C}_I
\end{equation}
consists of physical excitation operators only, and $\hat{D}$ has at most two unphysical
operators. Let us now write the $\hat{T}$ operator according to  
\begin{equation}
\hat{T} = \sum \hat{T}_{\mu}
\end{equation}
in terms of individual class operators $\hat{T}_{\mu},\, \mu = 1,2,\dots$. 
The $T$-amplitudes, being themselves subject of a well-defined 
(diagrammatic) perturbation theory, exhibit
the order relations (see Hubbard~\cite{hub57:539})
\begin{equation}
\label{eq:tor}
\hat{T}_{\mu} \sim O(\mu - 1),\,\,\,\, \mu > 1
\end{equation}
This means, for example, that the
PT expansions of the $T_2$ amplitudes,
\begin{equation} 
\hat{T}_{2} =  \sum t_{abij} \hat{C}_{abij}
\end{equation}
begin in first order. The $T_1$ amplitudes ($\mu = 1$), being of 2nd order, are an
exception reflecting Brioullin's theorem. 

What are the consequences of the expansion (\ref{eq:BHexp}) and the order
relations (\ref{eq:tor})? Since the 
BH expansion (\ref{eq:BHexp}) begins with $\hat{D}$,
there will be non-vanishing zeroth-order contributions to $D_{IJ}$ for $[I] = [J]$ and
$[I] = [J]+ 1$.  Now suppose that $I$ and $J$ differ by more than one class, that is,
$[I] \geq [J]+ 2$. In that case non-vanishing contributions in $D_{IJ}$ will 
arise only if there are terms in the BH expansion that are at least of rank 
$r = [I] - [J]$. Here, the rank of an operator is the number of its $c^{\dagger}$ 
(or $c$) factors. For example, $\hat{D}$ is of rank 1 and the $\hat{T}_{\mu}$ 
operators are of rank $\mu$. Now it is readily established that the commutators
$[\hat{D}, \hat{T}_{\mu}]$ and $[[\hat{D}, \hat{T}_{\mu}],\hat{T}_{\nu}]$ are of
rank $\mu$ and $\mu + \nu -1$, respectively. (A commutator of two operators 
$\hat{A}$ and $\hat{B}$ with definite ranks, $a$ and $b$, respectively, is of rank 
$a + b - 1$.) 
To determine the lowest (non-vanishing) PT contribution to the $D_{IJ}$ matrix 
elements, one has to inspect the terms of the BH expansion (\ref{eq:BHexp}) having 
rank $r = [I]-[J]$ (which is the lowest rank allowing for non-vanishing matrix elements)
and find the lowest PT order of those terms.   
For example, $[\hat{D}, \hat{T}_{2}]$ is of rank 2 and PT order 1, which means that
for $[I] = [J] + 2$ the PT order of $D_{IJ}$ is 1. In the general case, 
$[I] = [J] + \mu, \, \mu \geq 3$,   
terms with the required rank $r = \mu$ and lowest PT order 
are due to the
$[\hat{D}, \hat{T}_{\mu}]$ commutators, being of rank $\mu$ and PT order
$\mu - 1$. Likewise, also the double commutator $[[\hat{D}, \hat{T}_{2}],\hat{T}_{\mu-1}]$
gives rise to terms with rank $\mu$ and order $\mu -1$, but there are no rank $\mu$
terms with PT order lower than $\mu -1$. This proves the order relations (\ref{eq:or4d}).     

Let us note that the order relations 
$\dirint{\overline{\Phi}_I}{\hat{D}}{\Psi^{cc}_0} \sim  O([I] - 1)$ for 
the left basis state transition moments (Eq.~\ref{eq:ltm3}) follow as special case ($[J]=0$).

In a similar way the canonical order relations
\begin{equation}
\label{eq:or4sm}
 M_{IJ} = O([I]-[J]),\;\; [I] \geq [J]
\end{equation}
for the bCC secular matrix (LL triangle) elements,
\begin{eqnarray}
\nonumber
M_{IJ} & = & \dirint{\overline{\Phi}_I}{\hat{H} - E_0}{\Psi^0_J}\\ 
      & =  & \dirint{\overline{\Phi}_I}{e^{-\hat{T}} [\hat{H}, \hat{C}_J]\, e^{\hat{T}}}{\Phi_0}
\end{eqnarray}
can be established. Now we have to consider the BH expansion involving   
 the commutator $\hat{K}_J = [\hat{H},\hat{C}_J]$ and check the
emerging transition matrix elements of the type 
$\dirint{\Phi_I}{\hat{O}}{\Phi_0}$.   
In contrast to the case of the transition operator considered above,
$\hat{K}_J$ is itself of PT order 1 and of rank $[J] + 1$
(regarding here only the relevant two-particle part of the Hamiltonian). The BH expansion
\begin{equation}
\label{eq:BHexpx}
e^{-\hat{T}} \hat{K}_J\, e^{\hat{T}} = 
\hat{K}_J + [\hat{K}_J, \hat{T}] + \half \, [[\hat{K}_J, \hat{T}],\hat{T}] +
\sixth \, [[[\hat{K}_J, \hat{T}],\hat{T}],\hat{T}]
\end{equation}
terminates after the triple commutator, since $K_J$ has not more than three
unphysical $c^{\dagger}$($c$) operators. 
Let us consider a secular matrix element $M_{IJ}$, where $[I] = [J] + \mu, \, \mu \geq 1$.
Obviously, terms of the BH expansion (\ref{eq:BHexpx}) do not contribute to  $M_{IJ}$
if their rank is smaller than $[I]$. As above, we may analyze the terms
of rank $r=[J] + \mu$ with respect to their  
PT order. For $\mu = 1$, the first term $\hat{K}_J$ on the rhs of 
Eq.~(\ref{eq:BHexpx}) is of rank $[J] + 1$ and order 1, thus giving rise
to a first order contribution to $M_{IJ}$.
For higher values of $\mu$, it suffices to consider the commutators 
$[\hat{K}_J, \hat{T}_\mu]$, being of the required rank $r = [J] + \mu$ and PT
order $\mu$. Again it is readily established that there are no rank $r = [J] + \mu$
terms of lower PT order.

\appendix
\renewcommand{\theequation}{B.\arabic{equation}}
\setcounter{equation}{0}
\section*{Appendix B: Order relations of CI and bCC eigenvector matrices}

The order structure of the CI and bCC secular matrices give rise to specific
order relations for the eigenvector matrices, which, in turn, imply the 
respective truncation errors in the excitation energies and transition moments. 
In the following we will first consider the CI eigenvector matrix (B.1),
and then turn to the left and right eigenvector matrices associated with the bCC 
representation (B.2). In the third subsection B.3 we shall show how  
order relations established only for a triangular part of a matrix
can be extended to the entire matrix as a consequence of unitarity.

\subsection{CI eigenvector matrix}

The order relations of the CI eigenvector matrix rely  
on perturbation theory (PT) for the exact states.
Let us first consider the familiar case of the ground-state, where
the well-known Rayleigh-Schr\"odinger PT can be cast in the compact 
form
\begin{equation}
\label{eq:RSPT}
\ket{\Psi_0} = \ket{\Phi_0} + 
    \sum_{\nu=1}^{\infty} \left[\frac{\hat{Q}_0}{E^{(0)}_0 - 
       \hat{H}_0} (\hat{H}_I - E_0 +E^{(0)}_0)\right]^{\nu} \ket{\Phi_0}
\end{equation}
Here, the usual M\o{}ller-Plesset decomposition of the Hamiltonian,
\begin{equation}
\hat{H} = \hat{H}_0 + \hat{H}_I 
\end{equation}
into an unperturbed (HF) part $\hat{H}_0$ and an interaction part  $\hat{H}_I$
is supposed; $\ket{\Phi_0}$ is the (HF) ground-state of $\hat{H}_0$ with the energy
$E^{(0)}_0$, and $\hat{Q}_0 = \hat{1} - \ket{\Phi_0}\bra{\Phi_0}$.
To determine the lowest (nonvanishing) PT order for a 
specific eigenvalue component,
\begin{equation}
X_{J0} = \bracket{\Phi_J}{\Psi_0}
\end{equation}
one has to analyze the contributions arising from 
the expansion on the rhs of Eq.~(\ref{eq:RSPT}). In $\nu$-th order the leading
operator term is $\hat{H}_I^{\nu}$. 
Due to the two-electron (Coulomb repulsion) part of
$\hat{H}_I$, the matrix element
$\dirint{\Phi_J}{\hat{H}_I^{\nu}}{\Phi_0}$ vanishes if the excitation class of $J$
exceeds the value $2\nu$. For the excitation classes $[J] = 2 \nu$ and $[J] = 2 \nu - 1$,
on the other hand, the matrix element gives rise to a non-vanishing $\nu$-th 
order contribution. Obviously, there is no lower-order coupling between
the HF ground-state and excitations of class $2 \nu$ and $2 \nu - 1$. This means 
that $X_{J0}$ is of PT order $\nu$ for $[J] = 2 \nu$ and $[J] = 2 \nu -1$. 

This result can also be written in the form
\begin{equation}
\label{eq:x0or}
O[\ul{X}_{\mu0}]  =  \begin{cases}
	 \half \mu, &  \mu \text{ even}\\
 	 \half (\mu + 1), &  \mu \text{ odd},\, > 1
      \end{cases}
\end{equation}
where $\mu$ denotes collectively the configurations of class $\mu$.
The $p$-$h$ excitation class ($\mu = 1$) is an exception, as here $X_{J0} \sim O(2)$
due to Brioullin's theorem. The ground-state component $X_{00}$ is of course
of zeroth order. In Fig.~1, the order structure of $\ul{X}_0$ is depicted.
%%%%%%%%%%%%%%%%%%%

Now we turn to the order relations of
excited states $\ket{\Psi_n}$. Rather than using individual PT expansions,
the following analysis will be based directly on the order structure of the CI secular 
matrix (Fig.~1). However, a remark concerning the 
significance of excited-state perturbation theory is appropriate.
As is well-known, PT expansions for excited states and
excited state energies are of little practical use because the 
possibility of
small or vanishing denominators (``dangerous denominators'') in the 
PT expansions prevents meaningful computational results. In a formal
sense, however, excited-state PT expansions can be generated
analogously to the ground-state case, which then 
can be used to analyze, e.g., truncation errors of 
excited-state energies and transition moments. Underlying such 
a formal PT is the concept that each excited state is related to
a specific CI state,  
\begin{equation}
\nonumber
\ket{\Psi_n} \leftarrow \ket{\Phi_J}
\end{equation}
from which it emerges when the scaled interaction, e.g., in the form $\lambda \hat{H}_I$,
is gradually increased from $\lambda = 0$ to $1$. 
For our purpose we do not need 
the individual PT descent of an exact excited state. It suffices to suppose
that the exact states can be classified 
according their derivation from the (unperturbed) CI excitation classes,  
$p$-$h$, $2p$-$2h$, $\dots$, etc.
Analogously to the notation used for the CI excitation classes, 
we will denote by $[n]$ the class of the exact state $\ket{\Psi_n}$,
that is, $[n] = \mu$ if the excited state $n$ derives from the
$\mu p$-$\mu h$ class of CI states.  
The classification of both the CI and the exact states allows one to
partition the CI eigenvector matrix $\bs{X}$ into subblocks $\bs{X}_{\mu \nu}$,
where $\mu$ and $\nu$ refer to the component and state classes, respectively.
Fig.~9 shows the partitioning and the associated order structure
of $\bs{X}$. A general expression for the order structure is as follows: 
\begin{equation}
\label{eq:xnor}
O[\bs{X}_{\mu \nu}] = \begin{cases}
	 \half |\mu - \nu|, &  \mu - \nu \text{ even}\\
 	 \half |\mu - \nu| + \half, & \mu - \nu  \text{ odd}
      \end{cases}
\end{equation}

%%%%%%%%%%%%%%%%%%%%%%%%%%%%%%%%%%%%%%%%%%%%%%%%%%%%%%
%%%%%%%%%%%%%%%%%
%%%%%%
The order relations of the eigenvector matrix reflect the 
underlying order structure of the CI secular matrix. 
We begin by considering  
the class of singly excited states ($[n] = 1$). Rather than dealing with individual eigenvectors,
we can treat the entire set of class-1 eigenvectors at the same time. Let us 
therefore denote by $\ul{\bs{X}}_{1}$ the rectangular matrix formed by all 
(column) eigenvectors $\ul{X}_n$ with $[n] = 1$. 
The eigenvalue equations for the eigenvectors of class 1 can be written
compactly as  
\begin{equation}
\label{eq:phev}
\bs{H} \ul{\bs{X}}_{1} =  \ul{\bs{X}}_{1} \bs{\Omega}_{1} 
\end{equation}
where $\bs{\Omega}_{1}$ denotes the diagonal matrix of the $p$-$h$ energy eigenvalues.
Since any eigenvalue $\omega_n$ has an orbital energy (zeroth-order) contribution,
that is, $\bs{\Omega}_{1} \sim O(0)$,  
Eq.~(\ref{eq:phev}) leads to the following order equation
\begin{equation}
\label{eq:phevor}
O[\bs{H} \ul{\bs{X}}_{1}] = O[\ul{\bs{X}}_{1}]
\end{equation}
This equation can be used to establish successively the individual orders
of the component blocks, $\bs{X}_{k,1}$. Obviously, the starting point is given by
$\bs{X}_{1,1} \sim \bs{1} + O(1)$, which merely reflects the fact
that the singly excited states derive from the 
$p$-$h$ CI configurations. To proceed we inspect the matrix-vector block products 
$\sum_j \bs{H}_{k,j} \bs{X}_{j,1}$ for successive values of the (row) index $k$.
(To visualize these products it is helpful to write the order structure  
of $\bs{H}$ (Fig.~1) alongside the $\ul{\bs{X}}_{1}$ column matrix and fill in the 
successively determined order entries here, starting with the entry 0 in the 
$\bs{X}_{1,1}$ subblock.) 
For $k=2$ we may readily conclude 
that     
\begin{equation}
\bs{X}_{2,1} \sim \sum_{j \geq 1} \bs{H}_{2,j} \bs{X}_{j,1} \sim O(1)  
\end{equation}
where the first-order behaviour comes from the first term in the sum, 
$\bs{H}_{2,1} \bs{X}_{1,1} \sim \bs{H}_{2,1} \sim O(1)$. (Note  
that the diagonal eigenvector block
behaves as $\bs{X}_{1,1} = \bs{1} + O(1)$.)    
In a similar way,
we may establish that also the next subblock is of first order,
$\ul{\bs{X}}_{3,1} \sim O(1)$. For the 4-th subblock the situation changes 
since the $ \bs{H}_{4,1}$  matrix block vanishes so that here and beyond
the zeroth-order $\bs{X}_{1,1}$ block drops out.
One here obtains     
\begin{equation}
\ul{\bs{X}}_{4,1} \sim \sum_{j \geq 2} \bs{H}_{4,j} \bs{X}_{j,1} \sim O(2)  
\end{equation}
where the second-order behaviour of $\bs{X}_{4,1}$
derives from the first two summands involving the two first-order
subblocks $\bs{X}_{2,1}$ and $\bs{X}_{3,1}$. In the next step, one
first-order block drops out of the matrix-vector product, 
but the second one, $\bs{X}_{3,1}$,
combined with the $\bs{H}_{5,3}$ block of the secular matrix, again leads to
2nd-order behavior of $\bs{X}_{5,1}$. Only for $k=6$  
the order jumps to 3, since now 
due to the structure of the secular matrix the first order blocks 
(and the zeroth-order block) no longer contribute to the matrix-vector product.
Continuing in this way, the order 
relations of successively higher-class subblocks can be obtained. The general 
pattern is that for each even $k$ the order of the  $\bs{X}_{k,1}$ block
increases by 1. Let us note that the procedure can readily 
be cast in the formally correct form of induction.    

Now we may proceed to the next higher class of $2p$-$2h$ states. 
Let $\ul{\bs{X}}_{2}$ denote the $2p$-$2h$ eigenvector matrix 
with the (sub)blocks $\bs{X}_{k,2}$. In accord with the PT origin of the 
$2p$-$2h$ states here the $k=2$ (diagonal) block is of zeroth order, more
specifically, $\bs{X}_{2,2} \sim \bs{1} + O(1)$. 
But the order of the first block, $\bs{X}_{1,2}$ is fixed as well. 
As will be demonstrated in subsection B.3, the orthogonality of 
the $p$-$h$ and $2p$-$2h$ eigenvectors requires that $\bs{X}_{1,2}$
is of first order. In a similar way as above, the orders of the
higher $k$ blocks can now be derived successively from the    
matrix-vector products
\begin{equation}
\bs{X}_{k,2} \sim \sum_{j \geq 1} \bs{H}_{k,j} \bs{X}_{j,2}   
\end{equation}
It is readily established that the 
$\bs{X}_{3,2}$ and  $\bs{X}_{4,2}$ blocks are of 1-st order, 
followed by two blocks of 2-nd order, and so forth. 

The procedure outlined here for the $p$-$h$ and $2p$-$2h$ states can
easily be extended to higher excitation classes, $\mu$. The diagonal block,
$\bs{X}_{\mu,\mu} \sim \bs{1} + O(1)$ is always of zeroth order, while the order relations for 
blocks above
the diagonal, $k < \mu$, are determined by the orthogonality between the
eigenvectors of class $\mu$ and those of the lower classes (see subsection
B.3 below). In the $3p$-$3h$ states, for example, the orthogonality 
constraint with respect to the 
$p$-$h$ and $2p$-$2h$ states requires the 
$\bs{X}_{1,3}$  and $\bs{X}_{2,3}$ blocks to be of 1st order.
The full procedure for establishing the order relations of $\bs{X}$,
ascending both to higher excitation classes and higher blocks 
within a given excitation class, can, of course, be reformulated in a
formally satisfactory way making use of induction.

%%%%%%%%%%%%%%%%%%%%%%%%%%%%%%%%%%%%%%%%%%%%%%%%%%%%%%%%%
Once the order structure of the eigenvectors has 
been established, it is straightforward to  
analyze the truncation errors of the excitation energies. 
For this purpose one has to express an eigenvalue according to
\begin{equation}
\label{eq:evto1}
E_n = \ul{X}^{\dagger}_n \bs{H} \ul{X}_n 
 \end{equation}
as an energy expectation value, involving the full secular matrix $\bs{H}$
and the exact eigenvector $\ul{X}_n$. This expectation value can be 
written more explicitly as
\begin{equation}
\label{eq:evto2} 
E_n = \sum _{\kappa,\lambda} \ul{X}^{\dagger}_{\kappa n} \bs{H}_{\kappa,\lambda} \ul{X}_{\lambda n}   
\end{equation}
where the greek subscripts are referring to excitation classes rather than to individual
configurations. To specify the error arising from truncating the CI manifold after 
class $\mu$, one has to inspect the PT order of the terms with 
$\kappa = \mu +1, \lambda \leq \mu + 1$ (or $\lambda = \mu +1, \kappa \leq \mu + 1$).
Due to the structure of $\bs{H}$ it suffices to consider the
diagonal contribution $\ul{X}^{\dagger}_{\mu+1,n} \bs{H}_{\mu + 1,\mu + 1} \ul{X}_{\mu + 1, n}$.
Since $\bs{H}_{\mu + 1,\mu + 1} \sim O(0)$ the latter term is of the order  
\begin{equation}
O[\ul{X}^{\dagger}_{\mu + 1,n} \ul{X}_{\mu + 1,n}]
                                         = 2 \, O[\ul{X}_{\mu +1,n}]
\end{equation}
Using     
Eq.~(\ref{eq:xnor}) this translates into 
the general truncation error formula
\begin{equation}
\label{eq:tecien}
O^{[n]}_{TE}(\mu) =  \begin{cases}
	 \mu - [n] + 2, & \mu - [n] \text{ even}\\
 	  \mu -[n] + 1, &  \mu - [n] \text{ odd}
      \end{cases}
\end{equation}
for the CI excitation energies of class $[n]$, 
where, of course, $\mu \geq [n]$ is supposed.
In a similar way one may analyze the expression~(\ref{eq:tm1x})
for the transition moments,  
yielding the following truncation error
formula: 
\begin{equation}
\label{eq:tecitm}
O^{[n]}_{TE}(\mu) =  \begin{cases}
	 \mu -  \frac{1}{2}[n] + 1, & [n] \text{ even}\\        
 	 \mu - \frac{1}{2}[n] + \frac{1}{2}, & [n] \text{ odd}\\            
        \end{cases}
\end{equation}
Note that for the case of single excitations, $[n] = 1$,
the general formulas~(\ref{eq:tecien}, \ref{eq:tecitm}) reduce 
to the simple expressions~(\ref{eq:teoci1}) and (\ref{eq:teoci2}), respectively,
given in Sec.~2.

For completeness, let us note that the simple expression
\begin{equation}
O^{[n]}_{TE}(\mu) =  \mu -[n] + 1 
\end{equation}
applies to the truncation error of property matrix elements, 
\begin{equation}
\label{eq:tecipm}
T_{nn} = \ul{X}^{\dagger}_n \bs{D} \ul{X}_n 
\end{equation}
Here $\bs{D}$ is the CI representation~(\ref{eq:cirep1}) of a one-particle operator.

\subsection{bCC eigenvector matrices}

The order structures of the right and left bCC eigenvector matrices 
$\bs{X}$ and $\bs{Y}$, respectively, are shown in Fig.~10. 
(Note that $\bs{X}$ now denotes the right bCC eigenvector matrix rather than
the CI eigenvector matrix as in
the preceeding subsection B.1.)
The bCC eigenvector matrices display both canonical and CI-type order structures.
The LL part of $\bs{X}$ and the UR part of $\bs{Y}$ are canonical,
\begin{equation}
\label{eq:canor}
O[\bs{X}_{\mu\nu}] = O[\bs{Y}_{\nu\mu}] =  \mu - \nu, \;\; \mu \geq \nu\\
\end{equation}
while
the UR part of $\bs{X}$ and the LL part of $\bs{Y}$ are of CI-type,
\begin{equation}
\label{eq:cinor}
O[\bs{X}_{\nu\mu}] = O[\bs{Y}_{\mu\nu}] = \begin{cases}
	 \half (\mu - \nu), &  \mu - \nu \text{ even}, \,\geq 0 \\
 	 \half (\mu - \nu) + \half, & \mu - \nu  \text{ odd},\, \geq 0 
      \end{cases}
\end{equation}
The order structures of the bCC eigenvector matrices can be deduced 
from the order structure of the bCC secular matrix by an obvious 
generalization of the procedure used for the CI eigenvector matrix above (Sec.~B1).
Here it is important that at each successive step associated with an excitation class
$\mu$ both the right and left eigenvector matrices $\ul{\bs{X}}_{\mu}$ and 
$\ul{\bs{Y}}_{\mu}$ must be treated in parallel, while the biorthogonality
to the respective eigenvectors of the lower excitation classes, $1,\dots , \mu -1$,
must successively be taken into account. A formally correct  
way of translating the structure of the bCC secular matrix into the
order structures of the eigenvector matrices can readily be spelled out.  

However, one may take also an alternative route, 
in which the order
structures of the UR parts of both $\bs{X}$ and $\bs{Y}$ are determined directly.
The order relations of the complementary LL parts are then obtained
as a result of the biorthogonality, $\bs{Y}^{\dagger}\bs{X} = \bs{1}$. 
Let us first consider an eigenvector component
\begin{equation}
X_{Jn} = \bracket{\lo{\Phi}_J}{\Psi^{(r)}_n}, \,\,\,[n] \geq [J]  
\end{equation}
from an UR block of  $\bs{X}$. Since $\ket{\Psi^{(r)}_n}$ differs from the (normalized) 
exact eigenstate $\ket{\Psi_n}$ only by a normalization constant ($N_n \sim 1 + O(2)$),
we may consider $\bracket{\lo{\Phi}_J}{\Psi_n}$ rather than $X_{Jn}$, that is,  
\begin{equation}
\label{eq:up2norm}
X_{Jn} \sim \bracket{\lo{\Phi}_J}{\Psi_n}
\end{equation}
Using the expansion (\ref{eq:lset}) for 
$\bra{\lo{\Phi}_J}$, it is apparent that for $[n] \geq [J]$
the bCC eigenvecor component is of the same order as the
corresponding CI eigenvector component:
\begin{equation} 
O[\bracket{\lo{\Phi}_J}{\Psi_n}] =  O[\bracket{\Phi_J}{\Psi_n}]
\end{equation}
This establishes the CI-type order relations for the UR part of  $\bs{X}$.

In analogy to Eq.~(\ref{eq:up2norm}), the left bCC eigenvector components,
\begin{equation}
Y_{Jn} = \bracket{\Psi^{(l)}_n}{\Psi^0_J} 
\end{equation}
can be related to the normalized eigenstates,
\begin{equation}
Y_{Jn} \sim \bracket{\Psi_n}{\Psi^0_J} = \dirint{\Psi_n}{\hat{C}_J}{\Psi^{cc}_0}
\sim \dirint{\Psi_n}{\hat{C}_J}{\Psi_0}
\end{equation}
where we have used the fact that the CC and the normalized ground-state
differ by a normalization constant of the order $1 + O(2)$.
So it remains to show the order relations
\begin{equation}
\label{eq:gentmor}
\dirint{\Psi_n}{\hat{C}_J}{\Psi_0} \sim O([n] - [J]), \,\,\, [n] \geq [J]
\end{equation}
for the ``generalized transition moments'' (GTM) $\dirint{\Psi_n}{\hat{C}_J}{\Psi_0}$.
Let us first note that these GTM order relations are non-trivial. For a 
triply excited state ($[n] = 3$) and a $p$-$h$ excitation operator ($[J] = 1$), for example,
the (canonical) order is 2, rather than 1, as one might expect in view of 
two individual first-order contributions associated with $\ket{\Psi^{(1)}_n}$ and
$\ket{\Psi^{(1)}_0}$, respectively. It is instructive to verify that these 
two first-order contributions, in fact, cancel each other.

To prove the GTM order relations we may rely on the canonical order relations
\begin{equation}
\label{eq:gtmor}
\bracket{\Psi_n}{\tilde{\Psi}_J} \sim O(|[n] - [J]|)
\end{equation}
as established for the ISR eigenvector components (see Eq.~\ref{eq:corevm})   
discussed in Sec.~V.  
Since the intermediate states $\ket{\tilde{\Psi}_K}$ (including $\ket{\Psi_0}$) 
form a complete and 
orthonormal set of states they can be used to expand the GTM 
$\dirint{\Psi_n}{\hat{C}_J}{\Psi_0}$ as follows
\begin{eqnarray}
\nonumber
\dirint{\Psi_n}{\hat{C}_J}{\Psi_0} & = & \sum_K \bracket{\Psi_n}{\tilde{\Psi}_K}  
                         \dirint{\tilde{\Psi}_K}{\hat{C}_J}{\Psi_0}\\
\label{eq:gentmor1}
        & = & \sum_{[K] \leq [J]} \bracket{\Psi_n}{\tilde{\Psi}_K}  
                         \dirint{\tilde{\Psi}_K}{\hat{C}_J}{\Psi_0}
\end{eqnarray}
where the second line is due to the fact that by construction the 
intermediate states $\ket{\tilde{\Psi}_K}$ are orthogonal to all states
$\hat{C}_J\ket{\Psi_0}$ with $[J] < [K]$, that is,  
$\dirint{\tilde{\Psi}_K}{\hat{C}_J}{\Psi_0} = 0$ for $[J] < [K]$. Now
for $[n] > [J]$, the order of  $\bracket{\Psi_n}{\tilde{\Psi}_K}$ decreases
with increasing $[K]$ and so does the order of $\dirint{\tilde{\Psi}_K}{\hat{C}_J}{\Psi_0}$.  
This means that the minimal order contributions in the sum
on the rhs of Eq.~(\ref{eq:gentmor1}) are due to those where  
$[K] = [J]$:
\begin{equation}
\dirint{\Psi_n}{\hat{C}_J}{\Psi_0} 
 \sim O[\bracket{\Psi_n}{\tilde{\Psi}_J}\dirint{\tilde{\Psi}_J}{\hat{C}_J}{\Psi_0}] 
 = O[\bracket{\Psi_n}{\tilde{\Psi}_J}]
\end{equation}
The last equality follows from the observation that 
$\dirint{\tilde{\Psi}_J}{\hat{C}_J}{\Psi_0}$ is of zeroth order.
This completes the proof of the canonical order relations for the 
eigenvector components in the UR blocks of $\bs{Y}$.

The biorthogonality of the left and right bCC eigenvector matrices 
exacts the order relations of the LL blocks of the  $\bs{X}$ and 
 $\bs{Y}$ matrices, as will be shown in the ensuing Section B.3.

%%%%%%%%%%%%%%%%%%%%%%%%%%%%%%
Like in the last paragraph of Section B.1,
the truncation error of the bCC excitation energies for 
general excitation classes $[n]$ can be
deduced from the eigenstate order relations. 
The starting point is the expression
\begin{equation}
\label{eq:evto3}
E_n = \ul{Y}^{\dagger}_n \bs{M} \ul{X}_n =
\sum _{\kappa,\lambda} \ul{Y}^{\dagger}_{\kappa n} \bs{M}_{\kappa,\lambda} \ul{X}_{\lambda n}    
\end{equation}
where, analogously to Eq.~(\ref{eq:evto2}), the second equation 
reflects the partitioning of the energy expectation value with respect to excitation classes.
To determine the TEO at a given truncation level $\mu$, one has to analyze the
contributions where $\kappa = \mu +1, [n] \leq \lambda \leq \mu +1$ (set $S_1$)
and $\lambda = \mu +1, [n] \leq \kappa \leq \mu$ (set $S_2$). Here $\mu \geq [n]$ is supposed.
The former set of contributions is given by
\begin{equation}  
S_1 = \ul{Y}^{\dagger}_{\mu + 1,n} \sum^{\mu +1}_{\lambda = [n]} 
 \bs{M}_{\mu + 1,\lambda} \ul{X}_{\lambda n}
\end{equation}
Due to the order relations in the LL parts of $\bs{X}$ (Eq.~\ref{eq:canor}) and $\bs{M}$ 
(Eq.~\ref{eq:or2}),
we find 
\begin{equation}   
O[\bs{M}_{\mu + 1,\lambda} \ul{X}_{\lambda n}] = \mu - [n] +1
\end{equation}
irrespective of $\lambda$. Together with the 
OR of $\ul{Y}_{\mu + 1,n}$ (Eq.~\ref{eq:cinor}) this leads to 
the following TEO formula       
\begin{equation}
\label{eq:teccen}
O^{[n]}_{TE}(\mu) =  \begin{cases}
         \frac{3}{2} (\mu -[n]) + 2, &  \mu - [n] \text{ even}\\
	 \frac{3}{2}(\mu - [n]) + \frac{3}{2}, & \mu - [n] \text{ odd}\\ 	
      \end{cases}
\end{equation}
The $S_2$ set consists only of two contributions,
\begin{equation} 
S_2 = \ul{Y}^{\dagger}_{\mu,n} \bs{M}_{\mu,\mu + 1} \ul{X}_{\mu + 1, n}+
        \ul{Y}^{\dagger}_{\mu -1 ,n} \bs{M}_{\mu -1,\mu + 1} \ul{X}_{\mu + 1, n}
\end{equation}
because $\bs{M}_{\kappa,\mu + 1} = \bs{0}$ for $\kappa < \mu -1$. The 
two involved subblocks of $\bs{M}$ are of first order, 
$O[\bs{M}_{\mu,\mu + 1}] =O[\bs{M}_{\mu - 1,\mu + 1}] = 1,$ and the TEOs of 
the $S_2$ contributions are seen to exceed those of $S_1$. This means that
Eq.~(\ref{eq:teccen}) is the final expression for the truncation errors
in the bCC excitation energies.

%%%%%%%%%%%%%%%%%%%%%%%%%
%%%%%%%%%%%%%%%%%%%%%%%%
%%%%%%%%%%%%%%%%%%%%%%%%
In a similar way, one can derive general TEO formulas for the left and right transition moments,
and the excited state properties. In case of the right transition moments,
the dual ground-state eigenvector $\underline{Y}_0$ comes into play.
The CI-type order relations of $\underline{Y}_0$ (see Fig.~3b) can
readily be established by analyzing the eigenvalue equation 
$\underline{Y}^{\dagger}_{0} \bs{M} =- \underline{v}^{t}$ (Eq.~\ref{eq:lgsvx})
in a similar way as in App.~B.1, using here that only the $p$-$h$ and $2p$-$2h$ 
components of  $\underline{v}^{t}$ are non-vanishing, the latter being of first order. 
As to be expected, both the dual and the CI ground-state have the same order 
relations, as specified by Eq.~(\ref{eq:x0or}) in App.~B.1.

The resulting TEO formulas are as follows.\\
\emph{(i) left transition moments}:
\begin{equation}
\label{eq:teccltm}
O^{[n]}_{TE}(\mu) =  \begin{cases}
	 \frac{3}{2}\mu - \frac{1}{2}[n], & \mu - [n] \text{ even}\\
 	 \frac{3}{2}\mu - \frac{1}{2}[n] +\frac{1}{2}, &  \mu - [n] \text{ odd}
      \end{cases}
\end{equation}
\emph{(ii) right transition moments}:
\begin{equation}
\label{eq:teccrtm}
O^{[n]}_{TE}(\mu) =  \begin{cases}
	 \frac{3}{2}\mu - [n] + 1, & \mu  \text{ even}\\
 	 \frac{3}{2}\mu - [n] + \frac{1}{2}, &  \mu  \text{ odd}
      \end{cases}
\end{equation}
\emph{(iii) property matrix elements}:
\begin{equation}
\label{eq:teccpm}
O^{[n]}_{TE}(\mu) =  \begin{cases}
         \frac{3}{2}(\mu -[n]) +1, &  \mu - [n] \text{ even}\\
	 \frac{3}{2}(\mu - [n]) + \frac{1}{2}, & \mu - [n] \text{ odd} 	
      \end{cases}
\end{equation}
As above, $[n]$ and $\mu$ denote the final state excitation class 
and the truncation level, respectively, 
where of course $\mu \geq [n]$.

The first expression (\emph{i}) is obtained by generalizing 
the derivation of Eq.~(\ref{eq:teo2x}) in Sec. IV.B.
Here the error associated
with the truncation of the excited state manifold follows     
from Eq.~(\ref{eq:ltm2}), using the order relations (\ref{eq:ortml}) 
of the left basis set transition moments together
with the CI-type order relations (\ref{eq:cinor})
in the LL block of the left eigenvector matrix. In addition, one has to account for
the error arising from the truncation of the ground state CC expansion,
assuming here the same truncation levels in the ground and excited states.
Supposing a sufficiently large or even complete ground state CC expansion,
the TEO increases by 1 for even values of $\mu -[n]$ 
(first line on the rhs of Eq.~\ref{eq:teccltm}).

In the case of the right transition moments (\emph{ii}) the TEOs
are determined by the second term on the rhs of Eq.~(\ref{eq:rtmor1}).
This term can be analyzed in an analogous way as the bCC eigenvalues
by using the bCC order relations of the transition operator matrix $\bs{D}$
given in Fig.~5a rather than those of the bCC secular matrix.
It should be noted that apart from the case of single excitations, $[n] = 1$,
the right transition moments have larger truncation errors (lower orders)
than the left ones.

In (\emph{iii}), finally, one has to analyze the
the vector$\times$matrix$\times$vector-product
in Eq.~(\ref{eq:esp1}).  
The order relations coming here into play are
those of $\bs{D}$ (Fig.~5a) and the
LL parts of left and right eigenvector matrices. It should be noted that
the expression (\emph{iii}) applies also to inter-state transition moments for
excited states of the same class, $[n] = [m]$.

In the latter two cases, there is each an additional contribution, 
arising from the admixture of the ground state in the right eigenstate 
expansion (Eq.~\ref{eq:rrxs}), specified by the respective (extended) 
eigenvector component $x_n = - \underline{Y}^{\dagger}_0 \underline{X}_n$ (Eq.~\ref{eq:zn2}). 
Using the order relations for $\underline{Y}^{\dagger}_0$ and the LL part of
the right eigenvector matrix, one can readily derive the following TEO
formula for the $x_n$ components:
\begin{equation}
\label{eq:teoxn}
O^{[n]}_{TE}(\mu) =  \begin{cases}
	  \frac{3}{2}\mu - [n] +2, & \mu  \text{ even}\\
 	 \frac{3}{2}\mu - [n] + \frac{3}{2}, & \mu  \text{ odd}
      \end{cases}
\end{equation} 
As discussed in Sec.~4.B and 4.E, the TEOs in the additional (ground-state admixture)
terms exceed those of the respective main contribution.   

%%%%%%%%%%%%%%%%%%%%%%%%%%%%%%%%%%

\subsection{Order relations for biorthogonal matrices}

In the preceeding subsection B.2 the order relations for the 
bCC eigenvector matrices have been established only
for the respective UR blocks, being of 
CI-type in $\bs{X}$ and canonical in $\bs{Y}$.    
Now we will show that the biorthogonality of $\bs{X}$ and $\bs{Y}$
requires canonical and CI-type behaviour in the 
LL blocks of $\bs{X}$ and $\bs{Y}$, respectively.

Let us first note that the biorthogonality relation 
\begin{equation}
\label{eq:biort1}
\bs{Y}^{\dagger} \bs{X} = \bs{1} 
\end{equation}
also implies that  
\begin{equation}
\label{eq:biort2}
\bs{X} \bs{Y}^{\dagger}= \bs{1} 
\end{equation}
which will be our starting point here. In this form 
the UR order relations of the first factor $\bs{X}$ (CI-type) match 
the LL order relations of the second factor $\bs{Y}^{\dagger}$ (canonical),
which for brevity will be denoted by $\bs{Y}'$ henceforth. 
For a graphical notion of the following procedure we recommend 
to place the $\bs{X}$ and $\bs{Y}'$ partitioning schemes next to each other
and fill in successively the emerging order relation entries.

For the first row of $\bs{X}$-blocks, $\bs{X}_{1,k},\, k = 1,2,\dots,$
and the first column of $\bs{Y}'$-blocks, $\bs{Y}'_{k,1},\, k = 1,2,\dots,$ 
the order relations are already given. In the second row of $\bs{X}$-blocks 
and the second column of $\bs{Y}'$-blocks there is one undetermined block each,
namely, $\bs{X}_{2,1}$ and $\bs{Y}'_{1,2}$, respectively.  
The orthogonality of the second $\bs{X}$ row and the first $\bs{Y}'$ column
can be expressed as follows:
\begin{equation} 
\sum_{k} \bs{X}_{2,k} \bs{Y}'_{k, 1} = \bs{0}
\end{equation}
What can be concluded from this with respect to the order of $\bs{X}_{2,1}$?
Let focus on the first two terms in the sum, the remainder being (at least) of 
the order 2:
\begin{equation} 
\bs{X}_{2,1} \bs{Y}'_{1,1} + \bs{X}_{2,2} \bs{Y}'_{2,1} + O(2) = \bs{0}  
\end{equation}
Since the diagonal blocks of the eigenvector matrices behave as   
$\bs{Y}'_{1,1} = \bs{1} + O(1)$ and $\bs{X}_{2,2} = \bs{1} + O(1)$, respectively, and
$\bs{Y}'_{2,1} \sim O(1)$, the following relation holds through first order: 
\begin{equation}
\bs{X}_{2,1} + O(1) =  \bs{0}  
\end{equation}
This means that 
$\bs{X}_{2,1}$ must cancel a (non-vanishing) first-order contribution  
and, thus, is itself of the first order (more accurately: the lowest non-vanishing
contribution in the PT expansion of $\bs{X}_{2,1}$ is of first order).
  In a similar way, we may conclude $\bs{Y}'_{1,2} \sim O(1)$, being a consequence of the
orthogonality of the second column of $\bs{Y}'$-blocks and the first row
of $\bs{X}$-blocks. 

After having completed the order relations 
in the second row and column of $\bs{X}$ and $\bs{Y}'$, respectively, we may proceed 
to the third row of $\bs{X}$-blocks. Here the orders of the first two blocks,
$\bs{X}_{3,1}$ and $\bs{X}_{3,2}$, have to be derived, which in turn can be achieved
by exploiting that this row is orthogonal to both the first and second column
of $\bs{Y}'$-blocks. Expanded explicitely through the first three terms these 
order relations read 
\begin{eqnarray} 
\bs{X}_{3,1} \bs{Y}'_{1,1} + \bs{X}_{3,2} \bs{Y}'_{2,1} + \bs{X}_{3,3} \bs{Y}'_{3,1}
                          +  O(4) & = & \bs{0}\\ 
\bs{X}_{3,1} \bs{Y}'_{1,2} + \bs{X}_{3,2} \bs{Y}'_{2,2} + \bs{X}_{3,3} \bs{Y}'_{3,2} 
              +  O(3) & = & \bs{0} 
\end{eqnarray}
The second equation allows us to determine the order of $\bs{X}_{3,2}$.
Using $\bs{Y}'_{1,2} \sim O(1)$, $\bs{Y}'_{3,2} \sim O(1)$, and  
$\bs{Y}'_{2,2} = \bs{1} + O(1)$, $\bs{X}_{3,3} = \bs{1} + O(1)$ (as diagonal 
eigenvector blocks), we may conclude that
the relation 
\begin{equation} 
\bs{X}_{3,2} + O(1) = \bs{0} 
\end{equation}
holds through first order, which, as above, implies that $\bs{X}_{3,2}$ is of first order. 
Using this result in the first orthogonality equation, together with
$\bs{Y}'_{1,1} = \bs{1} + O(1)$, $\bs{Y}'_{2,1} \sim O(1)$, and $\bs{Y}'_{3,1} \sim O(2)$,  
yields through 2nd order
\begin{equation} 
\bs{X}_{3,1} + O(2) = \bs{0} 
\end{equation}
which means that $\bs{X}_{3,1} \sim O(2)$, consistent with the canonical order relations.
In a completely analogous way, the first two blocks in the third column of $\bs{Y}'$-blocks
can be treated, yielding the expected CI-type order results, $\bs{Y}'_{1,3} \sim O(1)$,
$\bs{Y}'_{2,3} \sim O(1)$ 

This brief demonstration has shown how the given CI-type UR order relations of $\bs{X}$  
and the canonical LL order relations of $\bs{Y}'$ ($=\bs{Y}^{\dagger}$)
impose canonical order relations in the LL part of $\bs{X}$ and CI-type 
order relations in the UR part $\bs{Y}^{\dagger}$ as a result of the biorthogonality
of right and left bCC eigenvector matrices. Of course, this derivation can readily be cast into 
a formally correct proof by induction (see Ref.~\cite{mer96:2140}).

In a related way, the unitarity of the CI eigenvector matrix 
(denoted $\bs{X}$ in App.~B.1) can be used to extend the CI order relations (\ref{eq:x0or}),
established in B.1 only for the LL part, to the entire matrix $\bs{X}$.
Writing the unitarity relation of the CI eigenvector matrix 
in the form $\bs{X}^{\dagger}\bs{X} = \bs{1}$, the order relations
of the UR part of $\bs{X}^{\dagger}$ combine with those of the LL part of $\bs{X} = \bs{1}$,
similar to the product (\ref{eq:biort2}) of the bCC eigenvector matrices.
The successive construction of the CI order relations in the UR part of $\bs{X}$
can be performed essentially as in the bCC case above.

%%%%%%%%%%%%%%%%%%%%%%%%%%%%%%%%%%%%%%%%%%%%%%%%%%%
\renewcommand{\theequation}{C.\arabic{equation}}
\setcounter{equation}{0}
\setcounter{subsection}{0}
\section*{Appendix C: Equivalence of CCLR and ordinary bCC
                 transition moments}

\subsection{Right transition moments}

In the exact (full) bCC treatment 
the ordinary right bCC transition moment (Eq.~\ref{eq:tramor})
\begin{equation}
\label{eq:appc1}
T^{(r)}_n  =  \dirint{\overline{\Psi}_0}{\hat{D}\hat{C}_n}{\Psi^{cc}_0} + 
           x_n \dirint{\overline{\Psi}_0}{\hat{D}}{\Psi^{cc}_0}
\end{equation}
and the separable CCLR expression (Eq.~\ref{eq:rtmlrf})
\begin{equation}
\label{eq:appc2}
T^{(r)}_n = \dirint{\overline{\Psi}_0}{[\hat{D}, \hat{C}_n]}{\Psi^{cc}_0}
- \sum_{I,J} \dirint{\overline{\Psi}_0}{[[\hat{H}, \hat{C}_I],\hat{C}_n]}{\Psi^{cc}_0}
            (\bs{M} + \omega_n)^{-1}_{IJ} \dirint{\overline{\Phi}_J}{\hat{D}}{\Psi^{cc}_0}
\end{equation}
are equivalent. Here, as in Sec.~4.D, 
\begin{equation}
\hat{C}_n = \sum X_{In} \hat{C}_I
\end{equation}
denotes the (right) excitation operator
associated with the $n$-th excited state,
$\ket{\Psi^{(r)}_n} = x_n \ket{\Psi^{cc}_0} + \hat{C}_n \ket{\Psi^{cc}_0}$.
This by no means obvious result was shown explicitly by 
Koch \emph{et al.}~\cite{koc94:4393}. 
The following is essentially a reformulation and slight extension of the
original proof, using
the more transparent wave function notations promoted here. 

Let us start from the the ordinary bCC expression~(\ref{eq:appc1})
and transform it successively into the CCLR expression~(\ref{eq:appc2}).
As a first step we make use of the commutator relation
$\hat{D} \hat{C}_n = [\hat{D},\hat{C}_n] + \hat{C}_n \hat{D}$, yielding 
\begin{equation}
\label{eq:appc3}
T^{(r)}_n  =  \dirint{\overline{\Psi}_0}{[\hat{D},\hat{C}_n]}{\Psi^{cc}_0} + 
        \dirint{\overline{\Psi}_0}{\hat{C}_n \hat{D}}{\Psi^{cc}_0}
         + x_n \dirint{\overline{\Psi}_0}{\hat{D}}{\Psi^{cc}_0}
\end{equation}
To proceed let us consider the (trivial) identity
\begin{equation}
\label{eq:trivid}
\dirint{\overline{\Psi}_0}{\hat{C}_n \hat{D}}{\Psi^{cc}_0}
= \dirint{\overline{\Psi}_0}{\hat{C}_n  (\hat{H} -E_0 + \omega_n)(\hat{H} -E_0 + \omega_n)^{-1}
\hat{D}}{\Psi^{cc}_0}
\end{equation}
and replace the inverse matrix operator on the rhs by its bCC representation.
Noting that $\bs{M}' + \omega_n$ is the bCC representation of $\hat{H} -E_0 + \omega_n$, 
where $\bs{M}'$ is the extended
bCC secular matrix given by Eq.~(\ref{eq:ccsmx}), the  
bCC representation of the inverse operator reads
\begin{equation}
\label{eq:invccsmx}
(\bs{M}' + \omega_n)^{-1}=\left( \begin{array}{cc}
\omega_n^{-1} & \ul{w}^{t}_n \\
\underline{0} & (\bs{M} + \omega_n)^{-1}
\end{array} \right).
\end{equation}
where 
\begin{equation}
\label{eq:invccsmxx}
\ul{w}^{t}_n = - \omega_n^{-1} \ul{v}^{t}(\bs{M} + \omega_n)^{-1} 
\end{equation}
This means that we can express $(\hat{H} -E_0 + \omega_n)^{-1}$ 
as follows:
\begin{equation}
\nonumber
(\hat{H} -E_0 + \omega_n)^{-1} = \sum_{I,J} \kett{\Psi^0_I}(\bs{M} + \omega_n)^{-1}_{IJ}
                  \brac{\lo{\Phi}_J}
             + \sum_J  w_{nJ} \ket{\Psi^{cc}_0} \brac{\lo{\Phi}_J} + \omega^{-1}_n \kett{\Psi^{cc}_0} 
\brac{\Phi_0}
\end{equation}
Inserting this expansion into Eq.~(\ref{eq:trivid}) yields
\begin{eqnarray}
\label{eq:cclrtm3}
\nonumber
\dirint{\overline{\Psi}_0}{\hat{C}_n \hat{D}}{\Psi^{cc}_0}  = 
\dirint{\overline{\Psi}_0}{\hat{C}_n  (\hat{H} -E_0 + \omega_n)}
  {\Psi^0_I}\,(\bs{M} + \omega_n)^{-1}_{IJ}\, \dirint{\overline{\Phi}_J}{\hat{D}}{\Psi^{cc}_0}\\
 -  x_n  \{ \dirint{\Phi_0}{\hat{D}}{\Psi^{cc}_0}  
+ \sum_J\omega_n w_{nJ}\dirint{\overline{\Phi}_J}{\hat{D}}{\Psi^{cc}_0}\} 
\end{eqnarray}
In deriving this result we have used that 
$(\hat{H} -E_0 + \omega_n) \kett{\Psi^{cc}_0} = \omega_n \kett{\Psi^{cc}_0}$ and 
$\dirint{\overline{\Psi}_0}{\hat{C}_n}{\Psi^{cc}_0} = - x_n$. 
Note that at this point another $x_n$-term comes into play,  
augmenting the 
third term on the rhs of Eq.~(\ref{eq:appc3}).  
The reformulation of Eq.~(\ref{eq:trivid}) is still not complete. To proceed
the matrix elements $\dirint{\overline{\Psi}_0}{\hat{C}_n (\hat{H} -E_0 + \omega_n)}
{\Psi^0_I}$ 
in the first term on the rhs of Eq.~(\ref{eq:cclrtm3}) 
can be expressed according to 
\begin{equation}
\label{eq:appc4}
\dirint{\overline{\Psi}_0}{\hat{C}_n (\hat{H} -E_0 + \omega_n) \hat{C}_I} {\Psi^{cc}_0} = 
- \dirint{\overline{\Psi}_0}{[[\hat{H}, \hat{C}_I],\hat{C}_n]}{\Psi^{cc}_0}
- x_n \omega_n Y^*_{I0}
\end{equation}
in terms of the double commutator $[[\hat{H}, \hat{C}_I],\hat{C}_n]$. Here
we have used the eigenvalue equation for the $n$-th excited state in the form 
\begin{equation}
\label{eq:cclrtm4}
(\hat{H} -E_0) \hat{C}_n \kett{\Psi^{cc}_0} = \omega_n \left (\hat{C}_n \kett{\Psi^{cc}_0} 
                    + x_n \kett{\Psi^{cc}_0} \right )  
\end{equation}
Moreover, recall that $\dirint{\overline{\Psi}_0}{\hat{C}_I}{\Psi^{cc}_0} = Y^*_{I0}$
and the operators $\hat{C}_n$ and $\hat{C}_I$ commute. 
Inserting Eq.~(\ref{eq:appc4}) on the rhs of Eq.~(\ref{eq:cclrtm3}) 
constitutes the final step in the reformulation  of 
$\dirint{\overline{\Psi}_0}{\hat{C}_n \hat{D}}{\Psi^{cc}_0}$. 
Using this result in Eqs.~(\ref{eq:cclrtm3}) and (\ref{eq:appc3}), respectively,
validates the second term in the 
CCLR expression~(\ref{eq:appc2}), but also introduces a third  $x_n$-term,
\begin{equation}
\label{eq:appc5}
- x_n \omega_n \sum_{I,J} Y^*_{I0}(\bs{M} + \omega_n)^{-1}_{IJ} 
\dirint{\overline{\Phi}_J}{\hat{D}}{\Psi^{cc}_0}
\end{equation}
due to the second term on the rhs of  Eq.~(\ref{eq:appc4}).
It remains to show that the three $x_n$ terms cancel each other, that is,
\begin{equation}
\dirint{\overline{\Psi}_0}{\hat{D}}{\Psi^{cc}_0} -  \dirint{\Phi_0}{\hat{D}}{\Psi^{cc}_0}
- \sum_J\omega_n w_{nJ}\dirint{\overline{\Phi}_J}{\hat{D}}{\Psi^{cc}_0}
-  \omega_n \sum_{I,J} Y^*_{I0}(\bs{M} + \omega_n)^{-1}_{IJ} 
\dirint{\overline{\Phi}_J}{\hat{D}}{\Psi^{cc}_0} = 0 
\end{equation}
where the contributions 1, 2+3, and 4 on the lhs arise from Eqs.~(\ref{eq:appc3}), 
(\ref{eq:cclrtm3}), and (\ref{eq:appc5}), respectively. 
The contributions 3 and 4 can be combined and further evaluated in a compact matrix 
notation as follows:   
\begin{eqnarray}
\omega_n (\ul{w}^{t} + \ul{Y}^{\dagger}_0) (\bs{M} + \omega_n)^{-1}
            & = & \omega_n (- \omega^{-1}_n \ul{v}^{t}(\bs{M} + \omega_n)^{-1} 
         - \ul{v}^{t} \bs{M}^{-1}(\bs{M} + \omega_n)^{-1})\\
            & = & - \omega_n \ul{v}^{t} (\omega^{-1}_n +  \bs{M}^{-1})(\bs{M} + \omega_n)^{-1}\\
            & = &  - \omega_n \ul{v}^{t} \omega^{-1}_n \bs{M}^{-1}(\bs{M} 
           + \omega_n)(\bs{M} + \omega_n)^{-1}\\
            & = & - \ul{v}^{t} \bs{M}^{-1} = \ul{Y}^{\dagger}_0  
\end{eqnarray}
As a result the sum of the three $x_n$ terms becomes
\begin{equation}
x_n \{\dirint{\overline{\Psi}_0}{\hat{D}}{\Psi^{cc}_0} -  \dirint{\Phi_0}{\hat{D}}{\Psi^{cc}_0} 
- \sum  Y^*_{J0} \dirint{\overline{\Phi}_J}{\hat{D}}{\Psi^{cc}_0} \}= 0
\end{equation}
where the cancellation now is obvious, as $\bra{\overline{\Psi}_0} = 
\bra{\Phi_0} + \sum  Y^*_{J0} \bra{\overline{\Phi}_J}$.

%%%%%%%%%%%%%%%%%%%%%%%%%%%%%%%%%%%%%%%%%%%%%%%%%%%%%%%%%%%5

\subsection{Excited state transition moments}
In a similar way one may show the equivalence of the
ordinary bCC and the CCLR expressions for excited state transition moments and
properties, the latter 
reading (Eq.~\ref{eq:estmcclr})
\begin{equation}
\label{eq:appctnm}
T_{nm} = \dirint{\overline{\Psi}^{(l)}_n}{[\hat{D},\hat{C}_m]}{\Psi^{cc}_0}
- \sum_{I,J} \dirint{\overline{\Psi}^{(l)}_n}{[[\hat{H}, \hat{C}_I],\hat{C}_m]}{\Psi^{cc}_0}
            (\bs{M} + \omega_{mn})^{-1}_{IJ} \dirint{\overline{\Phi}_J}{\hat{D}}{\Psi^{cc}_0}
        + \delta_{nm} \dirint{\overline{\Psi}_0}{\hat{D}}{\Psi^{cc}_0}
\end{equation}
where $\omega_{mn} = \omega_{m} - \omega_{n}$. As above, we 
start from the ordinary bCC expression (\ref{eq:esp1}) 
\begin{equation}
T_{nm} = \dirint{\overline{\Psi}^{(l)}_n}{\hat{D} \hat{C}_m}{\Psi^{cc}_0} + 
         x_m \dirint{\overline{\Psi}^{(l)}_n}{\hat{D}}{\Psi^{cc}_0}
\end{equation}
and use the commutator relation
$\hat{D}\hat{C}_m = [\hat{D},\hat{C}_m] + \hat{C}_m \hat{D}$
yielding
\begin{equation}
\label{eq:appcy}
T_{nm} = \dirint{\overline{\Psi}^{(l)}_n}{[\hat{D},\hat{C}_m]}{\Psi^{cc}_0}
    + \dirint{\overline{\Psi}^{(l)}_n}{\hat{C}_m \hat{D}}{\Psi^{cc}_0} 
    + x_m \dirint{\overline{\Psi}^{(l)}_n}{\hat{D}}{\Psi^{cc}_0}
\end{equation}
Analogously to Eq.~(\ref{eq:trivid}), we consider the identity
\begin{equation}
\dirint{\overline{\Psi}^{(l)}_n}{\hat{C}_m \hat{D}}{\Psi^{cc}_0}
= \dirint{\overline{\Psi}^{(l)}_n}{\hat{C}_m (\hat{H} -E_0 + \omega_{mn})(\hat{H} -E_0 + \omega_{mn})^{-1}
\hat{D}}{\Psi^{cc}_0}
\end{equation}
and use the 
the bCC representation of $(\hat{H} - E_0 + \omega_{mn})^{-1}$ to further evaluate
the rhs. Note that the only difference to Eqs.~(\ref{eq:invccsmx}) and (\ref{eq:invccsmxx})
is the replacement of $\omega_n$ by  $\omega_{mn}$. This leads to
\begin{eqnarray}
\nonumber
\dirint{\overline{\Psi}^{(l)}_n}{\hat{C}_m \hat{D}}{\Psi^{cc}_0}  = \sum_{I,J}
\dirint{\overline{\Psi}^{(l)}_n}{\hat{C}_m  (\hat{H} -E_0 + \omega_{mn})}
  {\Psi^0_I}(\bs{M} + \omega_{mn})^{-1}_{IJ} \dirint{\overline{\Phi}_J}{\hat{D}}{\Psi^{cc}_0}\\
\label{eq:appcx}
 + \delta_{mn} ( \dirint{\Phi_0}{\hat{D}}{\Psi^{cc}_0}  
 - \sum_{I,J} v_I (\bs{M} + \omega_{mn})^{-1}_{IJ} \dirint{\overline{\Phi}_J}{\hat{D}}{\Psi^{cc}_0})
\end{eqnarray}
using here
$\dirint{\overline{\Psi}^{(l)}_n}{\hat{C}_m}{\Psi^{cc}_0} = \delta_{mn}$.
Let us consider the 2nd term on the rhs, which is non-vanishing only for $n=m$ due
to the Kronecker symbol. Since $\omega_{mn} = 0$ for $m=n$,
and 
\begin{equation}
\ul{v}^t\bs{M}^{-1} = - \ul{Y}^{\dagger}_0 
\end{equation}
the latter term becomes $ \delta_{mn} \dirint{\overline{\Psi}_0}{\hat{D}}{\Psi^{cc}_0}$,
thus reproducing the 3rd term in the CCLR expression (\ref{eq:appctnm}).
As in Eq.~(\ref{eq:appc4}), we now may introduce a
double commutator according to 
\begin{equation}
\dirint{\overline{\Psi}^{(l)}_n}{\hat{C}_m (\hat{H} - E_0 + \omega_{mn}) \hat{C}_I} {\Psi^{cc}_0} = 
- \dirint{\overline{\Psi}^{(l)}_n}{[[\hat{H}, \hat{C}_I],\hat{C}_m]}{\Psi^{cc}_0}
- x_m \omega_m Y^*_{In}
\end{equation}
Here we have used the eigenvector equation 
for the $m$-th excited state in the form 
  \begin{equation}
(\hat{H} -E_0) \hat{C}_m \kett{\Psi^{cc}_0} = \omega_m (\hat{C}_m \kett{\Psi^{cc}_0} 
                    + x_m \kett{\Psi^{cc}_0} )  
\end{equation}
and the relations
$\dirint{\overline{\Psi}^{(l)}_n}{\hat{C}_I}{\Psi^{cc}_0} =  Y^*_{In}$
for the left eigenvector components.
Using this result in Eq.~(\ref{eq:appcx}) gives
\begin{eqnarray}
\nonumber
\dirint{\overline{\Psi}^{(l)}_n}{\hat{C}_m \hat{D}}{\Psi^{cc}_0}
 = - \sum_{I,J} \dirint{\overline{\Psi}^{(l)}_n}{[[\hat{H}, \hat{C}_I],\hat{C}_m]}{\Psi^{cc}_0}
            (\bs{M} + \omega_{mn})^{-1}_{IJ} \dirint{\overline{\Phi}_J}{\hat{D}}{\Psi^{cc}_0}\\
\label{eq:appcz}
- x_m \omega_m \sum_{I,J} Y^*_{In} (\bs{M} + \omega_{mn})^{-1}_{IJ} \dirint{\overline{\Phi}_J}{\hat{D}}{\Psi^{cc}_0}
\end{eqnarray}
The first term on the rhs is seen to reproduce, via Eq.~(\ref{eq:appcy}),
the 2nd term of the CCLR expression 
(\ref{eq:appctnm}). It remains to inspect the 2nd term on the rhs, containing the factor $x_m$.
Since $\ul{Y}_n$ is a left eigenvector of $\bs{M}$ it follows that 
\begin{equation}
\ul{Y}^{\dagger}_n (\bs{M} + \omega_{mn})^{-1} = (\omega_{n} + \omega_{mn})^{-1} \ul{Y}^{\dagger}_n 
= \omega^{-1}_m  \ul{Y}^{\dagger}_n   
\end{equation}
Thus
\begin{equation}
- x_m \omega_m \sum_{I,J} Y^*_{In} (\bs{M} + \omega_{mn})^{-1}_{IJ} \dirint{\overline{\Phi}_J}{\hat{D}}{\Psi^{cc}_0}
= - x_m \dirint{\overline{\Psi}^{(l)}_n}{\hat{D}}{\Psi^{cc}_0}
 \end{equation}
which cancels the original $x_m$ term on the rhs of Eq.~(\ref{eq:appcy}). This concludes 
our proof.

%\include{figs}
%\bibliographystyle{jcp}

%\newpage

\section*{Tables}

%----------------------------------------------------------------------%
\begin{table}[h]
  \centering
  \caption{Truncation errors (PT order) for excitation energies, transition moments, and excited state 
    properties of singly excited states: Comparison of CI, bCC, and ADC approaches at 
       the lowest 6 truncation levels.}
  \begin{ruledtabular}\begin{tabular}{cccccccccccc}
  truncation & \multicolumn{3}{c}{excitation energies\phantom{XX}} & \, 
   & \multicolumn{3}{c}{transition moments\phantom{XX}} & \,
& \multicolumn{3}{c}{properties\phantom{XX}} \\
      \cline{2-4}\cline{6-8}\cline{10-12}
  level &\phantom{X}CI\phantom{XX} & bCC\phantom{X} & ADC\phantom{XXX} 
               &&\phantom{X}CI\phantom{X} &\phantom{X}bCC\phantom{X} & ADC\phantom{XXX}  
               &&\phantom{X}CI\phantom{X} &\phantom{X}bCC\phantom{X} & ADC\phantom{XXX} \\
      \hline
      1 & \phantom{X}2\phantom{XX} & 2\phantom{X} & 2\phantom{XXX} 
       && \phantom{X}1\phantom{X} & \phantom{X}1\phantom{X} & 2\phantom{XXX}
       && \phantom{X}1\phantom{X} & \phantom{X}1\phantom{X} & 1\phantom{XXX}\\
      2 & \phantom{X}2\phantom{XX} & 3\phantom{X} & 4\phantom{XXX} 
       && \phantom{X}2\phantom{X} & \phantom{X}3\phantom{X} & 4\phantom{XXX} 
       && \phantom{X}2\phantom{X} & \phantom{X}2\phantom{X} & 3\phantom{XXX} \\
      3 & \phantom{X}4\phantom{XX} & 5\phantom{X} & 6\phantom{XXX} 
       && \phantom{X}3\phantom{X} & \phantom{X}4\phantom{X} & 6\phantom{XXX} 
       && \phantom{X}3\phantom{X} & \phantom{X}4\phantom{X} & 5\phantom{XXX} \\
      4 & \phantom{X}4\phantom{XX} & 6\phantom{X} & 8\phantom{XXX} 
       && \phantom{X}4\phantom{X} & \phantom{X}6\phantom{X} & 8\phantom{XXX} 
       && \phantom{X}4\phantom{X} & \phantom{X}5\phantom{X} & 7\phantom{XXX} \\
      5 & \phantom{X}6\phantom{XX} & 8\phantom{X} & 10\phantom{XXX} 
       && \phantom{X}5\phantom{X} & \phantom{X}7\phantom{X} & 10\phantom{XXX} 
       && \phantom{X}5\phantom{X} & \phantom{X}7\phantom{X} & 9\phantom{XXX} \\
      6 & \phantom{X}6\phantom{XX} & 9\phantom{X} & 12\phantom{XXX} 
       && \phantom{X}6\phantom{X} & \phantom{X}9\phantom{X} & 12\phantom{XXX} 
       && \phantom{X}6\phantom{X} & \phantom{X}8\phantom{X} & 11\phantom{XXX} \\         
  \end{tabular}\end{ruledtabular}
  \label{tab1x}
\end{table}
%----------------------------------------------------------------------%

\section*{Figure captions}
\noindent
Fig. 1: (a) Order structure of the CI secular matrix $\bs{H}$. The 
subblocks correspond to the partitioning with respect to
excitation classes of $\mu p$-$\mu h$ ($\mu$-particle-$\mu$-hole) 
excitations, $\mu = 1,2, \dots$. The entries $0,1$ indicate 
the ``PT order'' of the blocks, being here simply 0 (in the diagonal blocks)
or  1 (linear in the electron repulsion integrals); vanishing blocks 
are indicated by \emph{dashes}.   
(b) Order relations of CI eigenvectors associated with singly excited states ($\bs{X}_{ph}$)
and the ground state ($\bs{X}_0$). The entries denote the (lowest)
PT order of the respective eigenvector segments.   
\\
Fig. 2: (a) Block structure of the CI secular matrix $\bs{H}$
corresponding to the partitioning associated with
the separate fragment model (see text).
(b) Structure of CI eigenvectors for a local excitation (on fragment $A$).
\\
Fig. 3: (a) Order relations of the bCC secular matrix $\bs{M}$. 
As in Fig.~1, the block structure
reflects the partitioning of $\bs{M}$ with respect to excitation classes $\mu = 1,2, \dots$. 
The entries denote the (lowest) PT order of the matrix elements
in the respective subblock; vanishing blocks are indicated by \emph{dashes}. 
(b) Order relations of bCC eigenvectors: $\bs{X}_{ph}$ and $\bs{Y}_{ph}$ denote
right and left eigenvectors for singly excited states; $\bs{Y}_{0}$ is
the ``dual'' CC ground state.  
\\
Fig. 4: (a) Block structure of the bCC secular matrix $\bs{M}$
corresponding to the partitioning associated with
the separate fragment model (see text).
(b) Structure of right and left bCC eigenvectors for a local excitation (on fragment $A$).
\\
Fig. 5: (a) Order relations of the bCC representation $\bs{D}$ of a one-particle 
operator $\hat{D}$. Block structure and entries as in Figs.~1 and 3.
(b) Order relations of the left and right basis set transition moments (see text).
\\
Fig. 6: (a) Block structure of the bCC representation $\bs{D}$ of a general  
operator $\hat{D}$ with respect to the separate fragment model (as in Figs.~2 and 4).
\\
Fig. 7: (a) Order relations of the ADC-ISR secular matrix $\bs{M}$. 
Block structure and entries as in Figs.~1 and 3.
(b) Order relations of ADC-ISR eigenvectors for singly excited states.
\\
Fig. 8: (a) Block structure of the ADC-ISR secular matrix $\bs{M}$ with
respect the separate fragment model.
(b) Structure of ADC-ISR eigenvectors for a local excitation (on fragment $A$).
\\
Fig.~9: Order relations of the CI eigenvector matrix $\bs{X}$. 
Block structure and entries as in Figs.~1 and 3.
\\  
Fig.~10: Order relations of the right and left bCC eigenvector matrices $\bs{X}$
and $\bs{Y}$, respectively. Block structure and entries as in Figs.~1 and 3.

\section*{Figures}

%\begin{center}
\includegraphics[height=9cm]{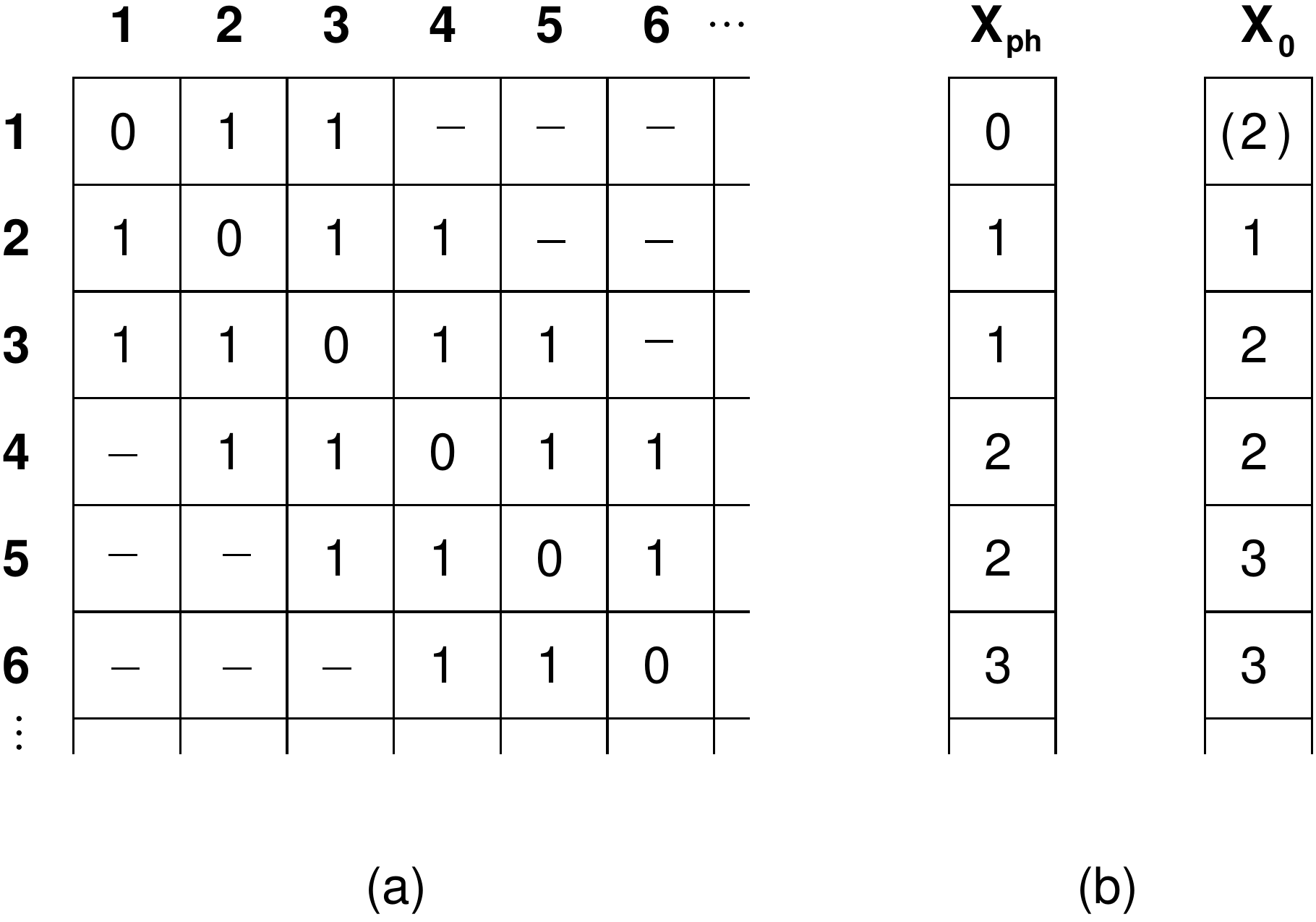}
\\
\textbf{Fig. 1}\\
\\
\\
%\end{center}
%\newpage
\includegraphics[height=9cm]{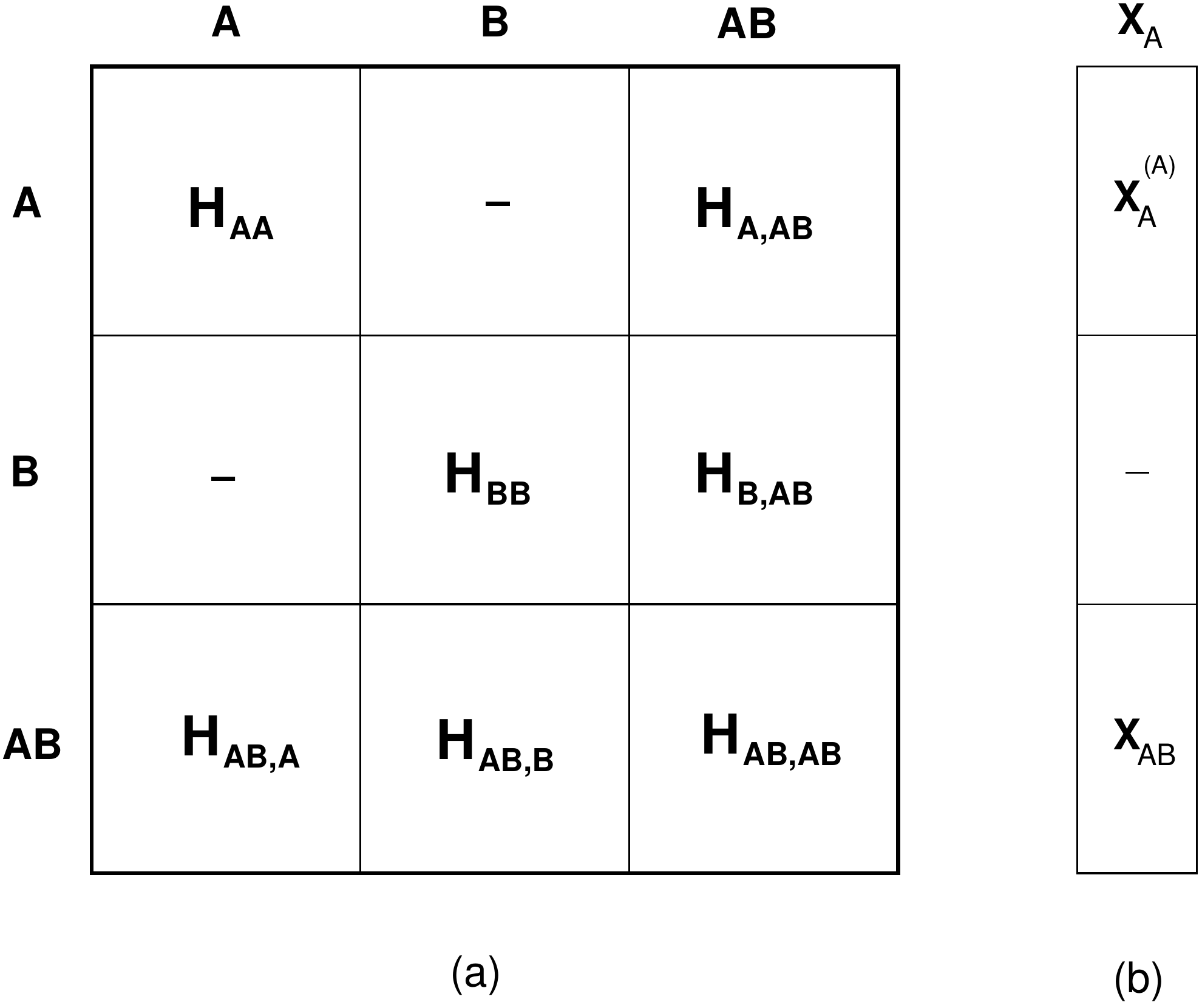}
\\
\textbf{Fig. 2}
\newpage
%\begin{center}
\includegraphics[height=9cm]{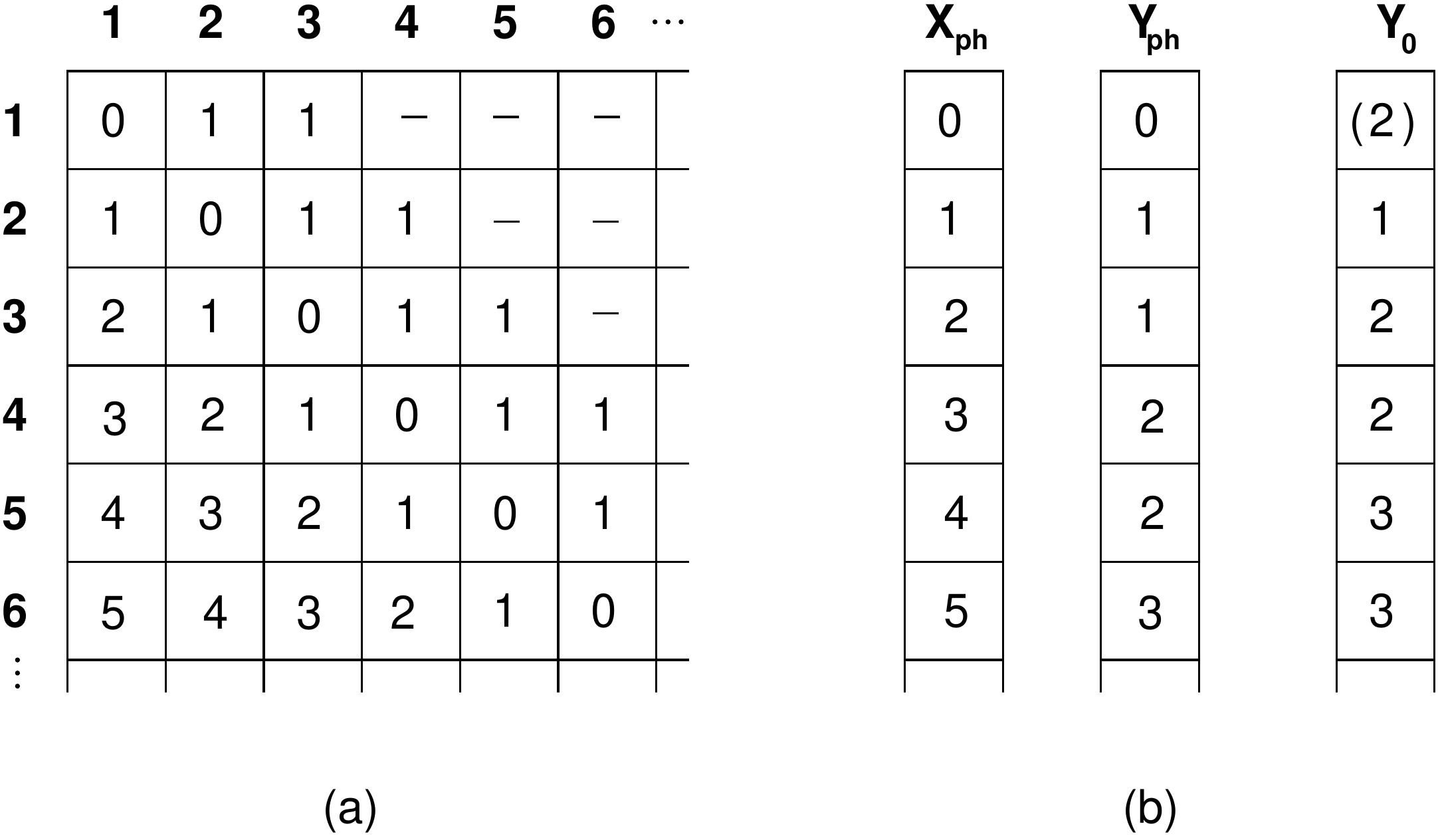}
%\end{center}
\\
\textbf{Fig. 3}
\\
\\
\\
%\newpage
\includegraphics[height=9cm]{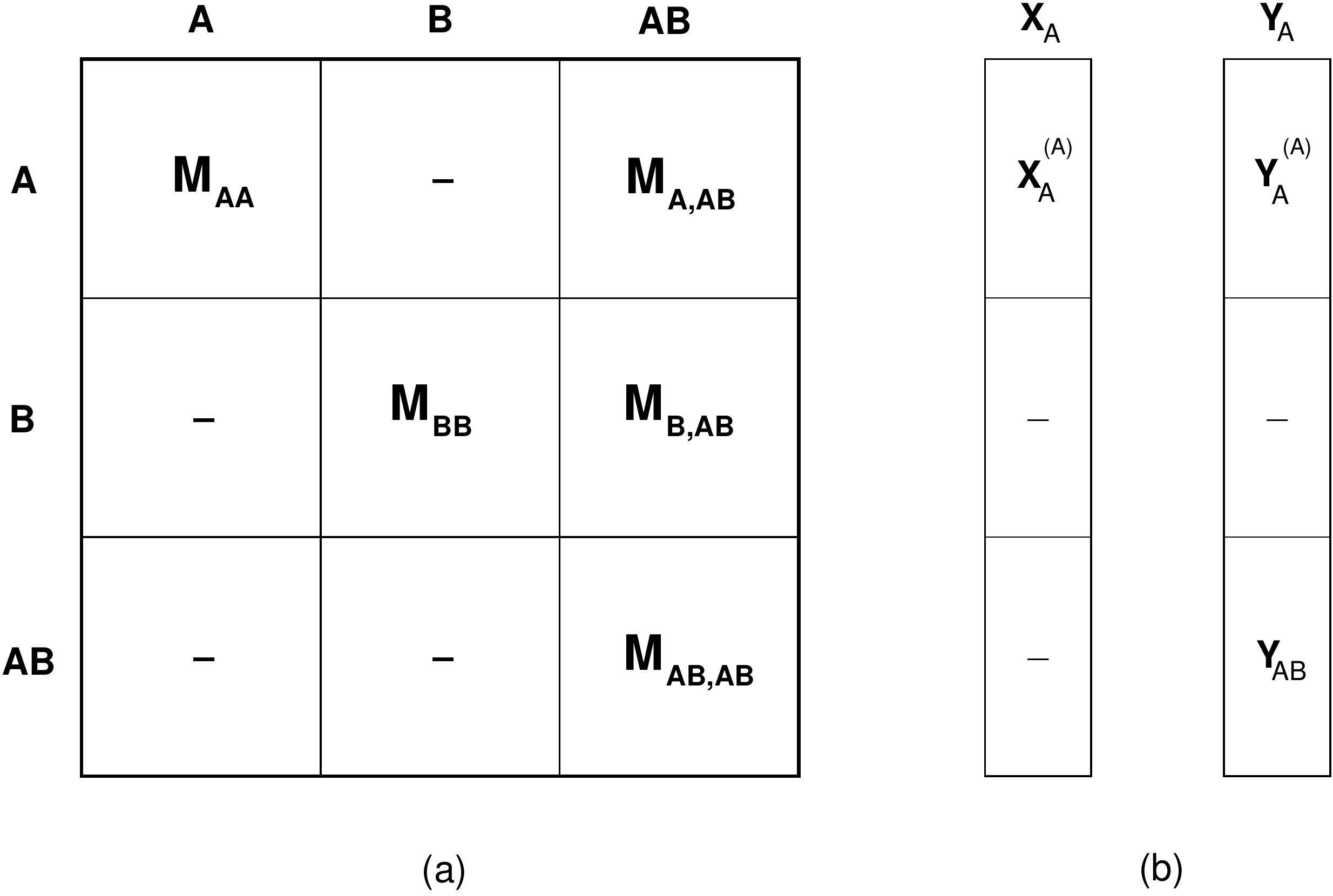}
\\
\textbf{Fig. 4}
\newpage
\includegraphics[height=10cm]{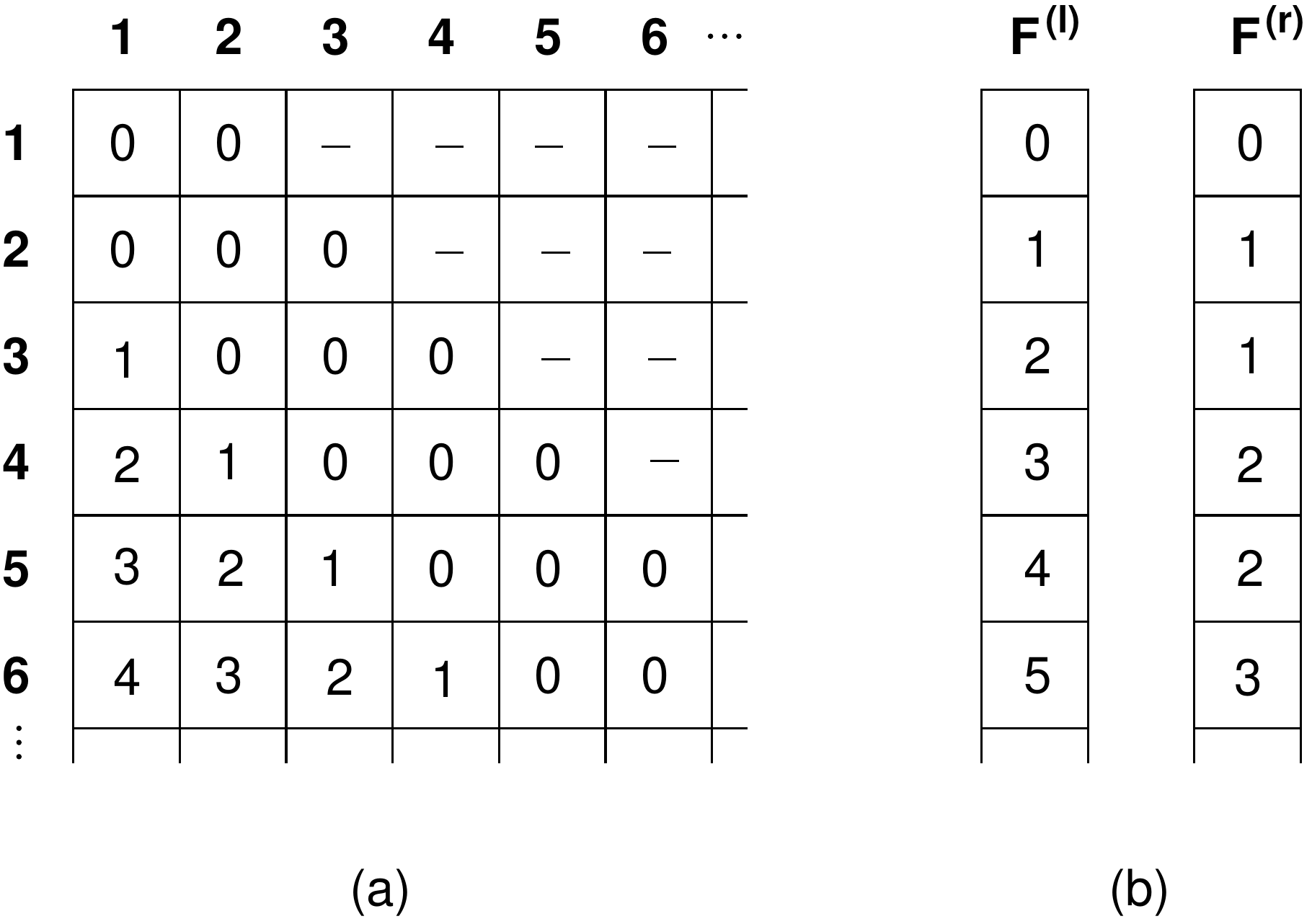}
\\
\textbf{Fig. 5}
\\
\\
\\
%\newpage
\begin{center}
\includegraphics[height=8cm]{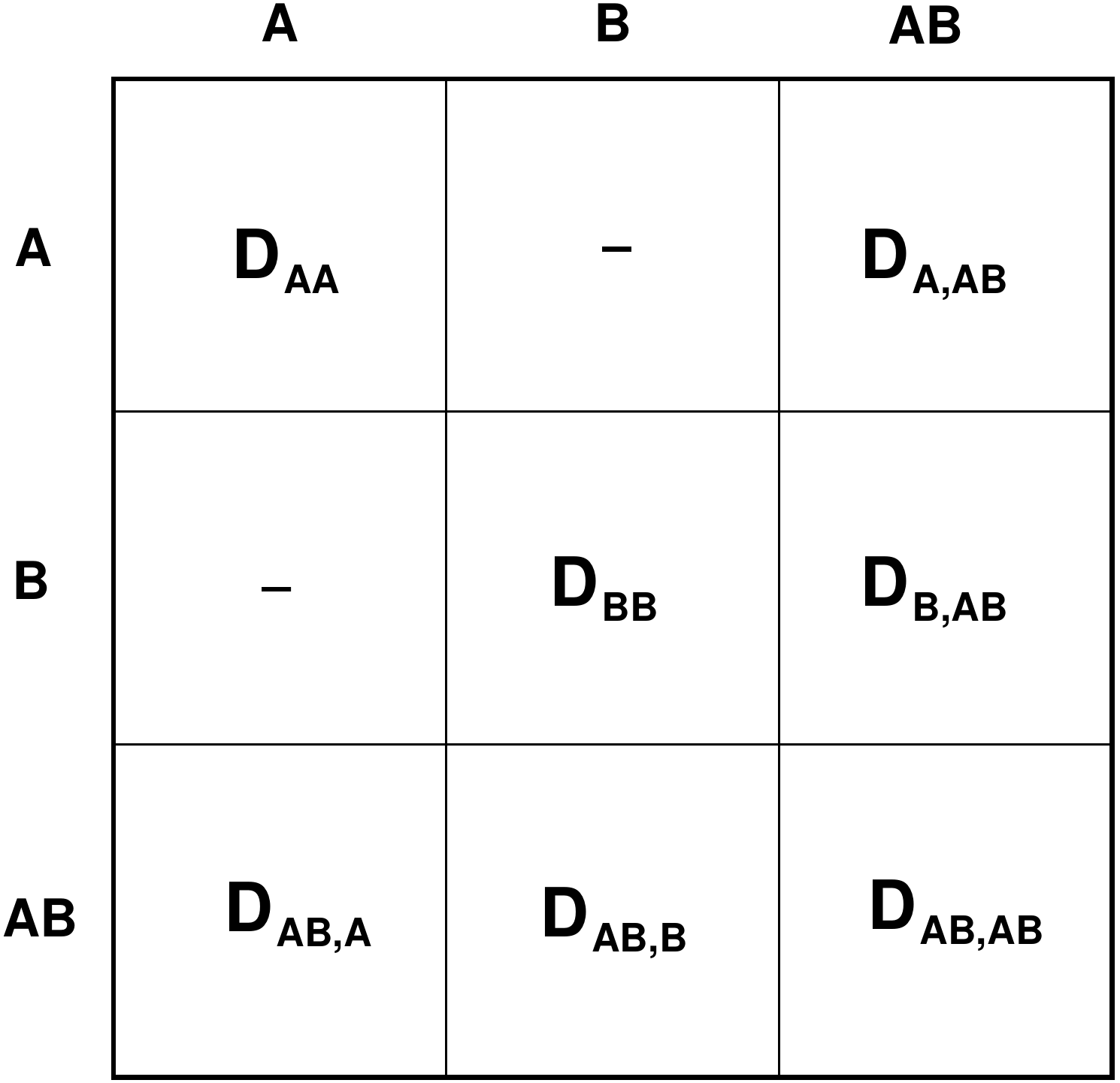}
\end{center}
\textbf{Fig. 6}
\newpage
\includegraphics[height=10cm]{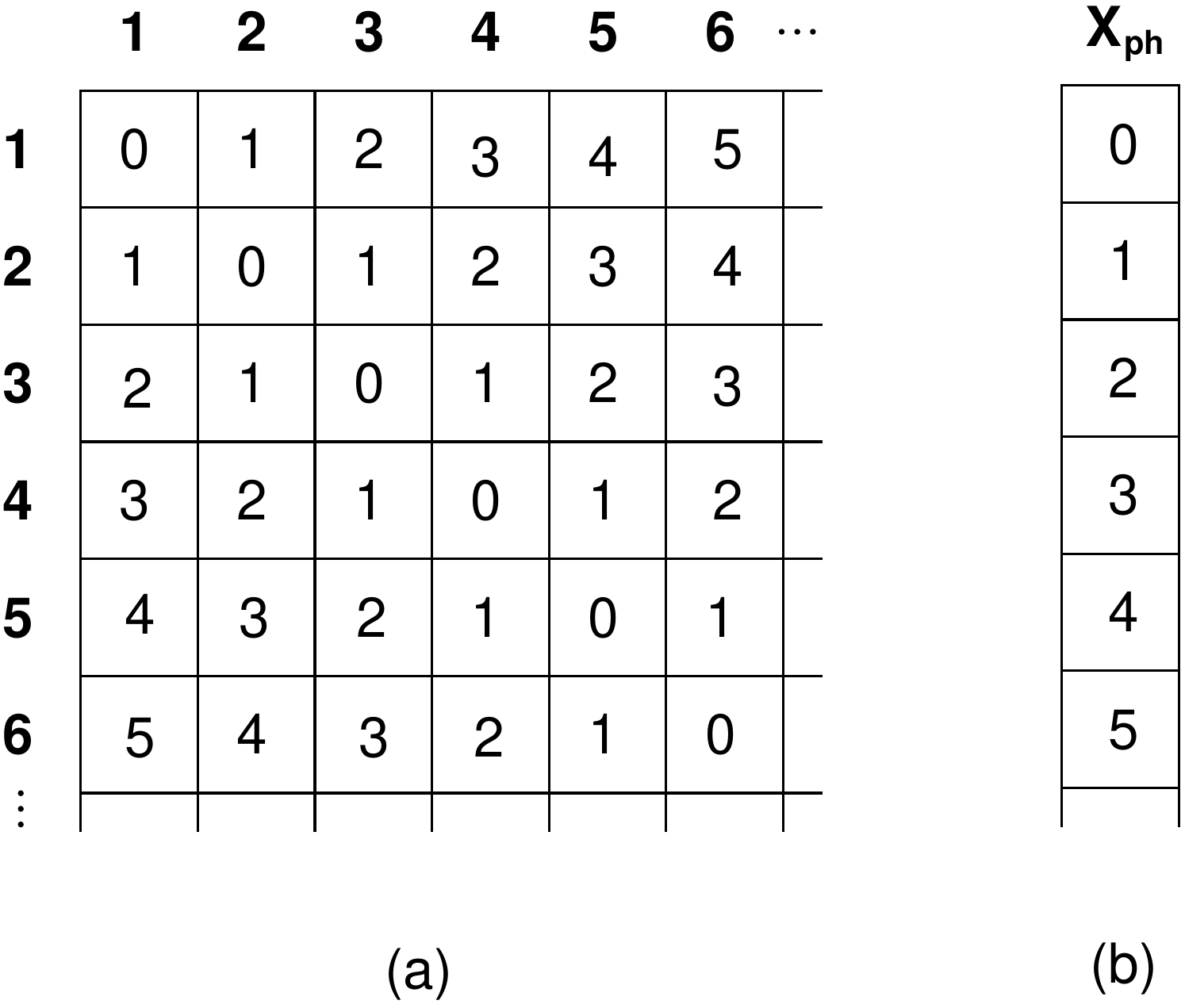}
\\
\textbf{Fig. 7}
\\
\\
\\
%\newpage
\includegraphics[height=10cm]{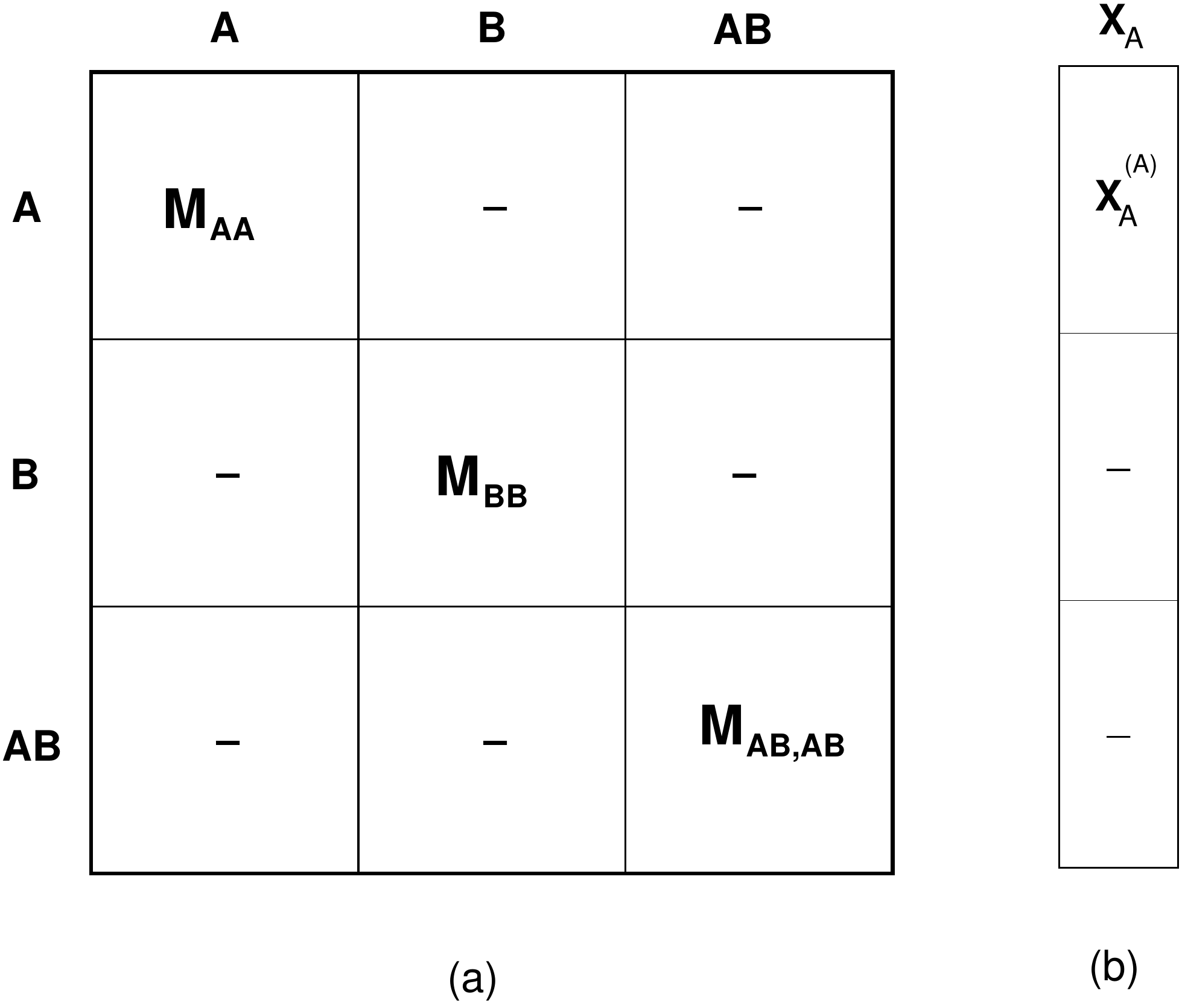}
\\
\textbf{Fig. 8}
\newpage
\includegraphics[height=9cm]{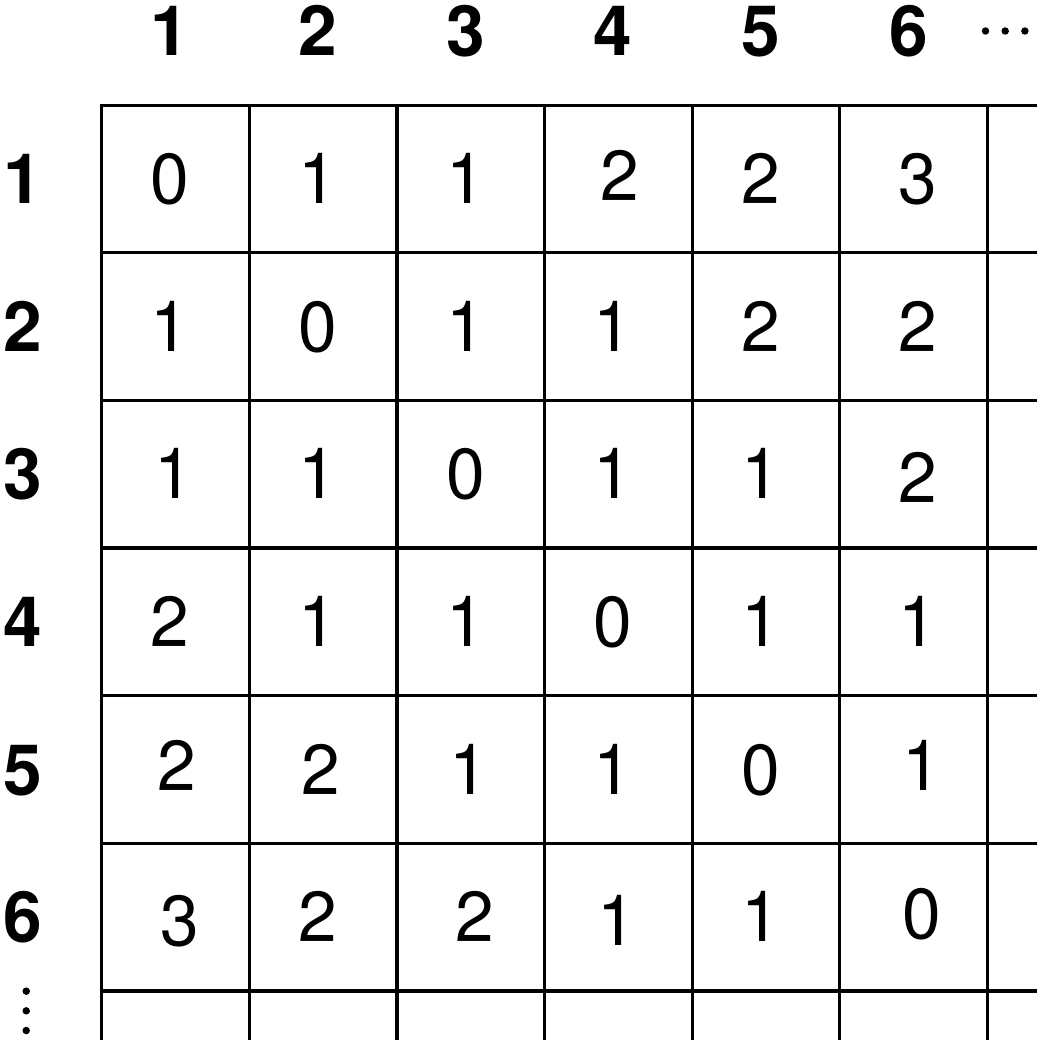}
\\
\\
\textbf{Fig. 9}
\\
%\newpage
\begin{center}
\includegraphics[height=9cm]{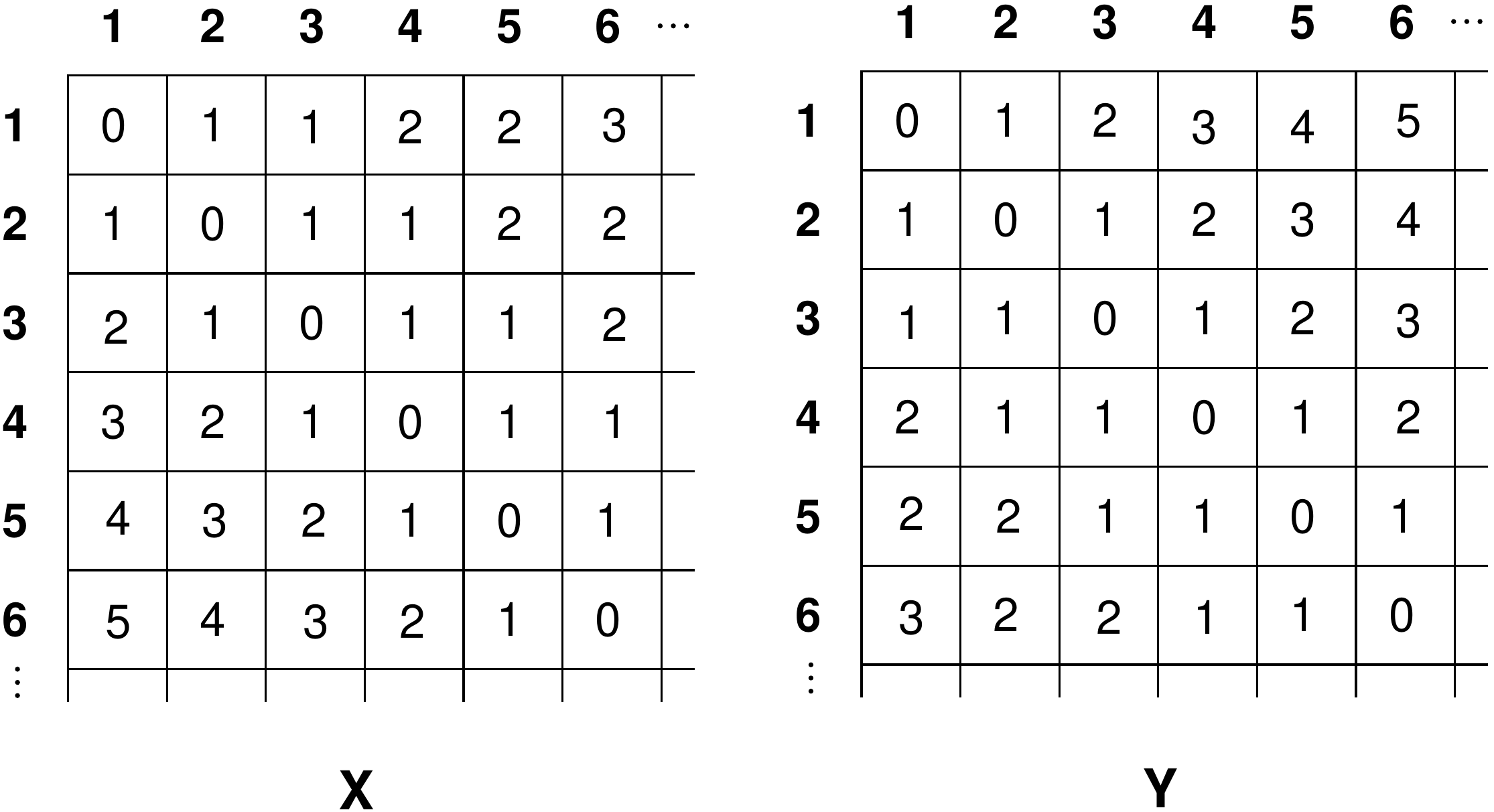}
\end{center}
\textbf{Fig. 10}

\end{document}